\documentclass[aps,prx,superscriptaddress,singlecolumn,showkeys,notitlepage,nofootinbib,11pt,tightenlines,longbibliography]{revtex4-2}

\usepackage{array}    
\usepackage[toc,page]{appendix}

\makeatletter
\renewcommand{\p@subsection}{}
\renewcommand{\p@subsubsection}{}
\makeatother

\usepackage[table]{xcolor}
\newcommand{\nocontentsline}[3]{}
\newcommand{\tocless}[2]{\bgroup\let\addcontentsline=\nocontentsline#1{#2}\egroup}
\usepackage{graphicx, color, graphpap}
\usepackage{enumitem}
\usepackage{amssymb}
\usepackage{amsthm}
\usepackage[OT1]{fontenc} 
\usepackage{multirow}
\usepackage[margin=0.9in]{geometry}
\usepackage[colorlinks=true,citecolor=blue,linkcolor=blue]{hyperref}
\usepackage[T1]{fontenc}

\usepackage{bbm}
\usepackage{thmtools,thm-restate}
\usepackage{verbatim}
\usepackage{mathtools}
\usepackage{titlesec}
\usepackage{amsmath}
\usepackage{tikz}
\usepackage{soul}
\usetikzlibrary{quantikz}
\usepackage[caption = false]{subfig}
\usepackage{float}

\usepackage{tabularx}
\usepackage{lmodern,adjustbox}
\usepackage[skins,breakable]{tcolorbox}
\tcbuselibrary{listings}
\tcbuselibrary{breakable}
\newtcolorbox{Code}{enhanced,fonttitle=\sffamily\bfseries\large,valign=center
, drop fuzzy shadow,sidebyside,lefthand ratio=0.4,lower separated=false}

\long\def\ca#1\cb{} 

\renewcommand{\leq}{\leqslant}

\newcolumntype{s}{>{\columncolor[HTML]{AAACED}} p{3cm}}

\renewcommand{\vec}[1]{\boldsymbol{#1}}  

\newcommand{\thv}{\vec{\theta}}

{}
{}

\begin{document}

\title{Graph Neural Networks on Quantum Computers}

\author{Yidong Liao}
\email{yidong.liao@student.uts.edu.au}
\affiliation{Centre for Quantum Software and Information, University of Technology Sydney, Sydney, NSW, Australia}
\affiliation{Sydney Quantum Academy, Sydney, NSW, Australia}

\author{Xiao-Ming Zhang}
\affiliation{Center on Frontiers of Computing Studies, School of Computer Science, Peking University, Beijing, China}
\affiliation{School of Physics, South China Normal University, Guangzhou, China}

\author{Chris Ferrie} \email{christopher.ferrie@uts.edu.au}
\affiliation{Centre for Quantum Software and Information, University of Technology Sydney, Sydney, NSW, Australia}

\begin{abstract}
Graph Neural Networks (GNNs) are powerful machine learning models that excel at analyzing structured data represented as graphs, demonstrating remarkable performance in applications like social network analysis and recommendation systems. However, classical GNNs face scalability challenges when dealing with large-scale graphs. This paper proposes frameworks for implementing GNNs on quantum computers to potentially address the challenges. We devise quantum algorithms corresponding to the three fundamental types of classical GNNs: Graph Convolutional Networks, Graph Attention Networks, and Message-Passing GNNs. A complexity analysis of our quantum implementation of the Simplified Graph Convolutional (SGC) Network shows potential quantum advantages over its classical counterpart, with significant improvements in time and space complexities. Our complexities can have trade-offs between the two: when optimizing for minimal circuit depth, our quantum SGC achieves logarithmic time complexity in the input sizes (albeit at the cost of linear space complexity). When optimizing for minimal qubit usage, the quantum SGC exhibits space complexity logarithmic in the input sizes, offering an exponential reduction compared to classical SGCs, while still maintaining better time complexity. These results suggest our Quantum GNN frameworks could efficiently process large-scale graphs. This work paves the way for implementing more advanced Graph Neural Network models on quantum computers, opening new possibilities in quantum machine learning for analyzing graph-structured data.
\end{abstract}
\maketitle

\setlength{\arrayrulewidth}{0.2pt} 

\begin{figure}[h!]
    \centering
\includegraphics[width=0.86\linewidth]{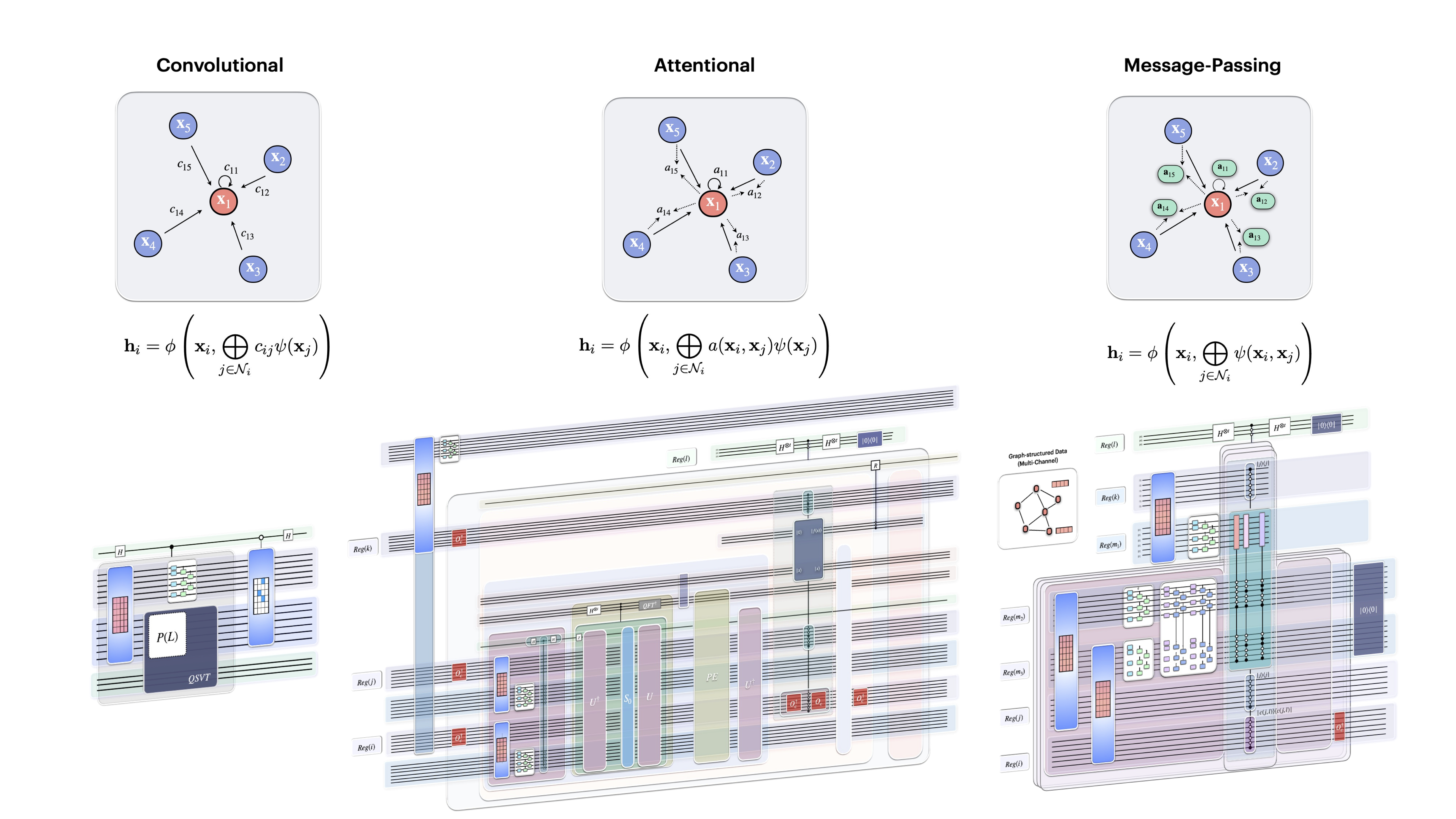}
   \caption{\textit{Overall circuit construction for the three Quantum GNN architectures along with the three ``flavours''of classical GNN layers}\cite{bronstein2021geometric}. }
    \label{intro2}
\end{figure}

\tableofcontents

\section{Introduction}\label{intro}
Graph Neural Networks (GNNs) are powerful machine learning models for analyzing structured data represented as graphs. They have shown remarkable success in various applications including social network analysis \cite{borisyuk2024lignn,fan2019graph}, recommendation systems \cite{jain2019food,de2024personalized}, drug discovery \cite{stokes2020deep,zitnik2018modeling}, and traffic prediction \cite{derrow2021eta}. From a theoretical perspective, GNNs have been posited as a universal framework for various neural network architectures: Convolutional Neural Networks (CNNs), Recurrent Neural Networks (RNNs), and Transformers, etc. can be viewed as special cases of GNNs \cite{bronstein2021geometric,Bronstein2023towards, joshi2020transformers}. \newline

Despite their success, classical GNNs face several challenges when dealing with large-scale graphs. One major challenge is the memory limitations that arise when handling giant graphs. Large and complex graphs become increasingly difficult to fit in the conventional memory used by most classical computing hardware~\cite{chiang2019cluster}. Another issue lies in the inherent sparse matrix operations of GNNs, which pose challenges for efficient computation on modern hardware like GPUs that are optimized for dense matrix operations\footnote{Customized hardware accelerators for sparse matrices can improve GNNs' latency and scalability, but their design remains an open question \cite{joshi2022efficientgnns}.} \cite{joshi2022efficientgnns}. Moreover, the common method of managing large graphs through graph subsampling techniques (e.g. dividing large graphs into smaller, more manageable subgraphs \cite{chiang2019cluster}) may encounter reliability issues, since it is challenging to guarantee that these subgraphs preserve the semantics of the entire graph and provide reliable gradients for training GNNs \cite{joshi2022efficientgnns}. In summary, the memory and computational requirements of processing large-scale graphs often exceed the capabilities of classical computing hardware, limiting the practical scalability of GNNs. The need for efficient and scalable graph learning has motivated ongoing efforts in developing specialized hardware accelerators for GNNs \cite{kiningham2020grip,auten2020hardware,abadal2021computing} as well as the exploration of utilizing alternative computing paradigms, such as quantum computing, to address these challenges.\newline

Quantum computers hold the promise of significantly improving machine learning by providing computational speed-ups or improved model scalability \cite{cerezo2022challenges,Perdomo_Ortiz_2018,coles2021seeking,gao2018quantum}. In the context of graph learning, quantum computing provides new opportunities to design quantum machine learning architectures tailored for graph-structured data \cite{beer2021quantum,skolik2023equivariant,verdon2019quantumgraph,mernyei2022equivariant}. Motivated by this potential, we propose quantum GNN architectures in accordance with the three fundamental types of classical GNNs: Graph Convolutional Networks (GCNs) (e.g.\cite{kipf2017semi}), Graph Attention Networks (GATs) (e.g.\cite{velickovic2017graph}), and Message-Passing Neural Networks (MPNNs) (e.g.\cite{gilmer2017neural}). \newline

The complexity analysis in our paper demonstrates that our quantum implementation of a Simplified Graph Convolution (SGC) network can potentially achieve significant improvements in time and/or space complexity compared to its classical counterpart, under certain conditions commonly encountered in real-world applications. When optimizing for the minimal number of qubits, the quantum SGC exhibits a space complexity that is logarithmic in the input sizes, offering an exponential reduction compared to the classical SGC\footnote{In this case, the quantum SGC still provides better time complexity than its classical counterpart, particularly for graphs with a large number of nodes and high-dimensional node features.}. On the other hand, when optimizing for minimal circuit depth, our quantum SGC provides a substantial improvement in time complexity, achieving logarithmic dependence on the input sizes\footnote{This improvement comes with a trade-off in space complexity, which is comparable to that of the classical SGC.}. These complexity results suggest that our quantum implementation of the SGC has the potential to efficiently process large-scale graphs, under certain assumptions that align with practical use cases. The trade-off between circuit depth and the number of qubits in the quantum implementation provides flexibility in adapting to specific quantum hardware constraints and problem instances, making it a promising approach for tackling complex graph-related machine learning tasks.\newline

The rest of the paper is organized as follows. In Section \ref{gnn}, we provide an overview of classical GNNs. In Section \ref{22}, \ref{23}, and \ref{24}, we present our quantum algorithms for GCNs, GATs, and MPNNs, respectively. We analyze the complexity of our Quantum implementation of two GCN variants in Section \ref{complexity} and discuss the potential advantages in Section \ref{sgc} and \ref{lgc}. Finally, we conclude the paper and outline future research directions in Section \ref{conclusion}.\newline

Before diving into our QNN architectures, it is worth noting that our work also falls within the emerging field of Geometric Quantum Machine Learning (GQML) \cite{ragone2022representation,larocca2022group, meyer2022exploiting, zheng2021speeding, sauvage2022building}, which aims to create quantum machine learning models that respect the underlying structure and symmetries of the data they process. To illustrate how our frameworks align with the principles of GQML, we present an overview of our approach for Quantum Graph Convolutional Networks, summarized in Fig.~\ref{frgnn}. This example demonstrates how our Quantum GNNs incorporate inductive biases to process graph-structured data, potentially leading to improvements compared to problem-agnostic quantum machine learning models. \newline

\section{Classical Graph Neural Networks}
\label{gnn}

Following Ref.~\cite{bronstein2021geometric,velivckovic2023everything,battaglia2018relational}, we provide a brief introduction to classical Graph Neural Networks, which serve as the foundation for the development of our quantum GNNs.\newline

Graphs are a natural way to represent complex systems of interacting entities. Formally, a graph $G = (V, E)$ consists of a set of nodes $V$ and a set of edges $E \subseteq V \times V$ that connect pairs of nodes. In many real-world applications, graphs are used to model relational structure, with nodes representing entities (e.g., users, proteins, web pages) and edges representing relationships or interactions between them (e.g., friendships, molecular bonds, hyperlinks). To enable rich feature representations, nodes are often endowed with attribute information in the form of real-valued feature vectors. Given a graph with $N = |V|$ nodes, we can summarize the node features as a matrix $\mathbf{X} \in \mathbb{R}^{N \times C}$, where the $u$-th row $\mathbf{x}_u \in \mathbb{R}^C$ corresponds to the $C$-dimensional feature vector of node $u$. The connectivity of the graph can be represented by an adjacency matrix $\mathbf{A} \in \mathbb{R}^{N \times N}$, where $a_{uv} = 1$ if there is an edge between nodes $u$ and $v$, and $a_{uv} = 0$ otherwise.\newline

Graph Neural Networks (GNNs) are a family of machine learning models that operate on the graph structure $(\mathbf{X}, \mathbf{A})$. The key defining property of GNNs is \textit{permutation equivariance}. Formally, let $\mathbf{P} \in \{0,1\}^{N \times N}$ be an permutation matrix. A GNN layer, denoted by $\textbf{F}(\mathbf{X}, \mathbf{A})$, is a permutation-equivariant function in the sense that:
$$\textbf{F}(\mathbf{P}\mathbf{X}, \mathbf{P}\mathbf{A}\mathbf{P}^\top) = \mathbf{P}\textbf{F}(\mathbf{X}, \mathbf{A})$$

Permutation equivariance is a desirable inductive bias for graph representation learning, as it ensures that the GNN output will be invariant to arbitrary reorderings of the nodes. This property arises naturally from the unordered nature of graph data, i.e., a graph is intrinsically defined by its connectivity and not by any particular node ordering.\newline

 In each GNN layer, nodes update their features by aggregating information from their local neighborhoods ((undirected) neighbourhood of node $u$ is defined as $\mathcal{N}_{u} = \{v | (u,v) \in E \textit{or} \ (v,u) \in E \}$). This local computation is performed identically (i.e., shared) across all nodes in the graph. Mathematically, a GNN layer computes a new feature matrix $\mathbf{H} \in \mathbb{R}^{N \times C'}$ from the input features $\mathbf{X}$ as follows:
\begin{equation}
\mathbf{H} = \textbf{F}(\mathbf{X}, \mathbf{A}) = [\phi(\mathbf{x}_1, \mathbf{X}_{\mathcal{N}_1}), \phi(\mathbf{x}_2, \mathbf{X}_{\mathcal{N}_2}), ..., \phi(\mathbf{x}_N, \mathbf{X}_{\mathcal{N}_N})]^{\top}
\end{equation}
where $\phi$ is a local function often called the \textit{neighborhood aggregation} or \textit{message passing} function, and $\mathbf{X}_{\mathcal{N}_{u}} = \{\!\!\{\mathbf{x}_v \ | \ v \in \mathcal{N}_{u}\}\!\!\} $ denotes the multiset of all neighbourhood features of node $u$. In other words, the new feature vector $\mathbf{h}_{u}:=\phi(\mathbf{x}_u, \mathbf{X}_{\mathcal{N}_u})$ of node $u$ is computed by applying $\phi$ to the current feature $\mathbf{x}_u$ and the features of its neighbors $\mathbf{X}_{\mathcal{N}_{u}}$. Since $\phi$ is shared across all nodes and only depends on local neighborhoods, it can be shown that if $\phi$ is permutation invariant in $\mathbf{X}_{\mathcal{N}_{u}}$, then $\textbf{F}$ will be permutation equivariant. Stacking multiple GNN layers allows information to propagate over longer graph distances, enabling the network to capture high-order interaction effects. \newline 

\begin{figure}[h!]
    \centering
    \includegraphics[width=\linewidth]{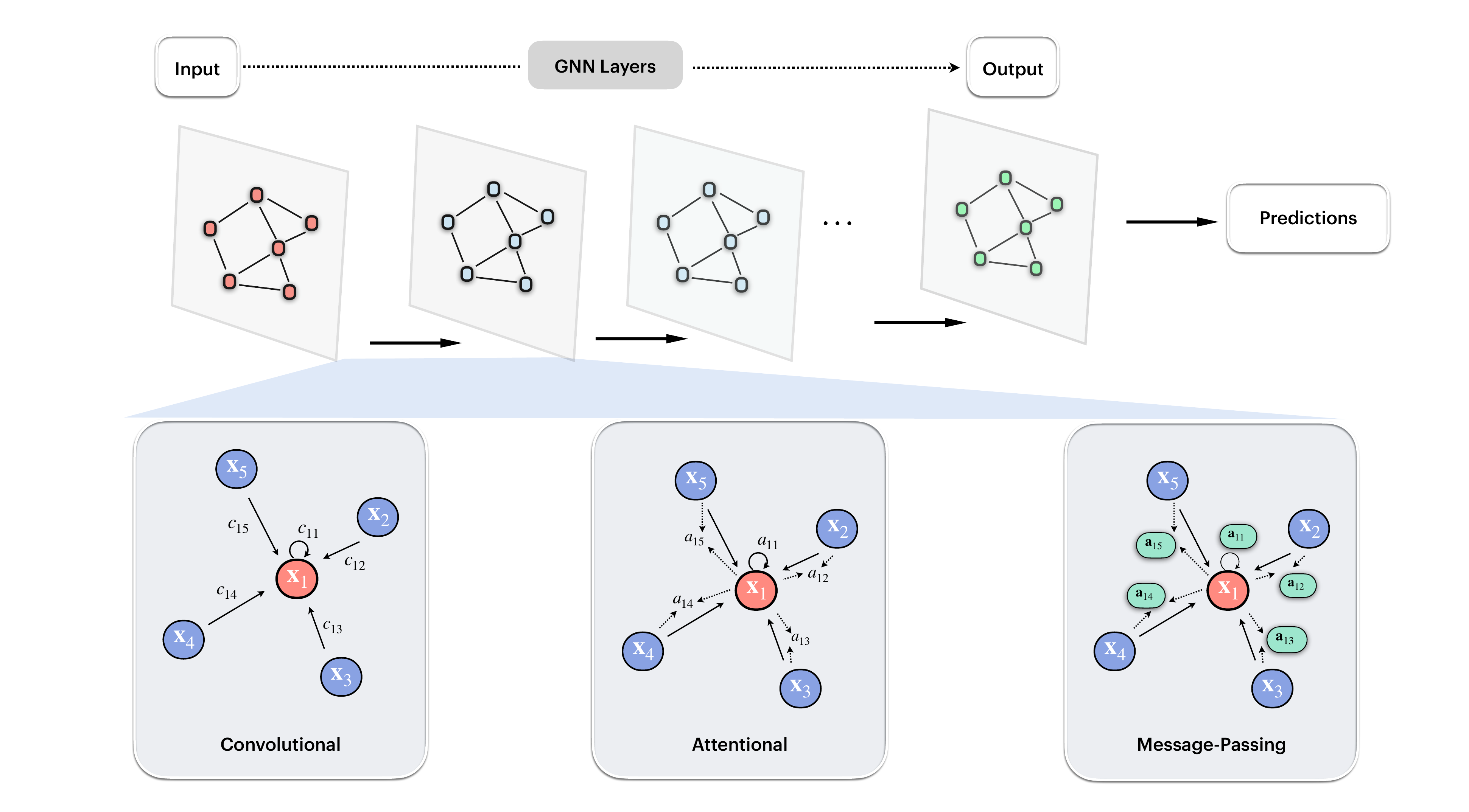}
   \caption{\textit{GNN pipeline and three ``flavours'' of GNN layers\cite{bronstein2021geometric}} GNN architectures are permutation equivariant functions $\mathbf{F}(\mathbf{X}, \mathbf{A})$ constructed by applying shared permutation invariant functions $\phi$ over local neighbourhoods. This local function $\phi$ is sometimes referred to as ``diffusion,'' ``propagation,'' or ``message passing,'' and the overall computation of such $\mathbf{F}$ is known as a ``GNN layer.'' These flavours determine the extent to which $\phi$ transforms the neighbourhood features, allowing for varying levels of complexity when modelling interactions across the graph.}
    \label{fig:gnns}
\end{figure}

 While the general blueprint of GNNs based on local neighborhood aggregation is quite simple and natural, there are many possible choices for the aggregation function $\phi$. The design and study of GNN layers is a rapidly expanding area of deep learning, and the literature can be divided into three ``flavours'' \cite{bronstein2021geometric}: convolutional, attentional, and message-passing (see Figure \ref{fig:gnns}). These flavours determine the extent to which $\phi$ transforms the neighbourhood features, allowing for varying levels of complexity when modelling interactions across the graph.\newline

In the convolutional flavour (e.g.\cite{kipf2017semi}), the features of the neighbouring nodes are directly combined with fixed weights, $$ \mathbf{h}_{u}=\phi\left(\mathbf{x}_{u}, \bigoplus_{v \in \mathcal{N}_{u}} c_{u v} \psi\left(\mathbf{x}_{v}\right)\right). $$ Here, $c_{u v}$ is a constant indicating the significance of node $v$ to node $u^{\prime}$ s representation. $\bigoplus$ is the aggregation operator which is often chosen to be the summation. $\psi$ and $\phi$ are learnable transformations\footnote{Note we omitted the activation function in the original definition in \cite{bronstein2021geometric}, the quantum implementation of the activation function is described in Section \ref{mgcn}. And for simplicity, we omit $\mathbf{b} $ in our quantum case.}: $\psi(\mathbf{x})=\mathbf{W} \mathbf{x}+\mathbf{b}$, $\phi(\mathbf{x}, \mathbf{z})=\mathbf{W} \mathbf{x}+\mathbf{U} \mathbf{z}+\mathbf{b}$.  \newline

In the attentional flavour (e.g.\cite{velickovic2017graph}), 
$$\mathbf{h}_{u}=\phi\left(\mathbf{x}_{u}, \bigoplus_{v \in \mathcal{N}_{u}} a\left(\mathbf{x}_{u}, \mathbf{x}_{v}\right)\psi\left(\mathbf{x}_{v}\right)\right),$$

a learnable self-attention mechanism is used to compute the coefficients $a\left(\mathbf{x}_{u}, \mathbf{x}_{v}\right)$. When $\bigoplus$ is the summation, the aggregation is still a linear combination of the neighbourhood node features, but the weights are now dependent on the features.\newline

Finally, the message passing flavour (e.g.\cite{gilmer2017neural}) involves computing arbitrary vectors (``messages'') across edges, $$ \mathbf{h}_{u}=\phi\left(\mathbf{x}_{u}, \bigoplus_{v \in \mathcal{N}_{u}} \psi\left(\mathbf{x}_{u}, \mathbf{x}_{v}\right)\right) . $$ Here, $\psi$ is a trainable message function, which computes the vector sent from $v$ to $u$, and the aggregation can be considered as a form of message passing on the graph.\newline

The three GNN flavors -- convolutional, attentional, and message-passing -- offer increasing levels of expressivity, albeit comes at the cost of reduced scalability. The choice of GNN flavor for a given task requires carefully considering this trade-off and prioritizing the most important desiderata for the application at hand.\newline

Classical GNNs have been shown to be highly effective in a variety of graph-related tasks including~\cite{wu2022graph,velivckovic2023everything}:\newline

1.\textsf{[Node classification]}, where the goal is to assign labels to nodes based on their attributes and the graph structure. For example, in a social network, the task could be to classify users into different categories by leveraging their profile information and social connections. In a biological context, a canonical example is classifying protein functions in a protein-protein interaction network~\cite{hamilton2017inductive}. \newline

2.\textsf{[Link prediction]}, where the objective is to predict whether an edge exists between two nodes in the graph, or predicting the properties of the edges. In a social network, this could translate to predicting potential interactions between users. In a biological context, it could involve predicting links between drugs and diseases—drug repurposing~\cite{morselli2021network}.\newline

3.\textsf{[Graph classification]}, where the goal is to classify entire graphs based on their structures and attributes. A typical example is classifying molecules in terms of their quantum-chemical properties, which holds significant promise for applications in drug discovery and materials science~\cite{gilmer2017neural}. \newline

As aforementioned in section \ref{intro}, despite their success, classical GNNs also face challenges in scalability. This motivates our exploration
of utilizing quantum computing to address the challenges. \newline

 In the following three sections of this paper, we will devise and analyze QNN architectures in accordance with the three major types of classical GNNs(corresponding to the three flavours): Graph Convolutional Networks, Graph Attention Networks, Message-Passing GNNs. We term our QNN architectures as Quantum Graph Convolutional Networks, Quantum Graph Attention Networks, and Quantum Message-Passing GNNs which fall into the research area of Quantum Graph Neural Networks.

 \section{Quantum Graph Convolutional Networks}\label{22}
 
 \subsection{Vanilla GCN and its Quantum version }

In this section, we present our quantum algorithm for the problem of node classification with Graph Convolutional Networks (GCN) \cite{kipf2017semi}. We start by restating some notations: Let ${G}=({V}, {E})$ be a graph, where ${V}$ is the set of nodes and ${E}$ is the set of edges. $A \in \mathbb{R}^{N \times N}$ is the adjacency matrix, with $N$ being the total number of nodes, and $X \in \mathbb{R}^{N \times C}$ is the node attribute matrix, with $C$ being the number of features for each node. The node representations at the $l$-th layer is denoted as $H^{(l)} \in \mathbb{R}^{N \times F_l}, l \in\{0,1,2, \cdots, K\}$, where $F_l$ being the dimension of node representation for each node. These notations are summarised in the following table.

\begin{center}
\begin{tabular}{c|c}
\hline\hline
Concept & Notation \\
\hline\hline
Graph & $ {G}=( {V},  {E})$ \\
\hline
Adjacency matrix & $A \in \mathbb{R}^{N \times N}$ \\
Node attributes & $X \in \mathbb{R}^{N \times C}$ \\
\hline
Total number of GCN layers & $\mathrm{K}$ \\
\hline
Node representations at the $l$-th layer & $H^{(l)} \in \mathbb{R}^{N \times F_l}, l \in\{0,1,2, \cdots, K\}$ \\
\hline\hline
\end{tabular}
    
\end{center}

The GNN layer (described in Section \ref{gnn}) in a Graph Convolutional Network, often termed ``Graph Convolution,'' can be carried out as\cite{kipf2017semi}:

\begin{equation}
H^{(l+1)}=\sigma\left(\hat{A} H^{(l)} W^{(l)}\right)
\label{gcneq}
\end{equation}

Here, $\hat{A}=\tilde{D}^{-\frac{1}{2}} \tilde{A} \tilde{D}^{-\frac{1}{2}}$ in which $\tilde{A}=A+I_{N}$ is the adjacency matrix of the graph $G$ with added self-connections($I_{N}$ is the identity matrix), $\tilde{D}_{i i}=\sum_{j} \tilde{A}_{i j}$, and $W^{(l)}$ is a layer-specific trainable weight matrix. $\sigma(\cdot)$ denotes an activation function. \newline

At the output of the last layer, the softmax function, defined as $\operatorname{softmax}\left(x_{i}\right)=\frac{1}{\mathcal{Z}} \exp \left(x_{i}\right)$ with $\mathcal{Z}=\sum_{i} \exp \left(x_{i}\right)$, is applied row-wise to the node feature matrix, producing the final output of the network:
\begin{equation}
    Z=\text{softmax}(\hat{A} H^{(K-1)} W^{(K-1)})
\end{equation}

\begin{figure}[h!]
    \centering
    \includegraphics[width=\linewidth]{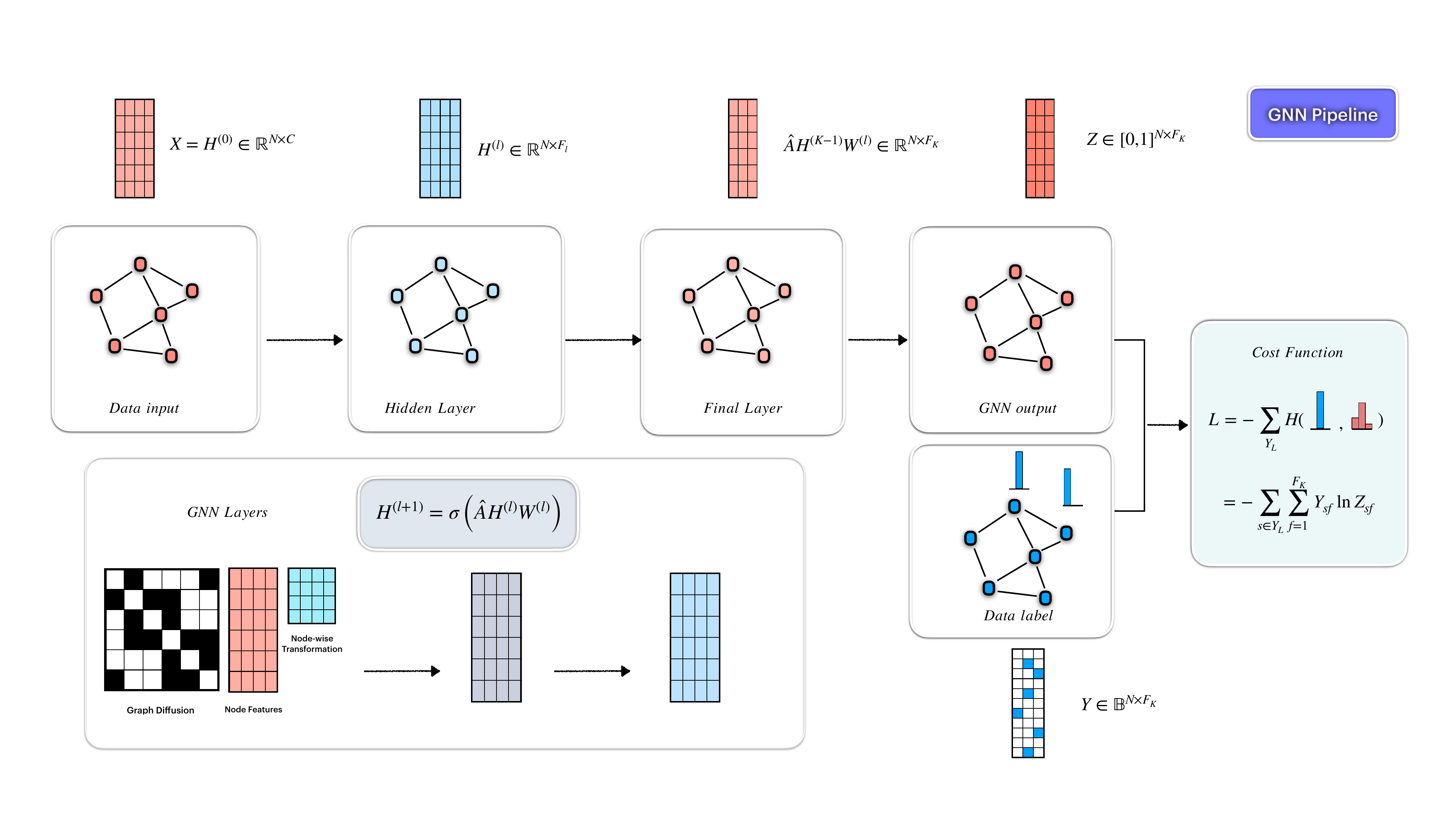}
    \caption{\textit{GCN Pipeline}. A GCN consists of a series of layers in which graph convolution and non-linear activation functions are applied to the node features. (Note that the schematics in this figure are for illustration purposes only, e.g. the normalized adjacency matrix depicted here does not include the added self-connections)  At the output of the last layer, softmax activation function, defined as $\operatorname{softmax}\left(x_{i}\right)=\frac{1}{\mathcal{Z}} \exp \left(x_{i}\right)$ with $\mathcal{Z}=\sum_{i} \exp \left(x_{i}\right)$, is applied row-wise to the node feature matrix, producing the final output of the network: $Z=\text{softmax}(\hat{A} H^{(K-1)} W^{(K-1)})$. For semi-supervised multi-class classification, the cost function is defined by the cross-entropy error over all labelled examples \cite{kipf2017semi}:$L=-\sum_{s \in {Y}_{L}} \sum_{f=1}^{F_K} Y_{s f} \ln Z_{s f}$, where ${Y}_{L}$ is the set of node indices that have labels, $Y\in \mathbb{B}^{N \times F_K}$ denotes the one-hot encoding of the labels. }
    \label{Pipeline}
\end{figure}

For semi-supervised multi-class classification, the cost function is defined by the cross-entropy error over all labelled examples \cite{kipf2017semi}:

\begin{equation}
   L=-\sum_{s \in {Y}_{L}} \sum_{f=1}^{F_K} Y_{s f} \ln Z_{s f},
   \label{cost}
\end{equation}

where ${Y}_{L}$ is the set of node indices that have labels, $Y\in \mathbb{B}^{N \times F_K}$ denotes the one-hot encoding of the labels. The GCN pipeline mentioned above is summarised in Fig.\ref{Pipeline}.\newline

Next, we present the Quantum implementation of GCN.

\subsubsection{Data Encoding}\label{de}

For GCN, the node features $X \in \mathbb{R}^{N \times C}$ of which the entries are denoted as $X_{ik}$, can be encoded in a quantum state $\left|\psi_{X}\right\rangle$ (after normalization)\footnote{Note throughout this paper we often omit the normalization factors in quantum states.} as follows:

\begin{equation}
\left|\psi_{X}\right\rangle=\sum_{i=1}^{N} |i\rangle\ket{\bold{x}_{i}}
\label{multistate}
\end{equation}

where $\ket{\bold{x}_{i}}=\sum_{k=1}^{C}X_{ik}|k\rangle$ , being the amplitude encoding of the features for node $i$ over the channels(indexed by $k$), is entangled with an address state $\ket{i}$. The entire state is prepared on two quantum registers hosting the channel index $k$ and node index $i$, which are denoted as $Reg(k)$ and $Reg(i)$, respectively. The data encoding, represented as the blue box in Fig.~\ref{Multi22}, can be achieved by various quantum state preparation procedures~\cite{Long.01,Grover.02,Mottonen.05,Plesch.11,Zhang.21,Sun.21,Rosenthal.21,zhang2022quantum,Clader.22,yuan2023optimal,Gui.23,zhang2024circuit}. We choose the method from Ref. \cite{zhang2024circuit} for our data encoding, as their work provides a tunable trade-off between the number of ancillary qubits and the circuit depth for state preparation.

\subsubsection{Layer-wise transformation}\label{mgcn}

The layer-wise linear transformation for multi-channel GCN (i.e. $H'^{(l)}=\hat{A} H^{(l)} W^{(l)}$ \footnote{Here we use $H'^{(l)}$ to denote the linearly transformed feature matrix.}), can be implemented by applying the block-encoding\footnote{Appendix \ref{blocke} provides a brief introduction of block-encoding.} of $\hat{A}$ and a parameterized quantum circuit implementing $W^{(l)}$ on the two quantum registers $Reg(i)$ and $Reg(k)$ respectively, as depicted in Fig.~\ref{Multi22}. This is proved in Appendix~\ref{proofappendix}.\newline

\begin{figure}[h!]
    \centering
    \includegraphics[width=\linewidth]{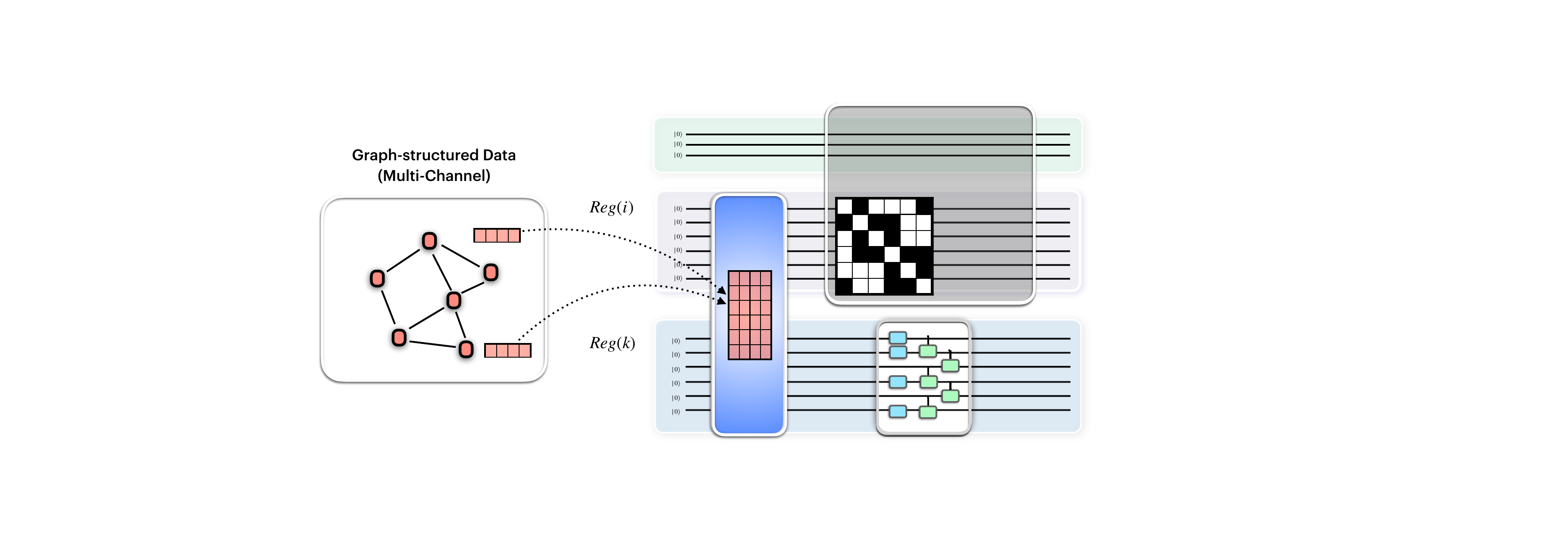}
   \caption{\textit{Quantum implementation of linear layer-wise transformation for multi-channel GCN} The linear layer-wise transformation for multi-channel GCN (i.e. the layer-specific trainable weight matrix and the normalized adjacency matrix multiplied on the node feature matrix), can be implemented by applying the block-encoding of the normalized adjacency matrix and a parametrized quantum circuit on the two quantum registers $Reg(i)$ and $Reg(k)$ respectively. Here we depicted the first layer of GCN --- the linear layer-wise transformation is applied on the state prepared by the data encoding procedure (the blue box) described in Section \ref{de}. Note that the schematics in this figure are for illustration purposes only, e.g. 1) the normalized adjacency matrix depicted here does not include the added self-connections; 2) the ancillary qubits used in the quantum state preparation for the data encoding is not depicted in this figure.}
    \label{Multi22}
\end{figure}

After the linear layer-wise transformation, the element-wise non-linear activation function can be applied using an established technique called Nonlinear Transformation of Complex Amplitudes (NTCA)\cite{guo2021nonlinear}. One can also potentially utilize the techniques from Ref.~\cite{rattew2023non} for applying the non-linear activation function and achieve better performance, we leave this and the detailed analysis for future work. Ref.~\cite{kipf2017semi} considered a two-layer\footnote{It has been observed in many experiments that deeper models does not always improve performance and can even lead to worse outcomes compared to shallower models.~\cite{zhou2020graph}} GCN where the non-linear activation function is applied only once and the forward model takes the following form:

\begin{equation}
Z = \text{softmax}\left(\hat{A} \sigma\left(\hat{A}XW^{(0)}\right) W^{(1)}\right). 
\end{equation}

Next, we present the quantum state evolution for the quantum version of this two-layer GCN. Denote the block-encoding of $\hat{A}$ as $U_{\hat{A}}$ and the parameterized quantum circuit for $W^{(0)}$ as $U_{W^{0}}$, applying these operations on the quantum state $\left|\psi_{X}\right\rangle$ results in the following state:

\begin{equation}
\left|\psi_{H'^{(0)}}\right\rangle \otimes \ket{0} + \ldots  = \left(U_{\hat{A}} \otimes U_{W^{(1)}}\right) \left|\psi_{X}\right\rangle\otimes \ket{0},
\end{equation}
where $\left|\psi_{H'^{(0)}}\right\rangle=\sum_{i=1}^{N} |i\rangle\ket{\bold{h'}^{(0)}_{i}}$ is on the two quantum registers $Reg(i)$, $Reg(k)$ and $\ket{\bold{h'}^{(0)}_{i}}=\sum_{k=1}^{C}H'^{(0)}_{ik}|k\rangle$ is the amplitude encoding of the linearly transformed features for node $i$ over the channels(indexed by $k$).
The term ``$+\ldots$''\footnote{Throughout this paper, the terms ``$+\ldots$'' in the quantum states are consistently used as defined here.} represents a quantum state that is orthogonal to the state before the ``$+$'' sign.\newline

The quantum state $\left|\psi_{H'^{(l)}}\right\rangle$ encodes the linearly transformed node features. The block-encoding of $\hat{A}$ performs the aggregation of neighboring node features, while the parameterized quantum circuit $U_{W^{(1)}}$ applies the trainable weight matrix to the node features.\newline

Then, using NTCA to implement a non-linear activation function on the amplitudes of the state $\left|\psi_{H'^{(0)}}\right\rangle \otimes \ket{0} + \ldots $, we obtain the state  $\left|\psi_{H^{(1)}}\right\rangle \otimes \ket{0} + \ldots $ in which $\left|\psi_{H^{(1)}}\right\rangle=\sum_{i=1}^{N} |i\rangle\ket{\bold{h}^{(1)}_{i}}$ and $\ket{\bold{h}^{(1)}_{i}}=\sum_{k=1}^{C}H^{(1)}_{ik}|k\rangle=\sum_{k=1}^{C}\sigma(H'^{(0)}_{ik})|k\rangle$.\newline

\begin{figure}[h!]
    \centering
    \includegraphics[width=0.97\linewidth]{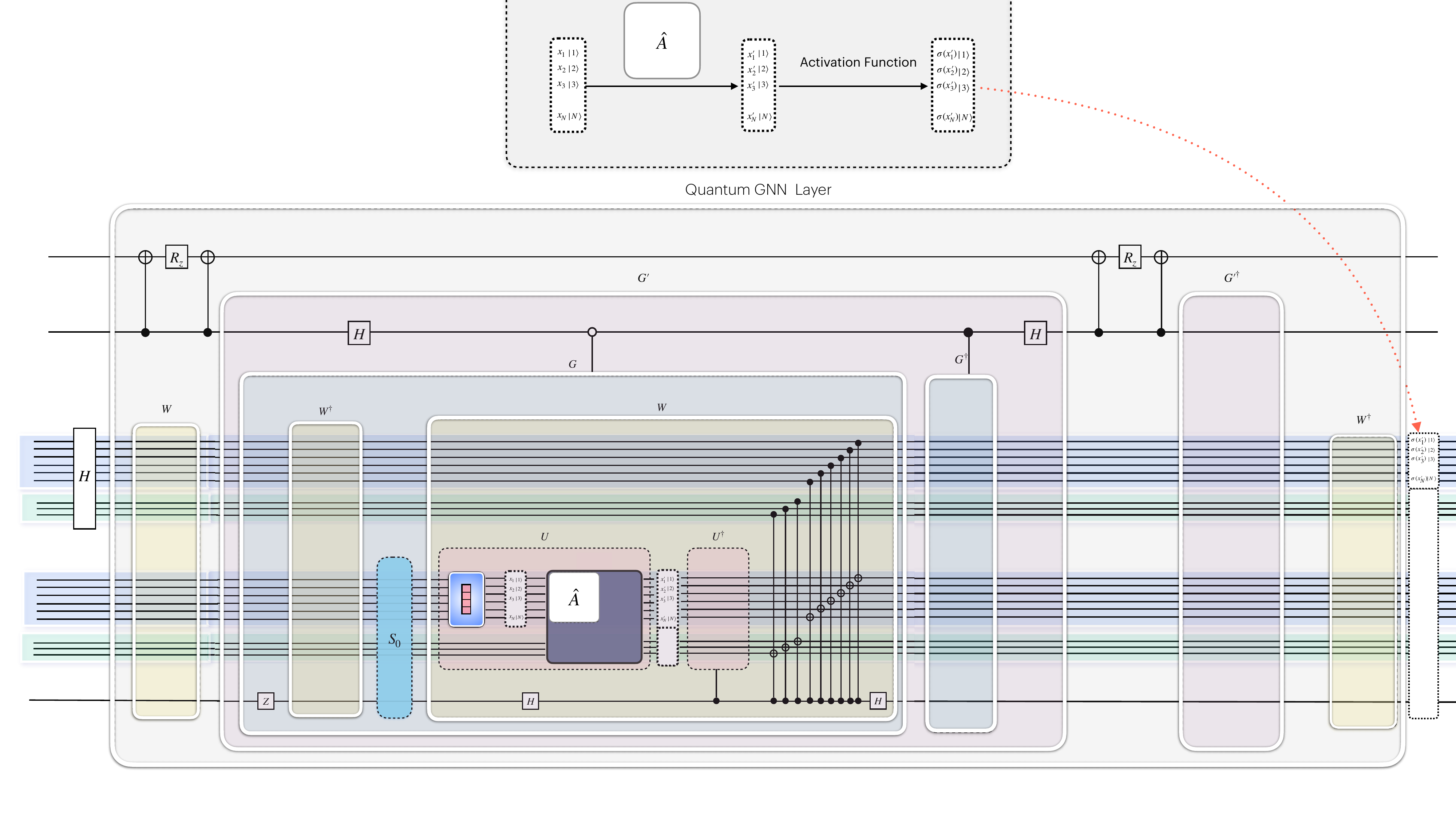}
    \caption{\textit{Example of the full quantum circuit for a GNN layer ($C=1$, single channel).} Utilising NTCA in our Quantum GCN to implement a non-linear activation function, we take the unitary of data encoding and graph convolution as components to build a new unitary that generates the desired state whose amplitudes are transformed by certain nonlinear functions. Note that the schematics in this figure are for illustration purposes only.}
    \label{Fullcircuit}
\end{figure}

An example of the full quantum circuit for the GNN layer ($C=1$, single channel) is depicted in Fig.\ref{Fullcircuit}.

Denote the parameterized quantum circuit for $W^{(1)}$ as $U_{W^{1}}$, applying $U_{\hat{A}}$ and $U_{W^{1}}$ on the quantum state the state  $\left|\psi_{H^{(1)}}\right\rangle \otimes \ket{0} + \ldots $ results in the state: $\left|\psi_{out}\right\rangle=\left|\psi_{H'^{(1)}}\right\rangle \otimes \ket{0} + \ldots $ in which $\left|\psi_{H'^{(1)}}\right\rangle=\sum_{i=1}^{N} |i\rangle\ket{\bold{h'}^{(1)}_{i}}$ and $\ket{\bold{h'}^{(1)}_{i}}=\sum_{k=1}^{C}H'^{(1)}_{ik}|k\rangle$.\newline

\subsubsection{Cost function}

For semi-supervised multi-class classification,  the cost function used in our QGCN is defined as the negative inner product between the output quantum state $\left|\psi_{out}\right\rangle$ and the target label state $\left|\psi_Y\right\rangle$:

\begin{equation}
L_{QGCN} = -\left\langle\psi_{out}|\psi_Y\right\rangle,
\end{equation}
where $\left|\psi_{out}\right\rangle = \sum_{i=1}^{N} \sum_{k=1}^{C}H'^{(1)}_{ik}|i\rangle|k\rangle \otimes \ket{0} + \ldots$ is the output state of the QGCN, with $H'^{(1)}_{ik}$ being the amplitude corresponding to node $i$ and class $k$, and $\left|\psi_Y\right\rangle := \sum_{s \in Y_L} \sum_{f=1}^{C} Y_{sf} \left|s\right\rangle \left|f\right\rangle\otimes \ket{0}$ represents the true labels of the labeled nodes as a quantum state. The cost function can be evaluated via the ``Modified Hadamard test'' \cite{PhysRevA.107.062424,Luongo2023}. \newline

The training of the QGCN involves optimizing the parameters of the quantum circuit to minimize the cost
function. This optimization can be performed using either classical or quantum techniques~\cite{liao2021quantum}. Once the QGCN is trained, the inference process involves applying the optimized QGCN circuit to an input graph to obtain the predicted node labels. To extract the predicted labels from the output state $\left|\psi_{out}\right\rangle$, quantum state tomography techniques are employed. After obtaining the tomographic estimates of the output state, post-processing steps are applied to convert the results into the final predicted node labels---softmax function is applied to the estimated amplitudes $H'^{(1)}_{ik}$(from the trained model) to obtain the normalized predicted probabilities for each class,  The class with the highest predicted probability is then assigned as the final predicted label for each node. \newline

In the following two subsections, we propose quantum versions of two GCN variants: the Simplified Graph Convolution (SGC) \cite{wu2019simplifying} and the Linear Graph Convolution (LGC) \cite{pasa2023empowering}.

\subsection{Simplified Graph Convolution (SGC) and its quantum version}\label{sgc}
The Simplified Graph Convolution (SGC) \cite{wu2019simplifying} reduces the complexity of Graph Convolutional Networks (GCNs) by removing nonlinearities(while exhibits comparable or even superior performance compared to vanilla GCN and other complicated GNN models \cite{wu2019simplifying,maekawa2022beyond}). For node classification, the prediction generated by SGC is, \begin{equation}
Y_{\text{SGC}} = \text{softmax}(S^K X \Theta),
\end{equation} where $S = \hat{A}$ is the normalized adjacency matrix with added self-loops, $X \in \mathbb{R}^{N \times C}$ is the node attribute matrix, $\Theta$ is a weight matrix, and $K$ is a positive integer (originally representing the number of layers in GCN, though this concept becomes irrelevant in the context of SGC). Importantly, the experimental results demonstrated that the simplifications do not affect the accuracy across various applications \cite{wu2019simplifying}. \newline

The quantum implementation of SGC is similar to that of the linear transformation in GCN, which comprises three key components: data encoding of the node attribute matrix $X$, quantum circuit for the block-encoding of $S^K$, and a parameterized quantum circuit for the weight matrix $\Theta$. \newline

The data encoding step(quantum state preparation) for SGC is identical to that of GCN. Extensive research has been conducted on the problem of quantum state preparation~\cite{Long.01,Grover.02,Mottonen.05,Plesch.11,Zhang.21,Sun.21,Rosenthal.21,zhang2022quantum,Clader.22,yuan2023optimal,Gui.23,zhang2024circuit}. We select the approach from Ref. \cite{zhang2024circuit} for our data encoding, as their work provides a tunable trade-off between the number of ancillary qubits and the circuit depth for the state preparation. This flexibility allows us to optimally encode our classical data as a quantum state $\left|\psi_X\right\rangle$ by selecting the appropriate number of ancillary qubits based on the capabilities of our quantum hardware, while minimizing the circuit depth overhead required to achieve the desired precision. According to Theorem 3 in \cite{zhang2024circuit}, with $n_{\text{anc}}$ ancillary qubits where $\Omega(\log(NC)) \leq n_{\text{anc}} \leq O(NC)$, the initial data state $\left|\psi_{X}\right\rangle$ can be prepared to accuracy $\varepsilon_1$ with $\tilde{O}(NC \log(1/\varepsilon_1) \log(n_{\text{anc}}) / n_{\text{anc}})$ depth of Clifford+T gates, where $\tilde{O}$ suppresses the doubly logarithmic factors of $n_{\text{anc}}$.\newline

The weight matrix $\Theta$ in SGC is implemented using a parameterized quantum circuit (PQC), similar to the approach used in the quantum GCN. We assume that the depth of this PQC is less than the depth of the circuit for $S^K$. This assumption is based on the flexibility in choosing the depth of the PQC, which allows for a trade-off between the circuit depth and its expressive power. The expressive power of a PQC is related to its ability to explore the unitary space in an unbiased manner, increasing the depth of the PQC can lead to higher expressive power \cite{Larocca2022,Ragone2023,Holmes2022,Friedrich2023}. By choosing the depth of the PQC for $\Theta$ to be less than that of the circuit for $S^K$, we prioritize the efficiency of the overall quantum SGC implementation while potentially sacrificing some expressiveness in the weight matrix. The interplay between the depth of the PQC (and its associated expressiveness) and the depth of the block-encoding circuit for $S^K$ is an interesting topic for future research, as it may reveal further opportunities for optimization in the quantum SGC implementation.\newline

In the quantum SGC, for \( K = 2 \),\footnote{ SGC with $K=2$ often achieves similar/better performance than that of $K>2$ in many downstream applications.\cite{wu2019simplifying}} we can efficiently implement \( S^K \) by leveraging the product of block-encoded matrices as stated in Lemma 53 of Ref.\cite{gilyen2019quantum}: if \( U \) is an \( (\alpha,   n_{\text{anc}}', \varepsilon_2) \)-block-encoding of \( S \), then the product \( (I \otimes U)(I' \otimes U) \), is an \( (\alpha^2, 2n_{\text{anc}}', 2\alpha\varepsilon_2) \)-block-encoding of \( S^2 \). For semi-supervised multi-class classification, similar to that of vanilla GCN, the cost function of our Quantum SGC is defined as the negative inner product of the outcome state of our quantum SGC and a target label state $\left|\psi_Y\right\rangle$ prepared as $\textit{vec}(Y^T)$. The cost function can be evaluated via the Modified Hadamard test \cite{PhysRevA.107.062424,Luongo2023}.\newline

The complexity of the quantum SGC depends on the choice of number of ancillary qubits in the data encoding procedure and the block-encoding procedure in the layer-wise linear transformation: there’s trade-off between circuit depth and the number of qubits in the quantum SGC implementation. We first consider the two extreme cases in the trade-off: Table \ref{sgctable} presents the complexity comparison between the quantum SGC and the classical SGC for a single forward pass and evaluation of the cost function, assuming fixed precision parameters. The details of the complexity analysis is given in Section \ref{complexity}.\newline

\begin{table}[h]
\centering
\resizebox{\textwidth}{!}{%
\begin{tabular}{|l|c|c|}
\hline
\textbf{Algorithm} & \textbf{Time Complexity} & \textbf{Space Complexity\footnote{space complexity in the quantum case refers to the number of qubits, including the ancilla qubits used by the circuit \cite{ragavan2024spaceefficient}.}} \\
\hline
Quantum SGC (Min. Depth) & $\tilde{O}(\log(1/\delta) \cdot(\log(NC) + \log (Ns)))$ 
& $O( NC + N \log N \cdot s \log s)$ \\
\hline
Quantum SGC (Min. Qubits) & $\tilde{O}(\log(1/\delta) \cdot(NC / \log(NC) + N s \log s) )$ & $O(\log(NC) )$ \\
\hline
Classical SGC & $O(|E|C + NC^2))=O(NdC + NC^2)$ & $O(|E|+NC + C^2)=O(Nd+NC + C^2)$ \\
\hline
\end{tabular}%
}
\caption{Complexity comparison between Quantum SGC and Classical SGC ($K=2$) for a single forward pass and cost function evaluation, assuming fixed precision parameters. $N$ is the number of nodes, $C$ is the number of features per node. $d$ is the average degree of the nodes in the graph. $s$ is the maximum number of non-zero elements in each row/column of $\hat{A}$. The quantum SGC provides a probabilistic result with a success probability of $1 - \delta$. Note that in the classical time complexity, at first glance, $O(NC^2)$ appears to be the dominating term, as the average degree d on scale-free networks is usually much smaller than C and hence $NC^2 > NdC$. However, in practice, node-wise feature transformation can be executed at a reduced cost due to the parallelism in dense-dense matrix multiplications. Consequently, $O (NdC) $ is the dominant complexity term in the time complexity of classical SGC and the primary obstacle to achieving scalability \cite{chen2020scalable}. }
\label{sgctable}
\end{table}

\textsf{[Trade-off] Case 1: Quantum SGC with Minimum Depth} -- \textsf{\textit{\underline {Unlocking Quantum Speedup}}} \newline

In this case, the quantum SGC prioritizes minimizing the circuit depth at the cost of requiring more ancillary qubits. The time complexity of the quantum SGC in the case is logarithmic in the input sizes, i.e., $\tilde{O}(\log(NC) + \log (Ns))$(assuming fixed success probability), this represents a significant improvement over the classical SGC's time complexity of $O(NdC + NC^2)$. However, the space complexity of quantum SGC in this case is comparable to that of the classical SGC. The quantum SGC's logarithmic time complexity in this scenario is particularly beneficial for time-efficient processing large-scale graphs with high-dimensional node features.\newline

\textsf{[Trade-off] Case 2: Quantum SGC with Minimum Qubits} -- \textsf{\textit{\underline {Tackling Memory Constraints}}} \newline

In this case, the quantum SGC focuses on minimizing the number of required qubits at the cost of increased circuit depth. This trade-off is particularly relevant when dealing with massive graph datasets that exceed the memory constraints of classical computing systems. The space complexity of the quantum SGC in the minimum qubits case is $O(\log(NC))$, which represents an exponential reduction compared to the classical SGC's space complexity of $O(Nd+NC + C^2)$. This logarithmic space complexity enables the quantum SGC to process graphs of unprecedented scale, even on quantum hardware with limited qubit resources.\newline

The ability to process large-scale graphs with limited quantum resources is particularly valuable in domains such as social network analysis, where the graph size can easily reach billions of nodes. Storing such a graph in the memory of a classical computing system becomes infeasible due to the space complexity. However, the quantum SGC's logarithmic space complexity allows for the efficient encoding and processing of the graph using only a logarithmic number of qubits. This capability enables the exploration and analysis of these massive graphs, uncovering insights and patterns that were previously computationally infeasible. Furthermore, the time complexity of the quantum SGC, in this case, still offers a computational advantage over the classical SGC. Although the speedup is less pronounced compared to the minimum depth case, it remains significant for graphs with a large number of nodes and high-dimensional node features. \newline

For the intermediate cases in the tradeoff, the quantum SGC seeks a balance between the circuit depth and the number of ancillary qubits, which could potentially lead to moderate improvements in both time and space complexity. For example, by choosing {\small $n_{\text{anc}} = \Theta(\sqrt{NC})$} and {\small $n_{\text{anc}}' = \Theta(\sqrt{N \log N \cdot s \log s})$}, we obtain a time complexity of {\small $\tilde{O}(\log(1/\delta) \cdot (\sqrt{NC} \log(NC) + \sqrt{N \log N \cdot s \log s} \log(N s)))$} and a space complexity of {\small $O(\sqrt{NC} + \sqrt{N \log N \cdot s \log s})$}. This moderate case could be advantageous for certain quantum hardware architectures or problem instances where neither the circuit depth nor the number of qubits is the sole limiting factor. This case demonstrates the flexibility of the quantum SGC implementation in adapting to specific resource constraints while still maintaining a potential quantum advantage over the classical SGC.\newline

The complexity comparison between the quantum and classical SGC highlights the potential for quantum advantage in terms of both time and space complexity. The trade-off between circuit depth and the number of qubits in the quantum SGC implementation offers flexibility in adapting to specific quantum hardware constraints and problem instances.

\subsection{Linear Graph Convolution (LGC) and its quantum version}\label{lgc}

The Linear Graph Convolution (LGC) proposed by Pasa et al. \cite{pasa2023empowering}  is a more expressive variant of SGC. The LGC operation is defined as:

\begin{equation}
H = \sum_{i=0}^K \alpha_i L^i X\Theta
\end{equation}

where $L$ is the Laplacian matrix of the graph, $X$ is the node feature matrix, $\alpha_i$ are learnable weights, $\Theta$ is a weight matrix, and $K$ is an integer(a hype-parameter). This is essentially a spectral graph convolution (e.g. \cite{defferrard2016convolutional}) without nonlinear activation function.\newline

\begin{figure}[h!]
    \centering
    \includegraphics[width=\linewidth]{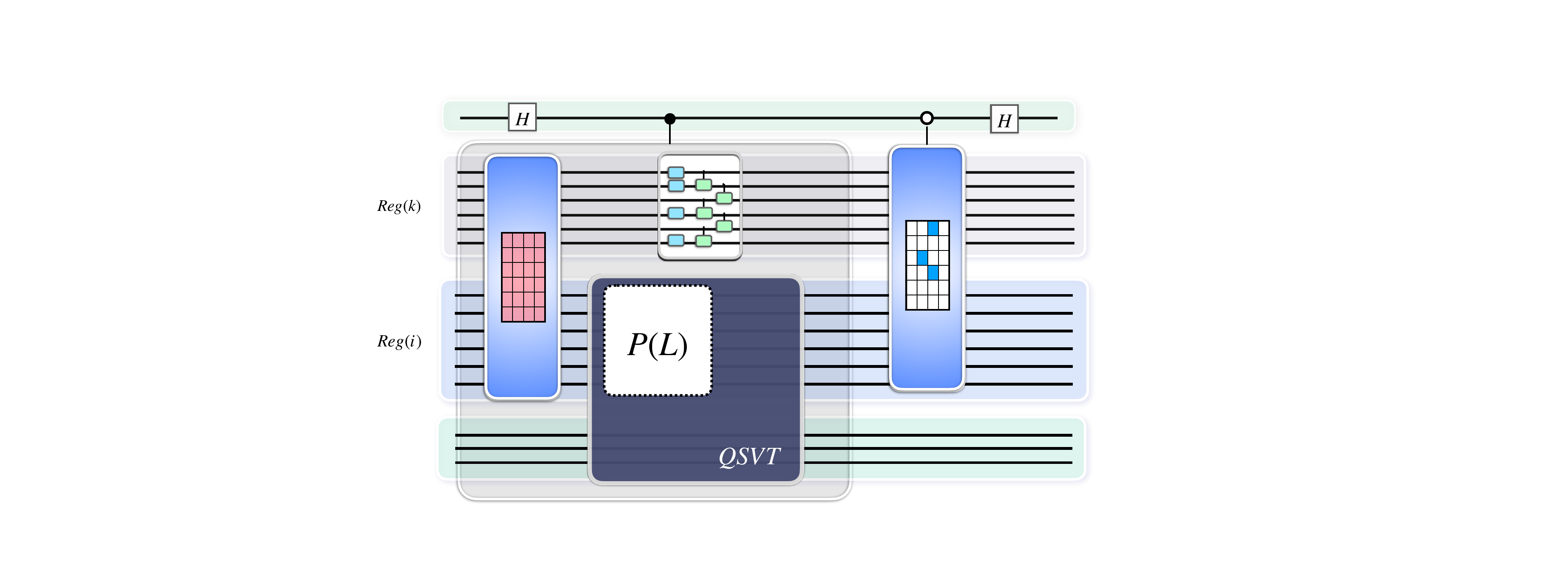}
   \caption{\textit{Schematic quantum circuit for the cost function evaluation procedure of our Quantum LGC}. The quantum implementation of LGC is similar to that of SGC, the major difference is the implementation of the aggregations of node features: we utilize the Polynomial eigenvalue transformation, a special instance of Quantum Singular Value Transformation (QSVT) (Theorem 56 in \cite{Gily_n_20192}), to implement $\sum_{i=0}^k \alpha_i L^i$. This requires a block-encoding of the Laplacian matrix $L$ and appropriate Pauli rotation angles in the QSVT circuit corresponding to the polynomial coefficients $\alpha_i$. The parametrization of the polynomial is equivalent to parametrization of the Pauli angles(phases) in the QSVT circuit, that is, the phases are the tunable weights to be trained. For semi-supervised multi-class classification, similar to that of vanilla GCN, the cost function of our Quantum LGC is defined as the inner product of the outcome state of our quantum LGC and a target label state $\left|\psi_Y\right\rangle$ prepared as $\textit{vec}(Y^T)$. The cost function can be evaluated via the Modified Hadamard test \cite{PhysRevA.107.062424,Luongo2023}.}
    \label{QLGC}
\end{figure}

By allowing multiple learnable weighting coefficients $\alpha_i$ for each $L^i$ up to order $K$, LGC can represent a much richer class of graph convolution filters compared to SGC. This increased expressiveness enables LGC to capture more complex graph structures and long-range dependencies, leading to improved performance on certain downstream tasks.\cite{pasa2023empowering} \newline

The quantum implementation of LGC is similar to that of SGC, the major difference is the implementation of the aggregations of node features: we utilize the ``Polynomial eigenvalue transformation,'' a special instance of the Quantum Singular Value Transformation (QSVT) (Theorem 56 in \cite{Gily_n_20192}), to implement $\sum_{i=0}^K \alpha_i L^i$. This requires a block-encoding of the Laplacian matrix $L$ and appropriate Pauli rotation angles in the QSVT circuit corresponding to the polynomial coefficients $\alpha_i$. The parametrization of the polynomial is equivalent to parametrization of the Pauli angles in the QSVT circuit, that is, the phases are the tunable weights to be trained. Figure \ref{QLGC} illustrates the schematic quantum circuit for the cost function evaluation procedure of our Quantum LGC.\newline

Similar to the quantum SGC, the complexity of the quantum LGC depends on the choice of number of ancillary qubits in the data encoding procedure and the block-encoding procedure, there’s trade-off between circuit depth and the number of qubits in the quantum LGC implementation. Here we focus on the case with minimum number of qubits in the trade-off: Table \ref{tablgc} summarizes the time and space complexities for classical LGC and quantum LGC with Min. Qubits, assuming fixed precision parameters and success probability(in the quantum case). The details of the complexity analysis is given in Section \ref{complexity}.\newline

\begin{table}[h]
\centering
\resizebox{\textwidth}{!}{%
\begin{tabular}{|l|c|c|}
\hline
Method & Time Complexity & Space Complexity \\

\hline

Quantum LGC (Min. Qubits) & $\tilde{O}(NC/\log(NC) + KN \cdot s \log s)$ & $O(\log(NC))$ \\
\hline
Classical LGC & $O(K|E|C + NC^2)=O(KNdC + NC^2)$ & $O(|E|+KNC + C^2)=O(Nd+KNC + C^2)$ \\
\hline
\end{tabular}%
}
\caption{Time and space complexity comparison for classical LGC and quantum LGC(Min. Qubits), for a single forward pass and cost function evaluation, assuming fixed precision parameters and success probability(in the quantum case). $N$ is the number of nodes, $C$ is the number of features per node. $d$ is the average degree of the nodes in the graph. $s$ is the maximum number of non-zero elements in each row/column of $L$. The quantum LGC provides a probabilistic result with a success probability of $1 - \delta$. }
\label{tablgc}
\end{table}

The Quantum LGC (Min. Qubits) case focuses on minimizing the space complexity. The space complexity of this case is $O(\log(NC))$, which is significantly lower than the space complexity of classical LGC. This logarithmic scaling of the space complexity in the Quantum LGC (Min. Qubits) case can enable the analysis of graphs that may be intractable for classical methods due to memory constraints. The time complexity of this case is $\tilde{O}(NC/\log(NC) + KN \cdot s \log s)$, which also provides a potential advantage over classical LGC's $O(KNdC + KNC^2)$ for certain problem instances.\newline

\section{Quantum Graph Attention Networks}\label{23}

 As mentioned in Section \ref{gnn}, the building block layer of Graph Attention Network achieves the following transformations, which we refer to as the ``Graph attention operation'':

\begin{equation}
\mathbf{h}_{j}=\phi\left(\mathbf{x}_{j}, \bigoplus_{i \in \mathcal{N}_{j}} a\left(\mathbf{x}_{i}, \mathbf{x}_{j}\right) \psi\left(\mathbf{x}_{i}\right)\right),
\label{atet}
\end{equation}
where $a\left(\mathbf{x}_{j}, \mathbf{x}_{i}\right) $ is a scalar that indicates the relationship strength between node $i$ and $j$, often referred as attention coefficients or attention scores \cite{ghojogh2020attention}. The following sections present our quantum implementation of Eq.~\eqref{atet}. Note we omit the activation function in the original definition of $\phi$ in \cite{bronstein2021geometric}, the quantum implementation of the activation function is described in Section \ref{mgcn}, here and in the next section \ref{24} we focus on the quantum implementation of this definition of $\phi$. In Appendix.~\ref{attentionsec}, we design a Quantum Attention Mechanism to evaluate and store attention score $a(\bold{x}_i,\bold{x}_j)$ on a quantum circuit, which serves as a crucial component for the subsequent construction described in Section \ref{attentionsec2}. \newline

The  Graph Attention operation defined in Eq.~\ref{atet} can also be described similar to the layer-wise linear transformation for multi-channel GCN in Section \ref{22} (i.e. $H'^{(l)}=\hat{A} H^{(l)} W^{(l)}$). Here in the Graph Attention operation, the non-zero elements in the normalized adjacency matrix $\hat{A}$ are modified to be the attention score of the corresponding node pairs \cite{velickovic2017graph}. On a quantum circuit, similar to the case of multi-channel GCN, the Graph Attention operation can be implemented by applying the block-encoding of the modified normalized adjacency matrix, which we refer to as the ``weighted adjacency matrix'' and a parameterized quantum circuit. In the following Section \ref{attentionsec2} we present how to achieve the Graph attention operation via quantum circuit. As a preliminary, the block-encoding of certain sparse\footnote{For many practical applications, the adjacency matrix of a graph is often sparse.} matrix is illustrated in Section \ref{be}.

\subsection{Block encoding of certain sparse matrices}\label{be}
The block encoding of a general sparse matrix (e.g.~\cite{sunderhauf2024block, lin2022lecture}) requires a certain oracle that is hard to construct for the Graph Attention operation. In this section, following Ref.~\cite{doi:10.1137/22M1484298, lin2022lecture}, we first investigate the sparse matrices that can be decomposed as the summation of 1-sparse matrices (A 1-sparse matrix is defined as there is exactly one nonzero entry in each row or column of the matrix\cite{lin2022lecture}). We start with the block encoding of 1-sparse matrices.\\ 

For each column $j$ of a 1-sparse matrix $A$, there is a single row index $c(j)$ such that $A_{c(j),j} \neq 0$, and the mapping $c$ is a permutation.~\cite{doi:10.1137/22M1484298, lin2022lecture} Therefore, $A$ can be expressed as the product of a diagonal matrix (whose diagonal entries are the non-zero entries of the 1-sparse matrix) and a permutation matrix. Ref.~\cite{doi:10.1137/22M1484298, lin2022lecture} showed that the block encoding of a 1-sparse matrix can be constructed by multiplying the block encoding of a diagonal matrix and the block encoding of a permutation matrix: the permutation matrix, denoted as $O_{c}$ act as,
$$
O_{c}|j\rangle=|c(j)\rangle,
$$
and the block encoding of the diagonal matrix, denoted as $O_{A}$, acts as:
$$
O_{A}|0\rangle|j\rangle=\left(A_{c(j), j}|0\rangle+\sqrt{1-\left|A_{c(j), j}\right|^{2}}|1\rangle\right)|j\rangle .
$$
$U_{A}=\left(I \otimes O_{c}\right) O_{A}$ is a block encoding of the 1-sparse matrix $A$ \cite{lin2022lecture}.\newline

Now, we consider the sparse matrices that can be decomposed as the summation of 1-sparse matrices (below, we also use $A$ to denote such a matrix). After the decomposition, we index the 1-sparse matrices by $l$. For the $l$-th 1-sparse matrix, the row index of the nonzero entry in each column $j$, is denoted by $c(j, l) $. There exist $O^l_{c}$ and $O^l_{A}$ and corresponding $U^l_{A}$ such that \cite{lin2022lecture},

\begin{equation}
    O^l_{c}|j\rangle=|c(j, l)\rangle ,
    \label{ocl}
\end{equation}
and,

\begin{equation}
O^l_{A}|0\rangle|j\rangle=\left(A_{c(j, l), j}|0\rangle+\sqrt{1-\left|A_{c(j, l), j}\right|^{2}}|1\rangle\right)|j\rangle.
\end{equation}

It can be shown that  $\sum_l U^l_{A}=\sum_l\left(I \otimes O^l_{c}\right) O^l_{A}$ is a block encoding of the sparse matrix $A$ \cite{lin2022lecture}. The summation over $l$ can be carried out by Linear Combination of Unitaries (LCU)\footnote{The concept of  \textit{LCU} was introduced in \cite{LongGui-Lu_2006,long2011duality}.} \cite{childs2012hamiltonian}.\newline

For the construction of $O^l_{A}$, assume that there is an oracle \cite{lin2022lecture},

$$
\widetilde{O}^l_{A}\left|0^{d^{\prime}}\right\rangle|j\rangle=\left|\widetilde{A}_{c(j, l), j}\right\rangle|j\rangle,
$$
where $\widetilde{A}_{c(j, l),  j}$ is a $d^{\prime}$-bit representation of $A_{c(j, l),  j}$. By arithmetic operations, we can convert this oracle into another oracle

$$
{O^l_{A}}^{\prime}\left|0^{d}\right\rangle|j\rangle=\left|\widetilde{\theta}_{c(j, l), j}\right\rangle|j\rangle,
$$
where $0 \leq \tilde{\theta}_{c(j, l),  j}<1$, and $\tilde{\theta}_{c(j, l),  j}$ is a $d$-bit representation of $\theta_{c(j, l),  j}=\arccos \left(A_{c(j, l),  j}\right) / \pi$. \newline

Next, using the ``Controlled rotation given rotation angles'' (Proposition 4.7 in Ref., denoted  as ``CR'' below) and  uncomputation of ${O^l_{A}}^{\prime}$ we can achieve the construction of $O^l_{A}$~\cite{lin2022lecture}:

\begin{align}
|0\rangle \left|0^{d}\right\rangle|j\rangle \stackrel{{O^l_{A}}^{\prime}}{\longrightarrow}|0\rangle\left|\tilde{\theta}_{c(j, l), j}\right\rangle|j\rangle ,\\
\stackrel{\mathrm{CR}}{\longrightarrow}\left(A_{c(j, l),  j}|0\rangle+\sqrt{1-\left|A_{c(j, l),  j}\right|^{2}}|1\rangle\right)\left|\widetilde{\theta}_{c(j, l),  j}\right\rangle|j\rangle ,\\
 \stackrel{\left({O^l_{A}}^{\prime}\right)^{-1}}{\longrightarrow}\left(A_{c(j, l),  j}|0\rangle+\sqrt{1-\left|A_{c(j, l),  j}\right|^{2}}|1\rangle\right)\left|0^{d}\right\rangle|j\rangle.
\end{align}

\subsection{Quantum Graph Attention operation}\label{attentionsec2}

 As mentioned in Section \ref{be}, in this paper we investigate certain sparse matrices (here, the weighted adjacency matrices) that can be decomposed as the summation of 1-sparse matrices. From the preliminary discussion in section \ref{be}, the block encoding of such matrices can be boiled down to the $\widetilde{O}^l_{A}$ for each 1-sparse matrix. That is, the core task is to construct the following operation for each 1-sparse matrix (indexed by $l$):
\begin{equation}
  O^{\textsf{diagonal}}_{l}:  \ket{j}\ket{0} \to \ket{j}\ket{A_{c(j,l),j}}.
\end{equation}
where $\ket{A_{c(j,l),j}}$ denotes $A_{c(j,l),j}$ being stored in a quantum register with some finite precision, and for simplicity we use $\ket{0}$ to represent a state of the register that all qubits in the register being in the state of $\ket{0}$. We also adopt this kind of notion in the rest of the paper: for a scalar $a$, $\ket{a}$ denotes $a$ being stored in a quantum register with some finite precision, and in contexts where there is no ambiguity, $\ket{0}$ represent a state of a quantum register that all qubits in the register being in the state of $\ket{0}$.\newline

In our case of Graph attention operation, the elements of the weighted adjacency matrix are the attention scores, i.e. $A_{i,j}=a(\bold{x}_{i},\bold{x}_j)$, and we have,
\begin{equation}
  O^{\textsf{diagonal}}_{l}:  \ket{j}\ket{0} \to \ket{j}\ket{a(\bold{x}_{c(j,l)},\bold{x}_j)}.
\end{equation}

In Appendix.~\ref{attentionsec} we have constructed a quantum oracle $O_{\textsf{attention}}$ such that,

\begin{equation}
  O_{\textsf{attention}}:  \ket{i}\ket{j}\ket{0} \to \ket{i}\ket{j}\ket{a(\bold{x}_i,\bold{x}_j)}.
\end{equation}

\begin{figure}[h!]
    \centering
    \includegraphics[width=\linewidth]{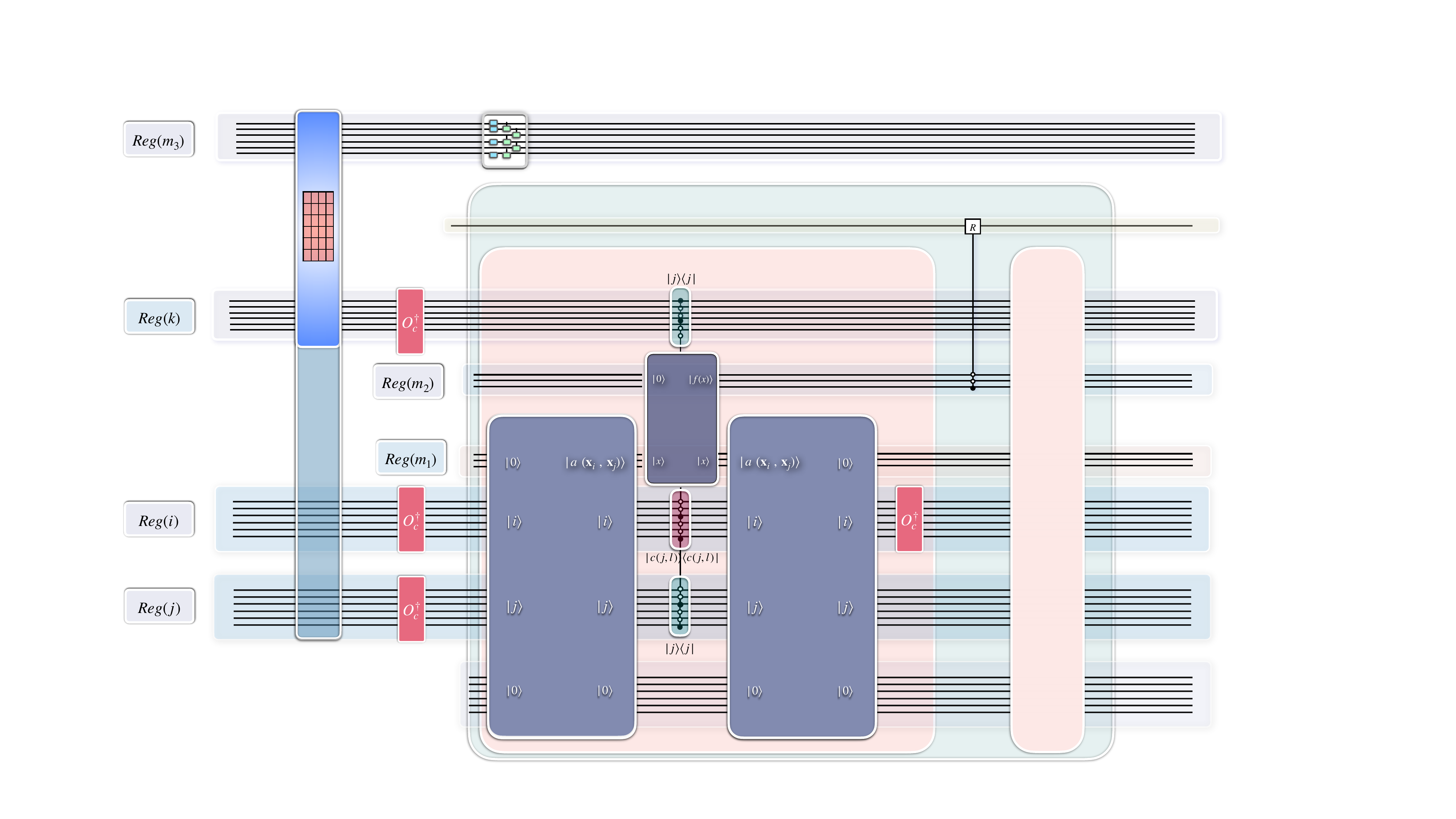}
   \caption{\textit{Quantum  implementation of linear layer-wise transformation for Graph Attention Networks.} The initial data state $\left|\psi^3_{X0}\right\rangle=\sum_i\ket{i}^{\otimes3}\ket{\bold{x}_i}$ is prepared by the blue box on the left. The QNN module, denoted as $U_w$, transform the state to $\left|\psi^3_{X}\right\rangle=\sum_i\ket{i}^{\otimes3}U_w\ket{\bold{x}_i}$. The pale green box together with the three red boxes which achieve $M'_l=\sum_j A_{c(j,l),j}\ket{j}^{\otimes3}\ket{0}\bra{c(j, l)}^{\otimes3}\bra{0}+... $, are then applied to the transformed initial data state, resulting $\sum_j A_{c(j,l),j}\ket{j}^{\otimes3}U_w\ket{\bold{x}_{c(j,l)}}\ket{0}$. The pale green box consists of the following Modules:  \textsf{Module 1}(the first pink box). $\mathcal{O}^{\textsf{diagonal}}_{l} $ \textsf{Module 2}. the ``Conditional Rotation'' (Theorem 3.5 in Ref. \cite{10.1088/1367-2630/abc9ef}) \textsf{Module 3} (the second pink box) is the uncomputation of Module 1.}
    \label{attetsum}
\end{figure}

In the following of this section, we present how to construct an alternative version\footnote{Note that we are not strictly constructing $O^{\textsf{diagonal}}_{l}$ here and the following operations do not strictly achieve a block-encoding of the weighted adjacency matrix, however the alternative versions do generate a quantum state that resembles the Graph attention operation.} of $O^{\textsf{diagonal}}_{l}$ utilising $O_{\textsf{attention}}$.\\

\begin{figure}[h!]
    \centering
    \includegraphics[width=\linewidth]{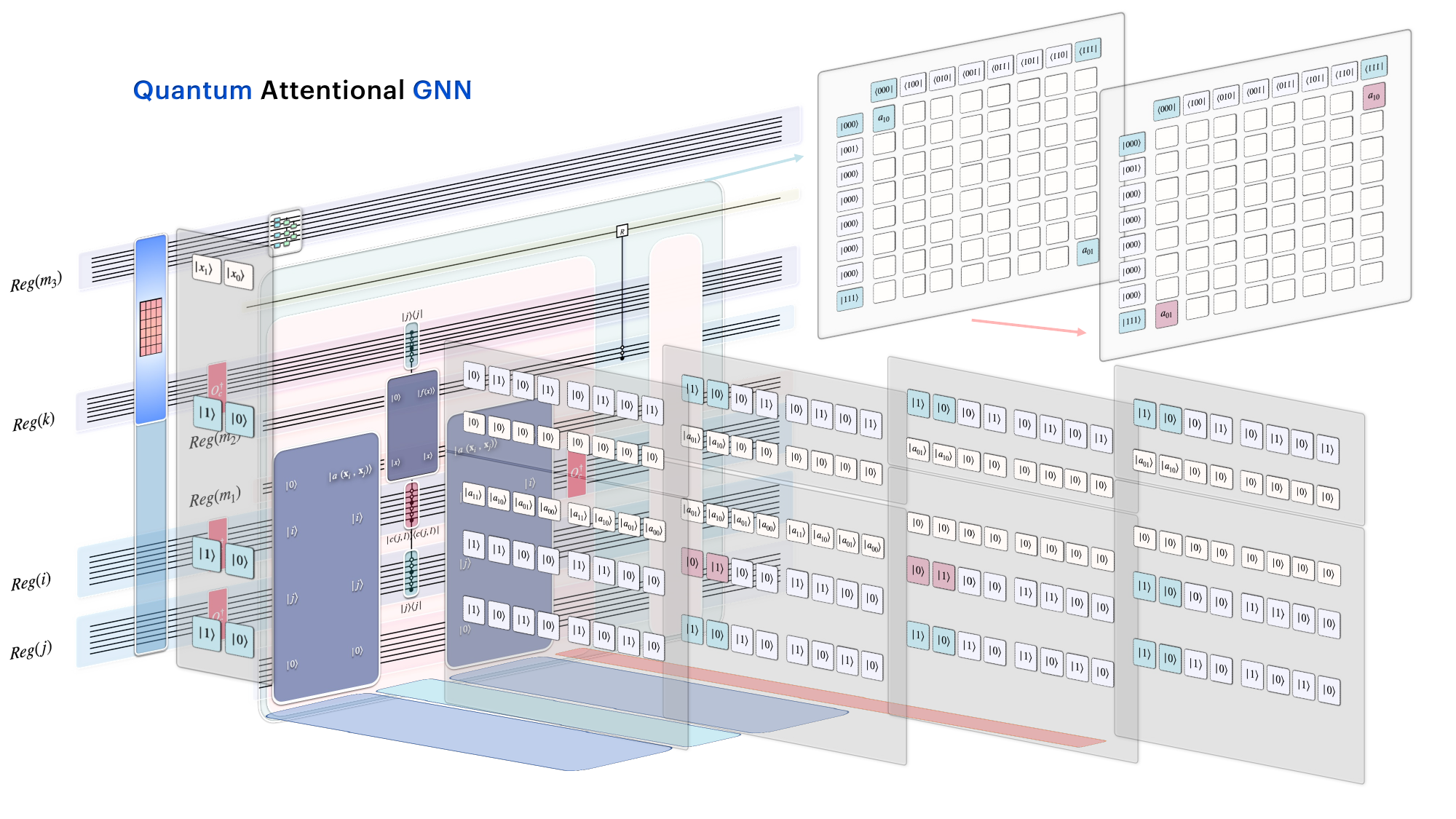}
   \caption{\textit{Quantum implementation of linear layer-wise transformation for Graph Attention Networks.} This figure provides a small example of the corresponding states and matrices in Fig.~\ref{attetsum}. The panels perpendicular to the circuit plane represent the quantum states, while the panels parallel to the circuit plane represent the corresponding matrices.}
   \label{attetsum2}
\end{figure}

 \textsf{Step 1: Attention oracle loading the attention scores $A_{i,j}=a(\bold{x}_{i},\bold{x}_j)$}
\newline

The first component  is the attention oracle $O_{\textsf{attention}}$, depicted as the navy box in Fig.\ref{attetsum}. When being applied onto the three address register $Reg(i)$, $Reg(j)$ and the corresponding memory register $Reg(m_1)$,  $O_{\textsf{attention}}$ loads the attention scores $A_{i,j}=a(\bold{x}_{i},\bold{x}_j)$ for each pair of the nodes $i,j$ into $Reg(m_1)$, while the other memory register $Reg(m_2)$ stays $\ket{0}$. $O_{\textsf{attention}}$ act as:

\begin{equation}
O_{\textsf{attention}} : \ket{i}\ket{j}\ket{0}\ket{0}\ket{k} \to \ket{i}\ket{j}\ket{a(\bold{x}_i,\bold{x}_j)}\ket{0}\ket{k}\end{equation}

If $O_{\textsf{attention}}$ is applied onto an input state as $\sum_i\sum_j\sum_k\ket{i}\ket{j}\ket{0}\ket{0}\ket{k}$, it transform the state as:

\begin{equation}
\sum_i\sum_j\sum_k\ket{i}\ket{j}\ket{0}\ket{0}\ket{k} \to \sum_i\sum_j\sum_k\ket{i}\ket{j}\ket{a(\bold{x}_i,\bold{x}_j)}\ket{0}\ket{k}\end{equation}

\textsf{Step 2: Selective copying of the attention scores $A_{i,j}=a(\bold{x}_{i},\bold{x}_j)$}
\newline

 The second component  is a multi-controlled unitary which performs the ``selective copying'' of the attention scores onto $Reg(m_2)$, depicted as the blue-navy-red-blue combo boxes following the attention oracle in Fig.\ref{attetsum}. The copying is implemented by a quantum oracle that acts as $\ket{0}\ket{x} \to \ket{f(x)}\ket{x}$ where $f$ can be a nonlinear activation function, however, we still name the operation as ``copying.'' \newline

 Consider the branches indexed by $i,j,k$ in the state $\sum_i\sum_j\sum_k\ket{i}\ket{j}\ket{a(\bold{x}_i,\bold{x}_j)}\ket{0}\ket{k}$, the copying is defined\footnote{For an implementation of the selective copying operation, see Appendix.~\ref{selective}.} to happen only for the branches $i=c(j,l); k=j$, that is, the selective copying operation transform the branches in the state $\sum_i\sum_j\sum_k\ket{i}\ket{j}\ket{a(\bold{x}_i,\bold{x}_j)}\ket{0}\ket{k}$ as follows:\newline

 For branches $i=c(j,l); k=j$:
\begin{equation}
\sum_j \ket{c(j,l)}\ket{j}\ket{a(\bold{x}_{c(j,l)},\bold{x}_j)}\ket{0}\ket{j} \to \sum_j  \ket{c(j,l)}\ket{j}\ket{a(\bold{x}_{c(j,l)},\bold{x}_j)}\ket{a(\bold{x}_{c(j,l)},\bold{x}_j)}\ket{j}.
\end{equation}

 For other branches:
 \begin{equation}
\sum_{i\neq c(j,l)} \sum_j \sum_{k\neq j}\ket{i}\ket{j}\ket{a(\bold{x}_i,\bold{x}_j)}\ket{0}\ket{k} \to \sum_{i\neq c(j,l)} \sum_j \sum_{k\neq j} \ket{i}\ket{j}\ket{a(\bold{x}_i,\bold{x}_j)}\ket{0}\ket{k}.
\end{equation}

Combining all branches, we have the selective copying of the attention scores $A_{i,j}=a(\bold{x}_{i},\bold{x}_j)$ as:
\begin{align*}
\sum_j \ket{c(j,l)}\ket{j}\ket{a(\bold{x}_{c(j,l)},\bold{x}_j)}\ket{0}\ket{j} +\sum_{i\neq c(j,l)} \sum_j \sum_{k\neq j}\ket{i}\ket{j}\ket{a(\bold{x}_i,\bold{x}_j)}\ket{0}\ket{k} \to 
\\
\sum_j  \ket{c(j,l)}\ket{j}\ket{a(\bold{x}_{c(j,l)},\bold{x}_j)}\ket{a(\bold{x}_{c(j,l)},\bold{x}_j)}\ket{j}+ \sum_{i\neq c(j,l)} \sum_j \sum_{k\neq j} \ket{i}\ket{j}\ket{a(\bold{x}_i,\bold{x}_j)}\ket{0}\ket{k}.
\end{align*}\newline

\textsf{Step 3: Uncomputation of attention oracle $O_{\textsf{attention}}^{\dagger}$}\newline

The third component  is the uncomputation of attention oracle $O_{\textsf{attention}}$ which act as

\begin{equation}
  O_{\textsf{attention}} ^{\dagger} : \ket{i}\ket{j}\ket{a(\bold{x}_i,\bold{x}_j)}\to\ket{i}\ket{j}\ket{0}.  
\end{equation}

When acting on the output state of Step 2, it transforms the state as follows:
\begin{align*}
\sum_j  \ket{c(j,l)}\ket{j}\ket{a(\bold{x}_{c(j,l)},\bold{x}_j)}\ket{a(\bold{x}_{c(j,l)},\bold{x}_j)}\ket{j}+ \sum_{i\neq c(j,l)} \sum_j \sum_{k\neq j} \ket{i}\ket{j}\ket{a(\bold{x}_i,\bold{x}_j)}\ket{0}\ket{k} 
\to\\
\sum_j  \ket{c(j,l)}\ket{j}\ket{0}\ket{a(\bold{x}_{c(j,l)},\bold{x}_j)}\ket{j}+ \sum_{i\neq c(j,l)} \sum_j \sum_{k\neq j} \ket{i}\ket{j}\ket{0}\ket{0}\ket{k}.
\end{align*}

\textsf{Step 4: Permutation of basis on register $Reg(i)$}\newline

The fourth component  is the permutation of basis in the register $Reg(i)$ by applying the unitary $O_{c}^{l^{\dagger}}$ (defined in Eq.\ref{ocl}) which act as, 

$$O_{c}^{l^{\dagger}}|c(j, l)\rangle=|j\rangle,$$

When acting on the output state of Step 3, it transforms the state as follows:
\begin{align*}
\sum_j  \ket{c(j,l)}\ket{j}\ket{0}\ket{a(\bold{x}_{c(j,l)},\bold{x}_j)}\ket{j}+ \sum_{i\neq c(j,l)} \sum_j \sum_{k\neq j}\ket{i}\ket{j}\ket{0}\ket{0}\ket{k}\to\\\sum_j  
\ket{j}\ket{j}\ket{0}\ket{a(\bold{x}_{c(j,l)},\bold{x}_j)}\ket{j}+ \sum_{i\neq c(j,l)} \sum_j \sum_{k\neq j} \ket{P(i)}\ket{j}\ket{0}\ket{0}\ket{k},
\end{align*}
where $\ket{P(i)}:=O_{c}^{l^{\dagger}}\ket{i}$.\newline

The state evolution during the four steps can be summarized as follows:

\begin{align*}
\sum_i \sum_j \sum_k \ket{i}\ket{j}\ket{0}\ket{0}\ket{k} =
\sum_j \ket{c(j,l)}\ket{j}\ket{0}\ket{0}\ket{j} +\sum_{i\neq c(j,l)} \sum_j \sum_{k\neq j}\ket{i}\ket{j}\ket{0}\ket{0}\ket{k} \to\\
\sum_j \ket{c(j,l)}\ket{j}\ket{a(\bold{x}_{c(j,l)},\bold{x}_j)}\ket{0}\ket{j} +\sum_{i\neq c(j,l)} \sum_j \sum_{k\neq j}\ket{i}\ket{j}\ket{a(\bold{x}_i,\bold{x}_j)}\ket{0}\ket{k} \to \\
\sum_j  \ket{c(j,l)}\ket{j}\ket{a(\bold{x}_{c(j,l)},\bold{x}_j)}\ket{a(\bold{x}_{c(j,l)},\bold{x}_j)}\ket{j}+ \sum_{i\neq c(j,l)} \sum_j \sum_{k\neq j} \ket{i}\ket{j}\ket{a(\bold{x}_i,\bold{x}_j)}\ket{0}\ket{k} 
\to\\
\sum_j  \ket{c(j,l)}\ket{j}\ket{0}\ket{a(\bold{x}_{c(j,l)},\bold{x}_j)}\ket{j}+ \sum_{i\neq c(j,l)} \sum_j \sum_{k\neq j} \ket{i}\ket{j}\ket{0}\ket{0}\ket{k}\to\\\sum_j  
\ket{j}\ket{j}\ket{0}\ket{a(\bold{x}_{c(j,l)},\bold{x}_j)}\ket{j}+ \sum_{i\neq c(j,l)} \sum_j \sum_{k\neq j} \ket{P(i)}\ket{j}\ket{0}\ket{0}\ket{k}.
\end{align*}

Gathering all four steps above, the pink box in Fig.~\ref{attetsum} implements an alternative version of $O^{\textsf{diagonal}}_{l}$ denoted as $\mathcal{O}^{\textsf{diagonal}}_{l}$ which act as:

\begin{equation}
\mathcal{O}^{\textsf{diagonal}}_{l}:  \ket{j}^{\otimes3}\ket{0}\to \ket{j}^{\otimes3}\ket{a(\bold{x}_{c(j,l)},\bold{x}_j)}.
\label{ol}
\end{equation}

Note that we neglected some registers that were unchanged. In terms of the elements of the weighted adjacency matrices, $\mathcal{O}^{\textsf{diagonal}}_{l}$ act as:
\begin{equation}
\mathcal{O}^{\textsf{diagonal}}_{l}  \ket{j}^{\otimes3}\ket{0} \to \ket{j}^{\otimes3}\ket{A_{c(j,l),j}}.
\end{equation}

Armed with $\mathcal{O}^{\textsf{diagonal}}_{l} $, we can then construct the Graph attention operation using the recipe discussed in the previous section \ref{be}, which is based on the following modules. \newline

\begin{figure}[h!]
    \centering
    \includegraphics[width=0.9\linewidth]{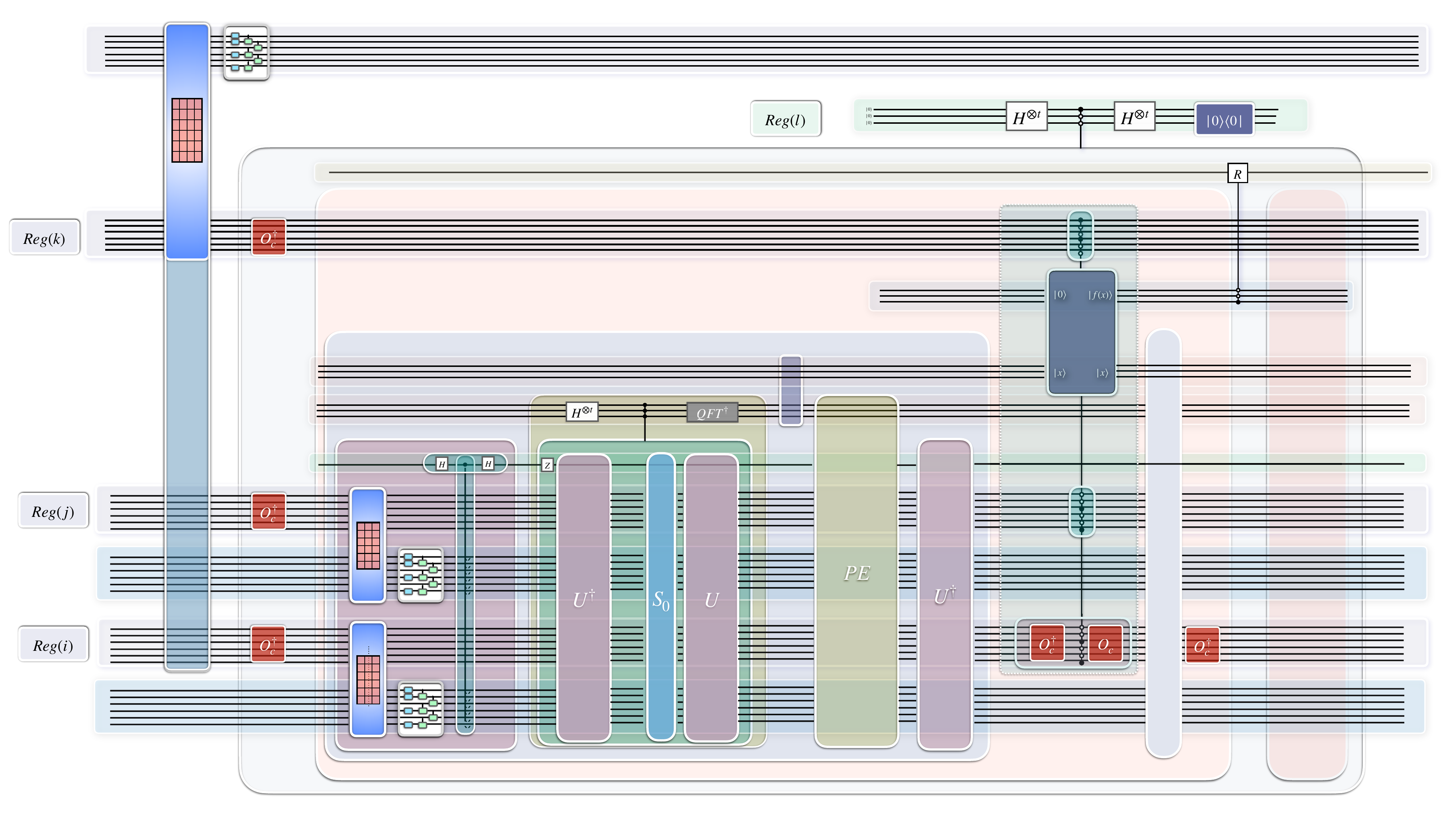}
\caption{\textit{Quantum implementation of linear layer-wise transformation for Graph Attention Networks} The initial data state $\left|\psi^3_{X0}\right\rangle=\sum_i\ket{i}^{\otimes3}\ket{\bold{x}_i}$ is prepared by the blue box on the left. The QNN module, denoted as $U_w$, transforms the state to $\left|\psi^3_{X}\right\rangle=\sum_i\ket{i}^{\otimes3}U_w\ket{\bold{x}_i}$. The transparent box which achieves $M'_l=\sum_j A_{c(j,l),j}\ket{j}^{\otimes3}\ket{0}\bra{c(j, l)}^{\otimes3}\bra{0}+... $, consist of four Modules:  \textsf{Module 1}(the first pink box) $\mathcal{O}^{\textsf{diagonal}}_{l} $.  \textsf{Module 2} the Conditional Rotation (Theorem 3.5 in Ref. \cite{landman2021quantum}), represented as the controlled-R gate between the two pink boxes. \textsf{Module 3} (the second pink box) Uncomputation of Module 1. \textsf{Module 4}(the three red boxes on the left of module 1) Permutation of basis. An overall LCU is then applied to the four modules, depicted in as the add-on register $Reg(l)$ controlling the transparent box, to achieve the addition over index $l$: $M=\sum_l  M'_l=\sum_l\sum_j A_{c(j,l),j}\ket{j}^{\otimes3}\ket{0}\bra{c(j, l)}^{\otimes3}\bra{0}+...$.
$M$ is then applied on $\left|\psi^3_{X}\right\rangle=\sum_i \ket{i}^{\otimes3}U_w\ket{\bold{x}_i}$, producing the outcome state
$\sum_j \ket{j}^{\otimes3}\sum_l A_{c(j,l),j} U_w\ket{\bold{x}_{c(j,l)}}\ket{0}$.}
    \label{Attenti}
\end{figure}

\textsf{Module 1}: $\mathcal{O}^{\textsf{diagonal}}_{l} .$\newline

\textsf{Module 2}: the Conditional Rotation (Theorem 3.5 in Ref. \cite{landman2021quantum}), to convert $A_{c(j,l),j}$ from basis to amplitude.\newline

\textsf{Module 3}: Uncomputation of Module 1.\newline

These three modules achieve the following unitary:
\begin{equation}
    M_l=\sum_j A_{c(j,l),j}\ket{j}^{\otimes3}\ket{0}\bra{j}^{\otimes3}\bra{0}+...
\end{equation}

\textsf{Module 4}: Permutation of basis.\newline

\begin{figure}[h!]
    \centering
    \includegraphics[width=\linewidth]{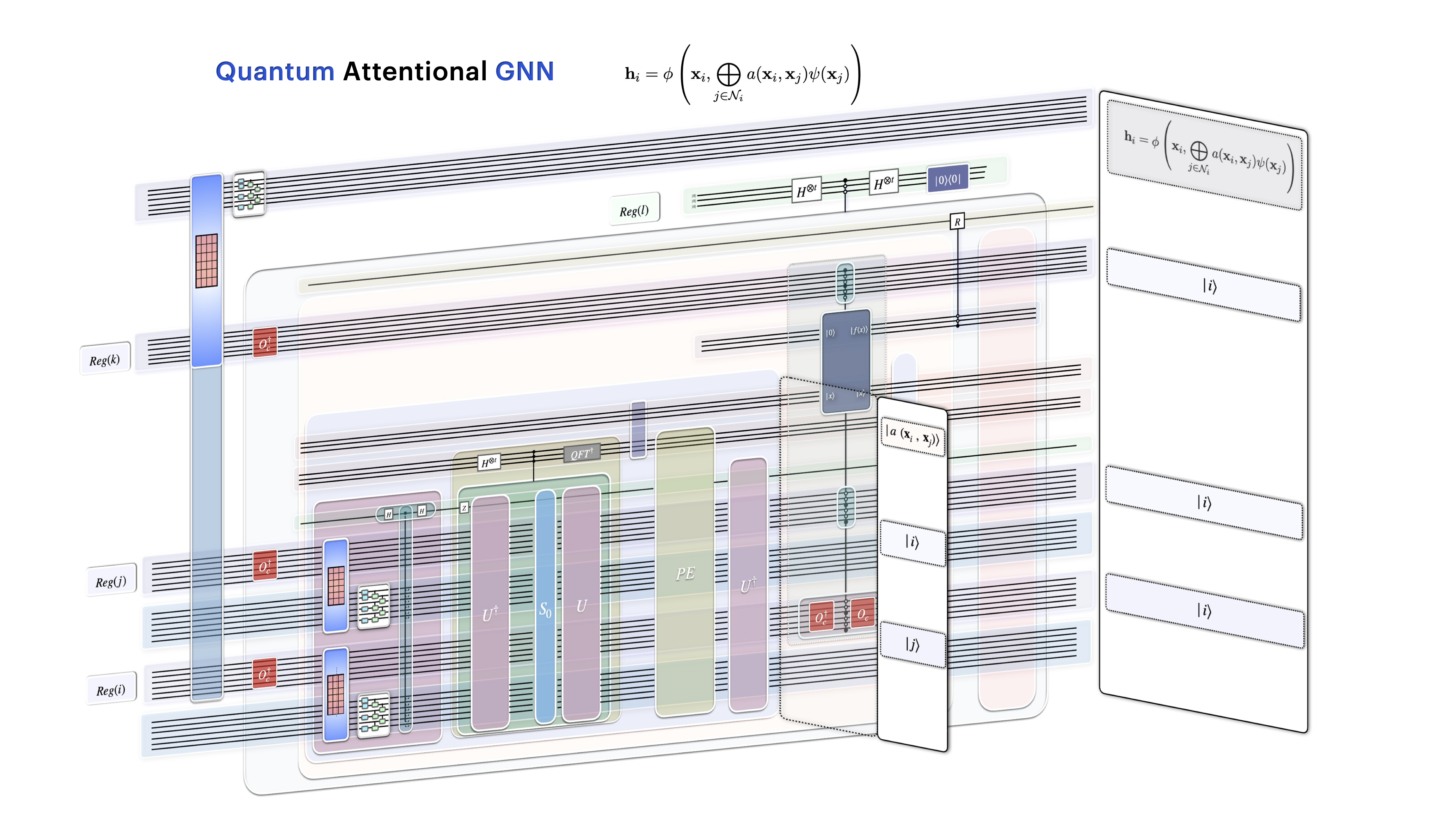}
\caption{\textit{Quantum implementation of linear layer-wise transformation for Graph Attention Networks}. This figure provides a 3D state-circuit view for Fig.~\ref{Attenti}. The panels perpendicular to the circuit plane represent the quantum states generated by corresponding circuits.}
    \label{Attenti2}
\end{figure}
Three $
O^{l_{c}^{\dagger}}$ which act as $\bra{j}O^{l_{c}^{\dagger}}=\bra{c(j, l)}
$ are applied before the previous three modules on the addresses, yield

\begin{equation}
    M'_l=\sum_j A_{c(j,l),j}\ket{j}^{\otimes3}\ket{0}\bra{c(j, l)}^{\otimes3}\bra{0}+...
\end{equation}

When $M'_l$ is  applied on the transformed data state $\left|\psi^3_{X}\right\rangle:=\sum_i\ket{i}^{\otimes3}U_w\ket{\bold{x}_i}$, prepared by the blue box and the QNN module (denoted as $U_w$) in Fig.~\ref{attetsum}, it act as follows

\begin{align}
    M'_l\left|\psi^3_{X}\right\rangle=(\sum_j A_{c(j,l),j}\ket{j}^{\otimes3}\ket{0}\bra{c(j, l)}^{\otimes3}\bra{0})\sum_i\ket{i}^{\otimes3}U_w\ket{\bold{x}_i}\ket{0}\\
    =\sum_j A_{c(j,l),j}\ket{j}^{\otimes3}U_w\ket{\bold{x}_{c(j,l)}}\ket{0}
\end{align}

The operations constructed so far can be summarised in Fig. \ref{attetsum}, Fig. \ref{attetsum2} provide a small example of the corresponding states and matrices.\newline

To achieve the addition over index $l$, an overall LCU is applied to the four modules, depicted in Fig.~\ref{Attenti} and \ref{Attenti2} as the add-on register $Reg(l)$ with the controlled unitaries in the transparent box, achieving the following operation: 

\begin{equation}
   M:=\sum_l  M'_l=\sum_l\sum_j A_{c(j,l),j}\ket{j}^{\otimes3}\ket{0}\bra{c(j, l)}^{\otimes3}\bra{0}+...
\end{equation}

When $M$ is applied on $\left|\psi^3_{X}\right\rangle=\sum_i \ket{i}^{\otimes3}U_w\ket{\bold{x}_i}$, it produces the outcome state as:

\begin{align}
  M\left|\psi^3_{X}\right\rangle=\sum_l  M'_l \left|\psi^3_{X}\right\rangle =\sum_l \sum_j A_{c(j,l),j}\ket{j}^{\otimes3}U_w\ket{\bold{x}_{c(j,l)}}\ket{0}\\=\sum_j \ket{j}^{\otimes3}\sum_l A_{c(j,l),j} U_w\ket{\bold{x}_{c(j,l)}}\ket{0}.
\end{align}

We can add an extra identity operator $I$ with coefficient $r$ in the LCU that produces $M$, yielding,

\begin{align}
  M'\left|\psi^3_{X}\right\rangle =(M+rI)\left|\psi^3_{X}\right\rangle =\sum_j \ket{j}^{\otimes3}\sum_l A_{c(j,l),j} U_w\ket{\bold{x}_{c(j,l)}}\ket{0}+r\sum_j\ket{j}^{\otimes3}U_w\ket{\bold{x}_j}\ket{0}\\=\sum_j \ket{j}^{\otimes3}(\sum_l A_{c(j,l),j} U_w\ket{\bold{x}_{c(j,l)}}+rU_w\ket{\bold{x}_j})\ket{0},\\
  =\sum_j \ket{j}^{\otimes3}(\sum_l a(\bold{x}_{c(j,l)},\bold{x}_j)U_w\ket{\bold{x}_{c(j,l)}}+rU_w\ket{\bold{x}_j})\ket{0}\\=\sum_j \ket{j}^{\otimes3} \ket{\bold{x}'_j}\ket{0}.
\end{align}
where $\ket{\bold{x}'_j}:=rU_w\ket{\bold{x}_j}+\sum_l a(\bold{x}_{c(j,l)},\bold{x}_j)U_w\ket{\bold{x}_{c(j,l)}}$ is the updated node feature in accordance with Eqn.~\ref{atet}, by identifying $U_w\ket{\bold{x}_i}$ is the amplitude encoding of $\psi\left(\mathbf{x}_{i}\right)$, setting $\phi(\mathbf{x}, \mathbf{z})=\mathbf{W} \mathbf{x}+\mathbf{z}$, and interpreting $c(j,l)$ as the node index for the $l$-th neighbourhood of a node indexed by $j$ in the graph.\newline

In summary, by the circuit construction described so far, we obtain the following state that resembles the Graph attention operation:

\begin{equation}
\sum_j \ket{j}^{\otimes3} \ket{\bold{h}_j}=\sum_j \ket{j}^{\otimes3} \ket{\phi\left(\mathbf{x}_{j}, \bigoplus_{i \in \mathcal{N}_{j}} a\left(\mathbf{x}_{i}, \mathbf{x}_{j}\right) \psi\left(\mathbf{x}_{i}\right)\right)}.
\label{atet2}
\end{equation}
\textbf{\textit{Multi-head attention}}  \ The preceding discussions have focused on implementing single-head attention in our Quantum Graph Attention Networks. The method described here could be extended to multi-head attention following the approach outlined in Ref~\cite{liao2024gpt}.

\section{Quantum Message-Passing GNN}\label{24}
Similar to the case of Graph Attention Networks, our Quantum Message-Passing GNN aims to evaluate and store the updated node features
\begin{equation}
\label{message}    \mathbf{h}_{j}=\phi\left(\mathbf{x}_{j}, \bigoplus_{i \in \mathcal{N}_{j}} \psi\left(\mathbf{x}_{j}, \mathbf{x}_{i}\right)\right), 
\end{equation}
 into a quantum state as $\sum_j \ket{j}^{\otimes3} \ket{\bold{h}_{j}}+...$, that is, to obtain the following state:\footnote{Note that we omit the activation function in the original definition of $\phi$ in \cite{bronstein2021geometric}, the quantum implementation of the activation function is described in Section \ref{mgcn}, here we focus on the quantum implementation of this definition of $\phi$.}

\begin{align}
\sum_j \ket{j}^{\otimes3} \ket{\phi(\bold{x}_j,\bigoplus_{i \in \mathcal{N}_{j}}\psi(\bold{x}_{i},\bold{x}_j))}+...
\label{mp1}
\end{align}

\begin{figure}[h!]
    \centering
    \includegraphics[width=\linewidth]{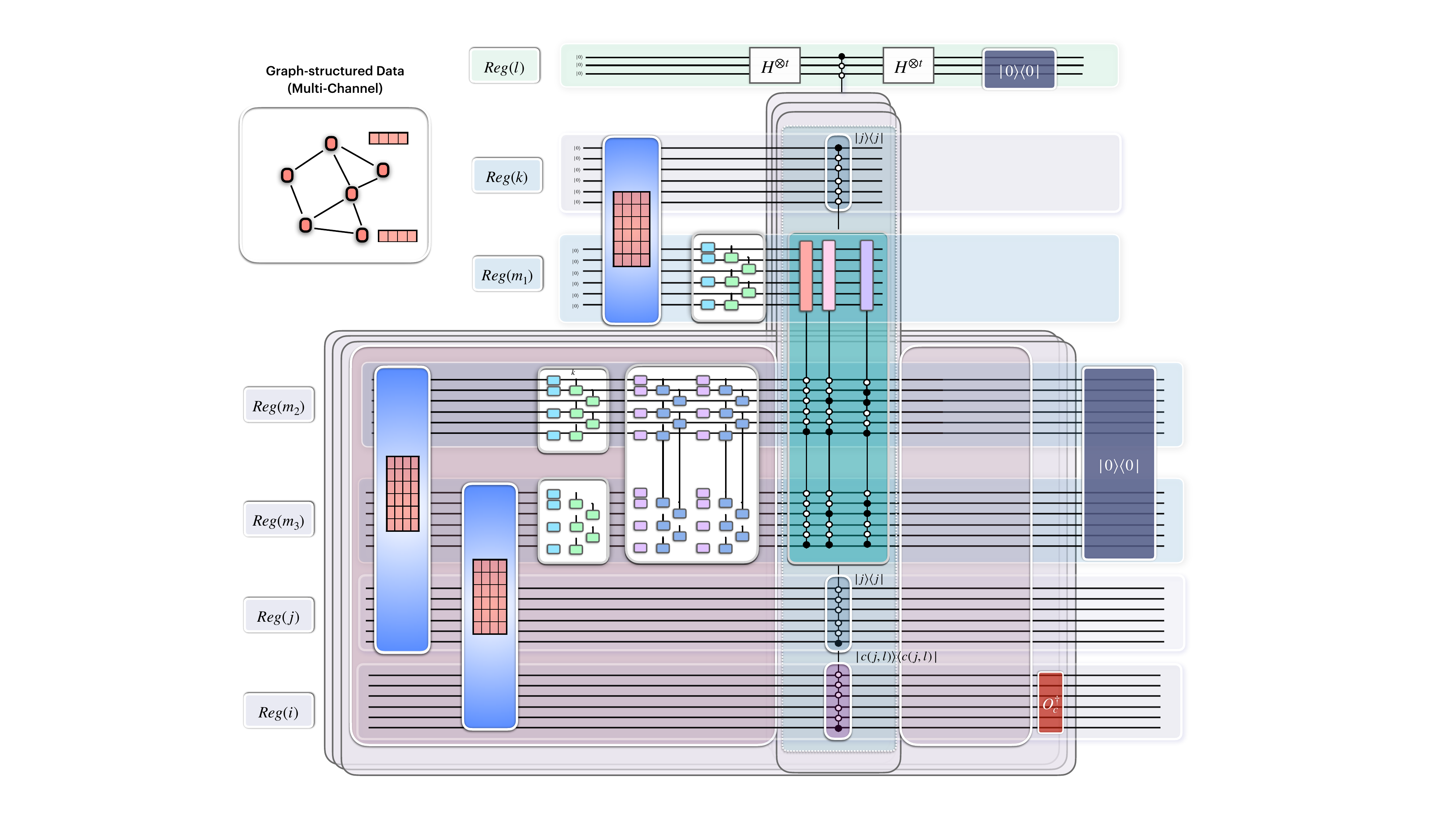}
   \caption{\textit{Quantum Algorithm for Message-Passing GNN}. Our Quantum Message-Passing GNN aims to evaluate and store the updated node features
$\mathbf{h}_{j}=\phi\left(\mathbf{x}_{j}, \bigoplus_{i \in \mathcal{N}_{j}} \psi\left(\mathbf{x}_{j}, \mathbf{x}_{i}\right)\right) 
$into a quantum state as $\sum_j \ket{j}^{\otimes3} \ket{\bold{h}_{j}}+...$
This can be achieved via the following steps: \textsf{Step 1: Data Loading of linearly transformed node features $\bold{x}_k$}; \textsf{Step 2: Selective LCU}; \textsf{Step 3: Permutation of basis}; Gathering all steps above, the Quantum Message-Passing GNN loads and transforms the node features as: $
\sum_i \sum_j \sum_k \ket{i}\ket{j}\ket{0}\ket{0}\ket{k} \to \sum_j \ket{j}^{\otimes3}\ket{\phi(\bold{x}_j,\psi(\bold{x}_{c(j,l)},\bold{x}_j))}+...$ \textsf{Step 4: Overall LCU}, we then apply the overall LCU module (depicted as the top add-on register $Reg(l)$ with the controlled unitaries in fainted blue box), to achieve the aggregation over different neighbours, obtaining the following state:$\sum_j \ket{j}^{\otimes3}\ket{\phi(\bold{x}_j,\sum_l\psi(\bold{x}_{c(j,l)},\bold{x}_j))}+...$, which can also be written as $\sum_j \ket{j}^{\otimes3}\ket{\phi(\bold{x}_j,\bigoplus_{v \in \mathcal{N}_{j}}\psi(\bold{x}_{v},\bold{x}_j))}+...$}
    \label{mpnni}
\end{figure}

\begin{figure}[h!]
    \centering
    \includegraphics[width=\linewidth]{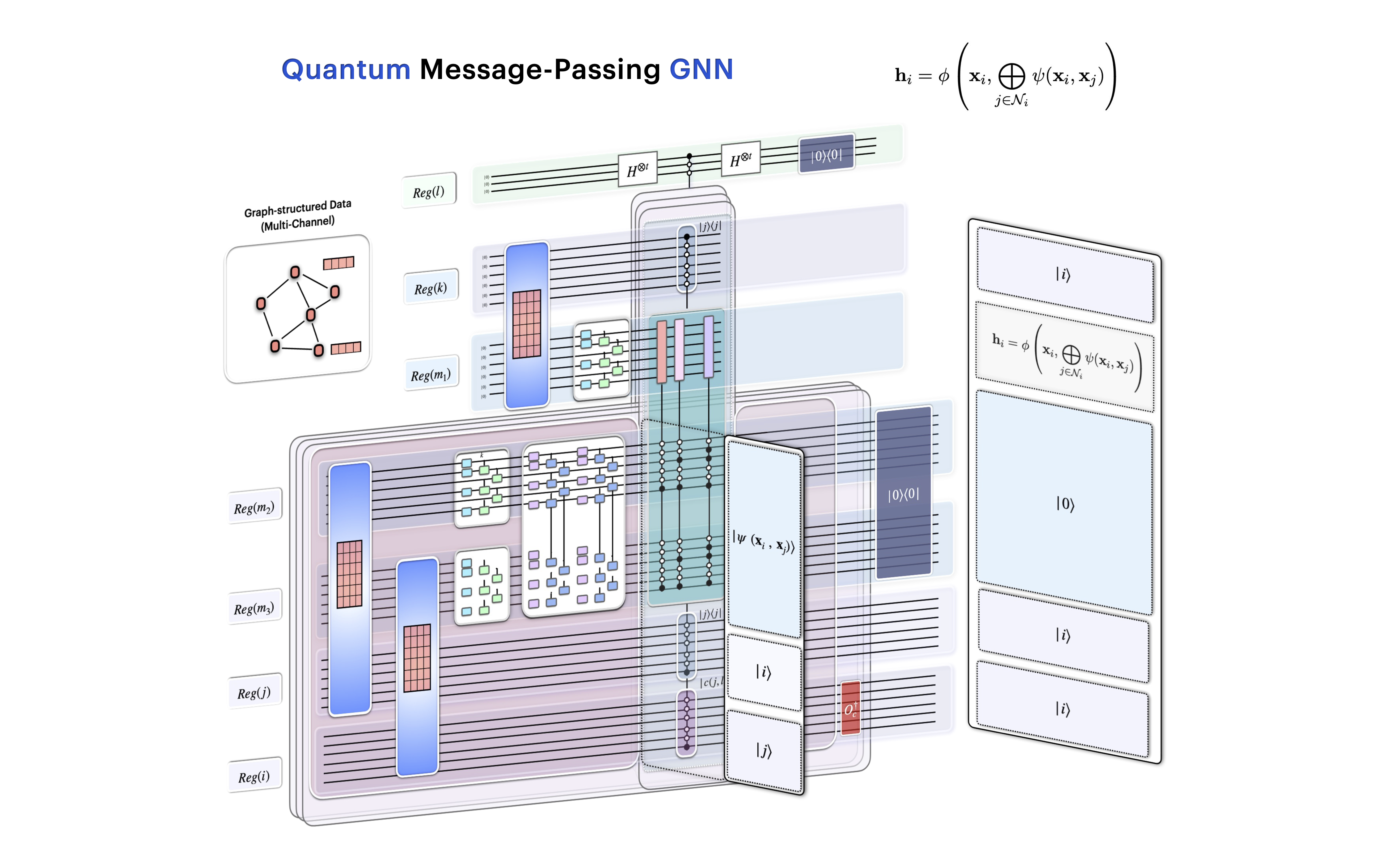}
   \caption{\textit{Quantum Algorithm for Message-Passing GNN.} This figure provides a 3D state-circuit view for Fig.~\ref{mpnni}. The panels perpendicular to the circuit plane represent the quantum states generated by corresponding circuits.}
    \label{mpnni2}
\end{figure}

This can be achieved via the following steps, as illustrated in Fig. \ref{mpnni} and \ref{mpnni2}.\newline

\textsf{Step 1: Data Loading of linearly transformed node features $\bold{x}_k$} \newline

The first step is to apply the data loading module described in Section \ref{de} on the address register $Reg(k)$ and the corresponding memory register $Reg(m_1)$ on which a parameterized quantum circuit module is then applied to linearly transform the node features. This step loads the linearly transformed node features $\bold{x}_k$ of each node into the memory register associated with address $\ket{k}$. Together with the other two address registers $Reg(i)$, $Reg(j)$ and corresponding memory registers $Reg(m_2)$, $Reg(m_3)$(which will be described in the following steps), the overall state transforms as:

\begin{equation}
\sum_i \sum_j \sum_k \ket{i}\ket{j}\ket{0}\ket{0}\ket{k} \to \sum_i \sum_j \sum_k \ket{i}\ket{j}\ket{0}\ket{\bold{x}_k}\ket{k}.
\end{equation}

\textsf{Step 2: Selective LCU}\newline 

The second step aims to implement updating each node’s feature  $\bold{x}_i$ from the vectors $\psi (\bold{x}_i,\bold{x}_j)$, as in Eq. \ref{message}. \newline

Similar to the case of Graph Attention Networks (mentioned in Section \ref{be}), in this section, we investigate the graphs with certain adjacency matrices that can be decomposed as the summation of 1-sparse matrices. After the decomposition, we index the 1-sparse matrices by $l$. For the $l$-th 1-sparse matrix, the row index of the nonzero entry in each column $j$, is denoted by $c(j, l) $. Interpreting $c(j,l)$ as the node index for the $l$-th neighbourhood of a node indexed by $j$ in the graph, aggregation over different neighbours can be formulated as summing over $l$, that is,

\begin{equation}
    \phi(\bold{x}_j,\bigoplus_{v \in \mathcal{N}_{j}}\psi(\bold{x}_{v},\bold{x}_j)):=\phi(\bold{x}_j,\sum_l\psi(\bold{x}_{c(j,l)},\bold{x}_j)).
\end{equation}

Since $\phi$ is linear in its arguments, we have,

\begin{equation}
\phi(\bold{x}_j,\sum_l\psi(\bold{x}_{c(j,l)},\bold{x}_j))=\sum_l\phi(\bold{x}_j,\psi(\bold{x}_{c(j,l)},\bold{x}_j)).
\end{equation}

This allows us to achieve the aggregation over different neighbours by the overall LCU module depicted in Fig. \ref{mpnni} and \ref{mpnni2} as the top add-on register $Reg(l)$ with the controlled unitaries in fainted blue box implementing $\phi(\bold{x}_j,\psi(\bold{x}_{c(j,l)},\bold{x}_j))$ for each $l$. For a node in the graph, we then first focus on the message-passing from one neighbour of the node represented as $\phi(\bold{x}_j,\psi(\bold{x}_{c(j,l)},\bold{x}_j))$.\newline

For each neighbour of a node, a ``selective LCU'' is performed to implement the node updating function $\phi(\bold{x}_i,\psi (\bold{x}_i,\bold{x}_j))$. This is achieved by applying the following  modules: \newline 

\textsf{Module 1}: A data loading+linear transformation module that evaluates the vector $\psi (\bold{x}_i,\bold{x}_j)$, depicted in Fig. \ref{mpnni} and \ref{mpnni2} as the pink box. This module comprises two data loading modules on address registers $\operatorname{Reg}(i)$, $Reg(j)$ and their corresponding data registers $Reg(m_2)$, $Reg(m_3)$, followed by two parametrized quantum circuits on $Reg(m_2)$, $Reg(m_3)$ respectively and an overall parametrized quantum circuits on $Reg(m_2)$, $Reg(m_3)$.\newline

This module acts as follows:

\begin{equation}
\sum_i \sum_j  \ket{i}\ket{j}\ket{0} \to \sum_i \sum_j \ket{i}\ket{j}\ket{\psi (\bold{x}_i,\bold{x}_j)}\end{equation}

\textsf{Module 2}: Selectively controlled unitaries on the three data registers, as gathered in the fainted blue box in Fig. \ref{mpnni} and \ref{mpnni2}. \newline

we can write out $\ket{\psi (\bold{x}_i,\bold{x}_j)}$ as:
\begin{equation}
\ket{\psi (\bold{x}_i,\bold{x}_j)}=\sum_p w^{ij}_p\ket{p}
\label{coeffi}
\end{equation}

and the controlled unitaries, depicted in Fig. \ref{mpnni} and \ref{mpnni2} as the multi-controlled red/purple boxes, can be written as

\begin{equation}
U_{\textsf{multi}}=\sum_p\ket{p}\bra{p} \otimes U_p
\end{equation}

where $U_p$ are some constant or trainable unitaries.\newline 

and the selective controlled unitaries are defined\footnote{The implementation of the ``Selective controlled unitaries'' can be achieved in the same way as the implementation of the ``selective copying'' operation described in Section \ref{attentionsec2}.} as:

\begin{equation}
U_{\textsf{Selective}}:=\sum_j\ket{j}\bra{j} \otimes\ket{j}\bra{j} \otimes\ket{c(j,l)}\bra{c(j,l)} \otimes U_{\textsf{multi}}.
\end{equation}

\textsf{Module 3}: Uncomputation of Module 1.\newline

\textsf{Module 4}: Projection onto zero state on $Reg(m_2)$, $Reg(m_3)$.\newline

For each specific combination of $i,j,k$, the above modules achieve LCU on $Reg(m_1)$ and act as,

\begin{equation}
\ket{\bold{x}_k} \to \sum_p |w^{ij}_p|^2 U_p\ket{\bold{x}_k}.
\label{transformed}
\end{equation}

Considering Eq.~\ref{coeffi} and the definitions of functions $\phi$ and $\psi$, We denote the transformed state in Eq.~\ref{transformed} as
\begin{equation}
\ket{\phi(\bold{x}_k,\psi(\bold{x}_{i},\bold{x}_j))}:=\sum_p|w^{ij}_p|^2 U_p\ket{\bold{x}_k}.
\end{equation}

Consider the branches indexed by $i,j,k$ in the overall state, according to the action of the selectively controlled unitaries defined in Module 2, the selective LCU only happens for the branches $i=c(j,l); k=j$.\newline

\textit{For branches} $i=c(j,l); k=j$:
\begin{equation}
\sum_j \ket{c(j,l)}\ket{j}\ket{0}\ket{\bold{x}_j}\ket{j} \to \sum_j \ket{c(j,l)}\ket{j}\ket{\psi(\bold{x}_{c(j,l)},\bold{x}_j)}\ket{\bold{x}_j}\ket{j} \to \sum_j  \ket{c(j,l)}\ket{j}\ket{0}\ket{\phi(\bold{x}_j,\psi(\bold{x}_{c(j,l)},\bold{x}_j))}\ket{j}\end{equation}
in which the node features transform as:

\begin{equation}
\ket{\bold{x}_j} \to \ket{\phi(\bold{x}_j,\psi(\bold{x}_{c(j,l)},\bold{x}_j))}.
\end{equation}

That is, the node features $\ket{\bold{x}_j}$ are updated by the ``message'' $\psi(\bold{x}_{c(j,l)},\bold{x}_j))$ from one of its neighbours indexed by $l$. \newline

 \textit{For other branches}:
 \begin{equation}
\sum_{i\neq c(j,l)} \sum_j \sum_{k\neq j}\ket{i}\ket{j}\ket{0}\ket{\bold{x}_k}\ket{k} \to \sum_{i\neq c(j,l)} \sum_j \sum_{k\neq j} \ket{i}\ket{j}\ket{\psi(\bold{x}_i,\bold{x}_j)}\ket{\bold{x}_k}\ket{k}\to \sum_{i\neq c(j,l)} \sum_j \sum_{k\neq j}\ket{i}\ket{j}\ket{0}\ket{\bold{x}_k}\ket{k}. \end{equation}

All branches combined together:
\begin{align*}
\sum_j \ket{c(j,l)}\ket{j}\ket{0}\ket{\bold{x}_j}\ket{j} +\sum_{i\neq c(j,l)} \sum_j \sum_{k\neq j}\ket{i}\ket{j}\ket{0}\ket{\bold{x}_k}\ket{k} \to 
\\
\sum_j  \ket{c(j,l)}\ket{j}\ket{0}\ket{\phi(\bold{x}_j,\psi(\bold{x}_{c(j,l)},\bold{x}_j))}\ket{j}+ \sum_{i\neq c(j,l)} \sum_j \sum_{k\neq j} \ket{i}\ket{j}\ket{0}\ket{\bold{x}_k}\ket{k}.
\end{align*}

\textsf{Step 3: Permutation of basis}\newline

We then apply a permutation of basis on register $Reg(i)$ via applying the unitary $O_{c}^{l^{\dagger}}$ (defined in Eq.\ref{ocl}) as, 

$$O_{c}^{l^{\dagger}}|c(j, l)\rangle=|j\rangle.$$

when acting on the output state of Step 2, it transforms the state as follows:
\begin{align*}
\sum_j \ket{c(j,l)}\ket{j}\ket{0}\ket{\phi(\bold{x}_j,\psi(\bold{x}_{c(j,l)},\bold{x}_j))}\ket{j}+ \sum_{i\neq c(j,l)} \sum_j \sum_{k\neq j} \ket{i}\ket{j}\ket{0}\ket{\bold{x}_k}\ket{k}\to\\\sum_j  \ket{j}\ket{j}\ket{0}\ket{\phi(\bold{x}_j,\psi(\bold{x}_{c(j,l)},\bold{x}_j))}\ket{j}+ \sum_{i\neq c(j,l)} \sum_j \sum_{k\neq j} \ket{P(i)}\ket{j}\ket{0}\ket{\bold{x}_k}\ket{k}.
\end{align*}
where $\ket{P(i)}:=O_{c}^{l^{\dagger}}\ket{i}$.\newline

The state evolution during the above steps can be summarized as follows:

\begin{align*}
\sum_i \sum_j \sum_k \ket{i}\ket{j}\ket{0}\ket{0}\ket{k} \to \sum_i \sum_j \sum_k \ket{i}\ket{j}\ket{0}\ket{\bold{x}_k}\ket{k}=\\\sum_j \ket{c(j,l)}\ket{j}\ket{0}\ket{\bold{x}_j}\ket{j} +\sum_{i\neq c(j,l)} \sum_j \sum_{k\neq j}\ket{i}\ket{j}\ket{0}\ket{\bold{x}_k}\ket{k} \to 
\\
\sum_j  \ket{c(j,l)}\ket{j}\ket{0}\ket{\phi(\bold{x}_j,\psi(\bold{x}_{c(j,l)},\bold{x}_j))}\ket{j}+ \sum_{i\neq c(j,l)} \sum_j \sum_{k\neq j} \ket{i}\ket{j}\ket{0}\ket{\bold{x}_k}\ket{k}\to\\\sum_j  \ket{j}\ket{j}\ket{0}\ket{\phi(\bold{x}_j,\psi(\bold{x}_{c(j,l)},\bold{x}_j))}\ket{j}+ \sum_{i\neq c(j,l)} \sum_j \sum_{k\neq j} \ket{P(i)}\ket{j}\ket{0}\ket{\bold{x}_k}\ket{k}.
\end{align*}

Gathering all the steps above, the Quantum message passing GNN load and transforms the node features as:

\begin{equation}
\sum_i \sum_j \sum_k \ket{i}\ket{j}\ket{0}\ket{0}\ket{k} \to \sum_j \ket{j}^{\otimes3}\ket{\phi(\bold{x}_j,\psi(\bold{x}_{c(j,l)},\bold{x}_j))}+...
\end{equation}
where we have neglected some registers that are unchanged.\newline

\textsf{Step 4: Overall LCU}\newline

We then apply the aforementioned overall LCU module (depicted in Fig. \ref{mpnni} and \ref{mpnni2} as the top add-on register $Reg(l)$ with the controlled unitaries in the faint blue box), to achieve the aggregation over different neighbours, obtaining the following state:

\begin{equation}
\sum_j \ket{j}^{\otimes3}\ket{\phi(\bold{x}_j,\sum_l\psi(\bold{x}_{c(j,l)},\bold{x}_j))}+...
\end{equation}
which can also be written as,
\begin{align}
\sum_j \ket{j}^{\otimes3}\ket{\phi(\bold{x}_j,\bigoplus_{v \in \mathcal{N}_{j}}\psi(\bold{x}_{v},\bold{x}_j))}+...
\end{align}

That is, through our Quantum Message passing GNN, we obtained the desired state in Eq. \ref{mp1}.

\section{Complexity Analysis}\label{complexity}

\subsection{Complexity of classical GCNs}

The Graph convolution described in Eq.\ref{gcneq} can be decomposed into three operations:
\begin{enumerate}
    \item $Z^{(l)} = H^{(l)} W^{(l)}$: node-wise feature transformation
    \item $H'^{(l)} = \hat{A} Z^{(l)}$: neighborhood aggregation
    \item $\sigma(\cdot)$: activation function 
\end{enumerate}

Operation 1 is a dense matrix multiplication between matrices of size $N \times F_{l}$ and $F_{l} \times F_{l+1}$. Assuming $F_{l} = F_{l+1} = C$ for all $l$, the time complexity for this operation is $O(NC^2)$. Considering $\hat{A}$ is typically sparse, Operation 2 has a time complexity of $O(|E|C)$ ($|E|$ is the number of edges in the graph) Considering $|E|=Nd$ where $d$ is the average degree of the nodes in the graph, we have $O(|E|C)=O(Nd C)$. Operation 3 is an element-wise function, and the time complexity is $O(N)$. For $K$ layers, the overall time complexity is $O(KNC^2 + K|E|C + KN) = O(KNC^2 + K|E|C)$. The space complexity is $O(|E|+KC^2 +KNC)$.~\cite{blakely2021time}\newline

\subsection{Complexity analysis of Quantum SGC}
In this section, we present the complexity results of the quantum implementation of a Simplified Graph Convolution (SGC) network\cite{wu2019simplifying} which reduces the complexity of Graph Convolutional Networks (GCNs) by removing nonlinearities while maintaining comparable or even superior performance. \newline

For node classification, the prediction generated by the SGC model is $\hat{Y}_{\text{SGC}} = \text{softmax}(S^K X \Theta)$ where $S = \hat{A}$ is the normalized adjacency matrix with added self-loops, $X \in \mathbb{R}^{N \times C}$ is the node attribute matrix, $\Theta$ is a weight matrix, and $K$ is the number of layers(a constant). Similar to quantum GCN, the quantum SGC comprises the following key components: data encoding of the node attribute matrix $X$, a quantum circuit for implementing $S^K$, a parameterized quantum circuit for the weight matrix $\Theta$, and cost function evaluation. Table \ref{tab:quantum_gcn_component_complexity} summarizes the quantum algorithmic techniques, number of ancillary qubits, and circuit depths for these components of the quantum SGC. \newline

\begin{table}[h]
\centering
\resizebox{\textwidth}{!}{%
\begin{tabular}{|l|c|c|c|c|}
\hline
\textbf{Component} & \textbf{Algorithmic Technique} & \textbf{Number of ancillary Qubits} & \textbf{Circuit Depth} & \textbf{References} \\
\hline
Data Encoding & State Preparation & $\Omega(\log(NC)) \leq n_{\text{anc}} \leq O(NC)$ & $\tilde{O}(NC \log(1/\varepsilon_1) \log(n_{\text{anc}}) / n_{\text{anc}})$ &  \cite{zhang2024circuit} \\
\hline
Simplified  & Block-Encoding & $\Omega(\log N) \leq n_{\text{anc}}' \leq O(N \log N \cdot s \log s)$ & $\tilde{O}(N \log N \cdot s \log s \log(1/\varepsilon_2) \log n_{\text{anc}}' / n_{\text{anc}}')$ & \cite{zhang2024circuit}\\
Graph Convolution & PQC  & & & \\

\hline
Cost Function & Modified Hadamard Test & 1 & $O(\log(1/\delta) / \epsilon^2)$ & \cite{PhysRevA.107.062424,Luongo2023}\\
Evaluation & & &(Query  complexity)  & \\
\hline
\end{tabular}%
}
\caption{Complexity overview for each component of the quantum SGC. Here, $N$ is the number of nodes, $C$ is the number of features per node, $s$ is the maximum number of nonzero elements at each row and column of $S$. $\varepsilon_1$ and $\varepsilon_2$ are the precision parameters, $n_{\text{anc}}$ and $n_{\text{anc}}'$ are the number of ancillary qubits for state preparation and block-encoding, respectively. $\delta$ and $\epsilon$ are the probability parameters and precision parameters for the Modified Hadamard Test. $\tilde{O}$ suppresses doubly logarithmic factors of $n_{\text{anc}}$ and $n_{\text{anc}}'$ \cite{zhang2024circuit}.}
\label{tab:quantum_gcn_component_complexity}
\end{table}

\tocless\subsubsection{Complexity for a single forward pass}

For a single forward pass, we analyze the complexity of the quantum SGC in terms of circuit depth, total number of qubits, and compare it with the classical SGC, assuming $K=2$.\newline

\textit{Circuit Depth.}

As aforementioned in section \ref{sgc}, we assume that the depth of the parameterized quantum circuit for the weight matrix $\Theta$ is less than the depth of the block-encoding circuit for $S^K$. The circuit depth of the quantum SGC is determined by the data encoding step and the block-encoding circuit to implement $S^2$. \newline

The circuit depth for the data encoding is $\tilde{O}(NC \log(1/\varepsilon_1) \log(n_{\text{anc}}) / n_{\text{anc}})$, where $\Omega(\log(NC)) \leq n_{\text{anc}} \leq O(NC)$.\cite{zhang2024circuit}\newline

The block-encoding of $S^2$ has a circuit depth of $\tilde{O}(N \log N \cdot s \log s \log(1/\varepsilon_2) \log n_{\text{anc}}' / n_{\text{anc}}')$, where $s$ is the sparsity of $S$, $\varepsilon_2$ is the precision parameter, and $n_{\text{anc}}'$ is the number of ancillary qubits used in the block-encoding with $\Omega(\log N) \leq n_{\text{anc}}' \leq O(N \log N \cdot s \log s)$.\cite{zhang2024circuit}\newline

The total circuit depth of the quantum SGC becomes:

\begin{align}
        \text{Depth}_{\text{Q-SGC}} = \tilde{O}(NC \log(1/\varepsilon_1) \log(n_{\text{anc}}) / n_{\text{anc}}+ N \log N \cdot s \log s \log(1/\varepsilon_2) \log n_{\text{anc}}' / n_{\text{anc}}').
\end{align}

The classical SGC($K=2$) has a time complexity of:

\begin{equation}
    \text{Time}_{\text{C-SGC}} = O(|E|C + NC^2) \log(1/\varepsilon')),
\end{equation}
where $\varepsilon'$ is the precision parameter.
\newline

\textit{Total Number of Qubits.} 

The total number of qubits required for the quantum SGC includes the qubits for encoding the node attributes ($\log (NC)$), the ancillary qubits for the data encoding ($n_{\text{anc}}$), and the ancillary qubits for the block-encoding of $S^2$ ($2n_{\text{anc}}'$). The total number of qubits is\footnote{One can potentially reuse the ancillary qubits in the data encoding for block-encoding of $S^2$, reducing the total number of qubits. We leave this for future work.}:

\begin{equation}
    \text{Qubits}_{\text{Q-SGC}} = O(\log (NC) + n_{\text{anc}}+n_{\text{anc}}')),
\end{equation}

where $\Omega(\log(NC)) \leq n_{\text{anc}} \leq O(NC)$ and $\Omega(\log N) \leq n_{\text{anc}}' \leq O(N \log N \cdot s \log s)$.\newline

The classical SGC has a space complexity\footnote{For all space complexities we assume fixed precisions.} of:

\begin{equation}
    \text{Space}_{\text{C-SGC}} = O(|E|+NC+C^2).
\end{equation}

\textit{Space-Time Trade-off}.

The quantum SGC offers a space-time trade-off depending on the choice of the number of ancillary qubits $n_{\text{anc}}$ and $n_{\text{anc}}'$ used in the data encoding and block-encoding, respectively. We first consider two extreme scenarios:\newline

1. Minimizing Circuit Depth (Time): To minimize the circuit depth, we choose the maximum number of ancillary qubits for the data encoding and block-encoding, i.e., $n_{\text{anc}} = O(NC)$ and $n_{\text{anc}}' = O(N \log N \cdot s \log s)$. Substituting these values into the total circuit depth equation, we get:

\begin{align}
    \text{Depth}_{\text{Q-SGC (Min. Depth)}} &= \tilde{O}(\log(1/\varepsilon_1) \log(NC)) + \tilde{O}(\log(1/\varepsilon_2) \cdot \log (N \log N \cdot s \log s)).
\end{align}

The total number of qubits in this case is $O( NC + N \log N \cdot s \log s)$.\newline

2. Minimizing Ancillary Qubits (Space): To minimize the number of ancillary qubits, we choose the minimum number of ancillary qubits for the data encoding and block-encoding, i.e., $n_{\text{anc}} = \Theta(\log(NC))$ and $n_{\text{anc}}' = \Theta(\log N)$. Substituting these values into the total circuit depth equation, we get:

\begin{align}
    \text{Depth}_{\text{Q-SGC (Min. Qubits)}} &= \tilde{O}(NC \log(1/\varepsilon_1) / \log(NC)) + \tilde{O}(N s \log s \log(1/\varepsilon_2)).
\end{align}

The total number of qubits in this case is $O(\log(NC))$.\newline

For moderate scenarios balancing circuit depth and the number of ancillary qubits, for example, we can choose $n_{\text{anc}} = \Theta(\sqrt{NC})$ and $n_{\text{anc}}' = \Theta(\sqrt{N \log N \cdot s \log s})$. Substituting these values into the total circuit depth equation, we get:

\begin{align}
\text{Depth}_{\text{Q-SGC (Moderate)}} &= \tilde{O}(\sqrt{NC} \log(1/\varepsilon_1)\log({NC})) + \sqrt{N \log N \cdot s \log s} \log(1/\varepsilon_2)\log({N  s})).
\end{align}

The total number of qubits in this case is $O(\sqrt{NC} + \sqrt{N \log N \cdot s \log s})$. \newline

\tocless\subsubsection{Cost Function Evaluation and its Complexity}

The cost function of the quantum SGC is evaluated using the Modified Hadamard test \cite{PhysRevA.107.062424,Luongo2023} as follows: \newline

\textbf{Modified Hadamard Test \cite{PhysRevA.107.062424, Luongo2023} } 
\textit{Assume to have access to a unitary \( U_1 \) that produces a state \( U_1|0\rangle = |\psi_1\rangle \) and a unitary \( U_2 \) that produces a state \( U_2|0\rangle = |\psi_2\rangle \), where \( |\psi_1\rangle, |\psi_2\rangle \in \mathbb{C}^N \) for \( N = 2^n, n \in \mathbb{N} \). There is a quantum algorithm that allows estimating the quantity \( \langle \psi_1 | \psi_2 \rangle \) with additive precision \( \epsilon \) using controlled applications of \( U_1 \) and \( U_2 \)    $O\left(\frac{\log(1/\delta)}{\epsilon^2}\right)$ times, with probability \( 1 - \delta \).}\newline

In our quantum SGC, $U_1$ is the circuit for SGC (described in the previous sections), and $U_2$ is a state preparation unitary for the target label state. The Modified Hadamard test allows us to estimate the quantity $(\psi_1|\psi_2)$ with additive precision $\epsilon$ using controlled applications of $U_1$ and $U_2$ $O(\frac{\log(1/\delta)}{{\epsilon}^2})$ times, with probability $1 - \delta$, where $\psi_1$ and $\psi_2$ are the states produced by $U_1$ and $U_2$, respectively.\newline 

The time complexity and space complexity of $U_2$ is smaller than those of $U_1$, as $U_2$ only needs to prepare the target label state, while $U_1$ performs the entire SGC computation. Therefore, the overall complexity of the cost function evaluation is dominated by the complexity of $U_1$.\newline

Assuming fixed precision $\epsilon$  for the Modified Hadamard test, the time complexity of the cost function evaluation for the quantum SGC is:

\begin{equation}
    \text{Time}_{\text{Q-SGC-Cost}} = \tilde{O}(\log(1/\delta) \cdot \text{Depth}_{\text{Q-SGC}}),
\end{equation}

where $\text{Depth}_{\text{Q-SGC}}$ is the circuit depth of the quantum SGC. Plugging in the results from the previous section, we get:

\begin{align}
    \text{Time}_{\text{Q-SGC-Cost (Min. Depth)}} &= \tilde{O}(\log(1/\delta) \cdot (\log(NC) + \log (Ns)) ), \\
    \text{Time}_{\text{Q-SGC-Cost (Min. Qubits)}} &= \tilde{O}(\log(1/\delta) \cdot (NC / \log(NC) + N s \log s )).
\end{align}

The space complexity of the cost function evaluation for the quantum SGC is the same as the space complexity of the quantum SGC itself, as the Modified Hadamard test requires only one  additional qubits. Therefore we have:

\begin{align}
    \text{Space}_{\text{Q-SGC-Cost (Min. Depth)}} &= O( NC + N \log N \cdot s \log s), \\
    \text{Space}_{\text{Q-SGC-Cost (Min. Qubits)}} &= O(\log(NC)).
\end{align}

For the classical SGC, the complexity of the cost function evaluation is less than that of in the forward pass, therefore can be omitted in the overall complexity.\newline

Table \ref{tab:quantum_vs_classical_sgc_cost_function_evaluation} summarizes the complexity comparison between the quantum SGC and the classical SGC for the cost function evaluation, considering the case of $K=2$, with  fixed precision parameters.

\begin{table}[h]
\centering
\resizebox{\textwidth}{!}{%
\begin{tabular}{|l|c|c|}
\hline
\textbf{Algorithm} & \textbf{Time Complexity} & \textbf{Space Complexity} \\
\hline
Quantum SGC (Min. Depth) & $\tilde{O}(\log(1/\delta) \cdot(\log(NC) + \log (Ns)))$ & $O( NC + N \log N \cdot s \log s)$ \\
\hline
Quantum SGC (Min. Qubits) & $\tilde{O}(\log(1/\delta) \cdot(NC / \log(NC) + N s \log s) )$ & $O(\log(NC) )$ \\
\hline
Classical SGC & $O(|E|C + NC^2))=O(NdC + NC^2)$ & $O(|E|+NC + C^2)=O(Nd+NC + C^2)$ \\
\hline
\end{tabular}%
}
\caption{Complexity comparison between Quantum SGC and Classical SGC ($K=2$) for a single forward pass and cost function evaluation, assuming fixed precision parameters. $N$ is the number of nodes, $C$ is the number of features per node. $d$ is the average degree of the nodes in the graph. $s$ is the maximum number of non-zero elements in each row/column of $\hat{A}$, and $d<s<N$. The quantum SGC provides a probabilistic result with a success probability of $1 - \delta$. Note that in the classical time complexity, at first glance, $O(NC^2)$ appears to be the dominating term, as the average degree d on scale-free networks is usually much smaller than C and hence $NC^2 > NdC$. However, in practice, node-wise feature transformation can be executed at a reduced cost due to the parallelism in dense-dense matrix multiplications. Consequently, $O (NdC) $ is the dominant complexity term in the time complexity of classical SGC and the primary obstacle to achieving scalability \cite{chen2020scalable}. }
\label{tab:quantum_vs_classical_sgc_cost_function_evaluation}
\end{table}

\subsection{Complexity analysis of Quantum LGC}
The complexity analysis of the quantum LGC is similar to that of the quantum SGC in the previous section.\newline

The data encoding step has a circuit depth of $\tilde{O}(NC \log(1/\varepsilon_1) \log(n_{\text{anc}}) / n_{\text{anc}})$ using $\Omega(\log(NC)) \leq n_{\text{anc}} \leq O(NC)$ ancillary qubits, where $N$ is the number of nodes, $C$ is the number of features per node, and $\varepsilon_1$ is the precision parameter \cite{zhang2024circuit}.\newline

The block-encoding of $L$, denoted as $U_L$, has a circuit depth of $\tilde{O}(N \log N \cdot s \log s \log(1/\varepsilon_2) \log n_{\text{anc}}' / n_{\text{anc}}')$, where $s$ is the maximum number of non-zero elements in the rows/columns of $L$, $\varepsilon_2$ is the precision parameter, and $n_{\text{anc}}'$ is the number of ancillary qubits used in the block-encoding with $\Omega(\log N) \leq n_{\text{anc}}' \leq O(N \log N \cdot s \log s)$ \cite{zhang2024circuit}.\newline

We utilize the ``Polynomial eigenvalue transformation'', a special instance of the Quantum Singular Value Transformation (QSVT) (Theorem 56 in \cite{Gily_n_20192}), to implement $\sum_{i=0}^K \alpha_i L^i$. The depth of the circuit for the block encoding of $P(L)$ is $K$ times the depth of the block-encoding $U_L$ plus $O((n_{\text{anc}}' + 1)K)$ for the additional one- and two-qubit gates. Therefore, the total circuit depth of the quantum LGC is:

\begin{align}
\text{Depth}_{\text{Q-LGC}} = \tilde{O}(NC \log(1/\varepsilon_1) \log(n_{\text{anc}}) / n_{\text{anc}} + kN \log N \cdot s \log s \log(1/\varepsilon_2) \log n_{\text{anc}}' / n_{\text{anc}}' + Kn_{\text{anc}}'),
\end{align}

where $\varepsilon_2$ is the precision parameter for the block-encoding, and $\Omega(\log N) \leq n_{\text{anc}}' \leq O(N \log N \cdot s \log s)$.\newline

The total number of qubits required for the quantum LGC is:

\begin{align}
\text{Qubits}_{\text{Q-LGC}} = O(\log(NC) + n_{\text{anc}}+ n_{\text{anc}}'),
\end{align}
where $\Omega(\log(NC)) \leq n_{\text{anc}} \leq O(NC)$ and $\Omega(\log N) \leq n_{\text{anc}}' \leq O(N \log N \cdot s \log s)$.\newline

The classical LGC time and space complexities are:

\begin{align}
\text{Time}_{\text{C-LGC}} &= O(K|E|C + NC^2), \\
\text{Space}_{\text{C-LGC}} &= O(|E|+KNC + C^2).
\end{align}

The quantum LGC offers a space-time trade-off depending on the choice of the number of ancillary qubits $n_{\text{anc}}$ and $n_{\text{anc}}'$ used in the data encoding and block-encoding, respectively. We consider the scenario Minimizing Ancillary Qubits (Space): Choosing $n_{\text{anc}} = \log(NC)$ and $n_{\text{anc}}' = \log N$:

\begin{align}
\text{Depth}_{\text{Q-LGC (Min. Qubits)}} &= \tilde{O}(NC \log(1/\varepsilon_1)/\log(NC) + KN \cdot s \log s \log(1/\varepsilon_2)+K \log N), \\
\text{Qubits}_{\text{Q-LGC (Min. Qubits)}} &= O(\log(NC)).
\end{align}

Similar to the analysis of SGC, Table \ref{tab:complexity_comparison_fixed_precision} summarizes the time and space complexities of a single forward pass and cost function evaluation, for classical LGC and quantum LGC with Min. Qubits, assuming fixed precision parameters and success probability(in the quantum cases).

\begin{table}[h]
\centering
\begin{tabular}{|l|c|c|}
\hline
Method & Time Complexity & Space Complexity \\
\hline
Classical LGC & $O(K|E|C + NC^2)$ & $O(|E|+KNC + C^2)$ \\
\hline
Quantum LGC (Min. Qubits) & $\tilde{O}(NC/\log(NC) + KN \cdot s \log s)$ & $O(\log(NC))$ \\
\hline
\end{tabular}
\caption{Time and space complexity comparison for classical LGC and quantum LGC with Min. Qubits, assuming fixed precision parameters and success probability(in the quantum cases).}
\label{tab:complexity_comparison_fixed_precision}
\end{table}

\section{Conclusion}\label{conclusion}

\tocless\subsection{Related works and our contributions }\label{related}

The area of Quantum Graph Neural Networks (QGNNs) has recently emerged as a promising avenue to leverage the power of quantum computing for graph representation learning. In this section, we provide an overview of relevant works in this area and highlight the key contributions of our work.\newline

Verdon et al. \cite{verdon2019quantum} proposed one of the first QGNN architectures, introducing a general framework based on Hamiltonian evolutions. While their work demonstrated the use of QGNNs for tasks like \textit{learning quantum dynamics, creating multipartite entanglement in quantum networks, graph clustering, and graph isomorphism classification}, our architectures are specifically designed for tasks like \textit{node classification} on classical graph-structured data. Furthermore, \cite{verdon2019quantum} suggests several future research directions for QGNNs, including quantum-optimization-based training and extending their QSGCNN to multiple features per node. Our work makes progress on both of these aspects: The design of our architectures enables quantum-optimization-based training \cite{liao2021quantum} for our quantum GNNs, and our quantum GNN architectures natively support multiple features per node. While \cite{verdon2019quantum} provides a general framework for QGNNs, our work delves into the specifics of designing quantum circuits that closely mimic the functionality of classical GNNs and analyzes their potential quantum advantages, thus advancing the field in a complementary direction.\newline

Beer et al. \cite{beer2021quantum} designed quantum neural networks specifically for graph-structured quantum data. In contrast, our QGNNs are primarily designed to handle classical graph-structured data. Skolik et al. \cite{skolik2023equivariant} proposed a PQC ansatz for learning on weighted graphs that respect equivariance under node permutations. In their ansatz, the node features are encoded in the rotation angles of the $R_x$ gates, whereas in our GNN architecture, the node features are encoded directly in the amplitudes of the quantum state, enabling the usage of quantum linear algebra for the subsequent transformation.\newline

 Ai et al. \cite{ai2022decompositional} proposed DQGNN, which decomposes large graphs into smaller subgraphs to handle the limited qubit availability on current quantum hardware, however as mentioned in section \ref{intro}, subsampling techniques have reliability issues--it is challenging to guarantee that the subgraphs preserve the semantics of the entire graph and provide reliable gradients for training GNNs\cite{joshi2022efficientgnns}. Tuysuz et al. \cite{tuysuz2021hybrid} introduced a hybrid quantum-classical graph neural network (HQGNN) for particle track reconstruction. Mernyei et al. \cite{mernyei2022equivariant} proposed equivariant quantum graph circuits (EQGCs) as a unifying framework for QGNNs. In the niche of a quantum graph \textit{convolutional} neural networks, detailed comparisons between our work (specifically, quantum GCN/SGC/LGC) and three other related works are provided in Appendix \ref{relatedgcn}.\newline
 
In summary, while these related works share the high-level goal of developing quantum neural networks for graph-structured data, our work makes distinct contributions in the following aspects:
\newline

First, we propose novel QGNN architectures that are specifically designed to mirror the structure and functionality of popular classical GNN variants, namely Graph Convolutional Networks (GCNs), Graph Attention Networks (GATs), and Message Passing Neural Networks (MPNNs). This allows us to leverage the proven effectiveness of these architectures while harnessing the power of quantum computing. \newline

Second, our quantum GNN architectures go beyond generic parameterized quantum circuits: For example, in our quantum graph convolutional networks, we employ Quantum Singular Value Transformation (QSVT) circuits to implement the spectral graph convolutions; in our quantum graph attention networks, we construct quantum circuits to evaluate and store attention scores, allowing the incorporation of self-attention mechanisms.\newline

Third, we provide a detailed theoretical analysis of the potential quantum advantages of our quantum SGC and LGC architectures in terms of time and space complexity. This analysis offers new insights into the scalability and efficiency benefits of our quantum SGC and LGC compared to their classical counterparts.\newline

To conclude, our work makes significant contributions to the field of QGNNs by introducing beyond-generic-parameterized-quantum-circuits architectures aligned with classical GNNs, and providing theoretical complexity analysis. These advances complement and extend the existing literature on Quantum Graph Neural Networks and lay the foundation for further research in this promising area.\newline

\tocless\subsection{Summary and Outlook}

In this paper, we have introduced novel frameworks for implementing scalable Graph Neural Networks on quantum computers, drawing inspiration from the three fundamental types of classical GNNs: Graph Convolutional Networks, Graph Attention Networks, and Message-Passing Neural Networks. Our Quantum GNN architectures have the potential to achieve significant improvements in time and space complexities compared to their classical counterparts, offering a promising solution to address the scalability challenges faced by classical GNNs.\newline

The complexity analysis of our quantum implementation of a Simplified Graph Convolution (SGC) network demonstrates the potential for quantum advantage: when optimizing for minimal qubit usage, the quantum SGC exhibits space complexity logarithmic in the input sizes, offering a substantial reduction in space complexity compared to classical GCNs, while still providing better time complexity; when optimizing for minimal circuit depth, the quantum SGC achieves logarithmic time complexity in the input sizes, albeit at the cost of linear space complexity. The trade-off between circuit depth and qubit usage in the quantum implementation provides flexibility in adapting to specific quantum hardware constraints and problem instances. These complexity results suggest that our quantum GNN frameworks have the potential to efficiently process large-scale graphs that are intractable for classical GNNs, opening new possibilities for analyzing graph-structured data. Furthermore, by incorporating inductive biases tailored to graph-structured data, our Quantum GNNs align with the principles of Geometric Quantum Machine Learning and have the potential to improve upon problem-agnostic quantum machine learning models. \newline

Future research directions include further analysis and empirical evaluations to assess the performance and scalability of our Quantum GNN architectures in more general settings and real-world applications. Additionally, investigating the integration of more advanced classical GNN architectures into the quantum domain could lead to even more powerful Quantum GNN models. \newline 

In conclusion, this work lays the foundation for harnessing the potential of quantum computing in graph representation learning. As quantum hardware continues to advance, we anticipate that our Quantum GNN frameworks will offer promising avenues for addressing the limitations of classical GNNs and pave the way for the development of scalable and efficient quantum-enhanced graph learning algorithms.\newline

\appendix

\renewcommand{\appendixpagename}{\textbf{APPENDIX}}
\begin{center}  
\appendixpagename
\end{center}

\section{Implementation of the ``selective copying'' operation}\label{selective}

\begin{figure}[h!]
    \centering
    \includegraphics[width=0.9\linewidth]{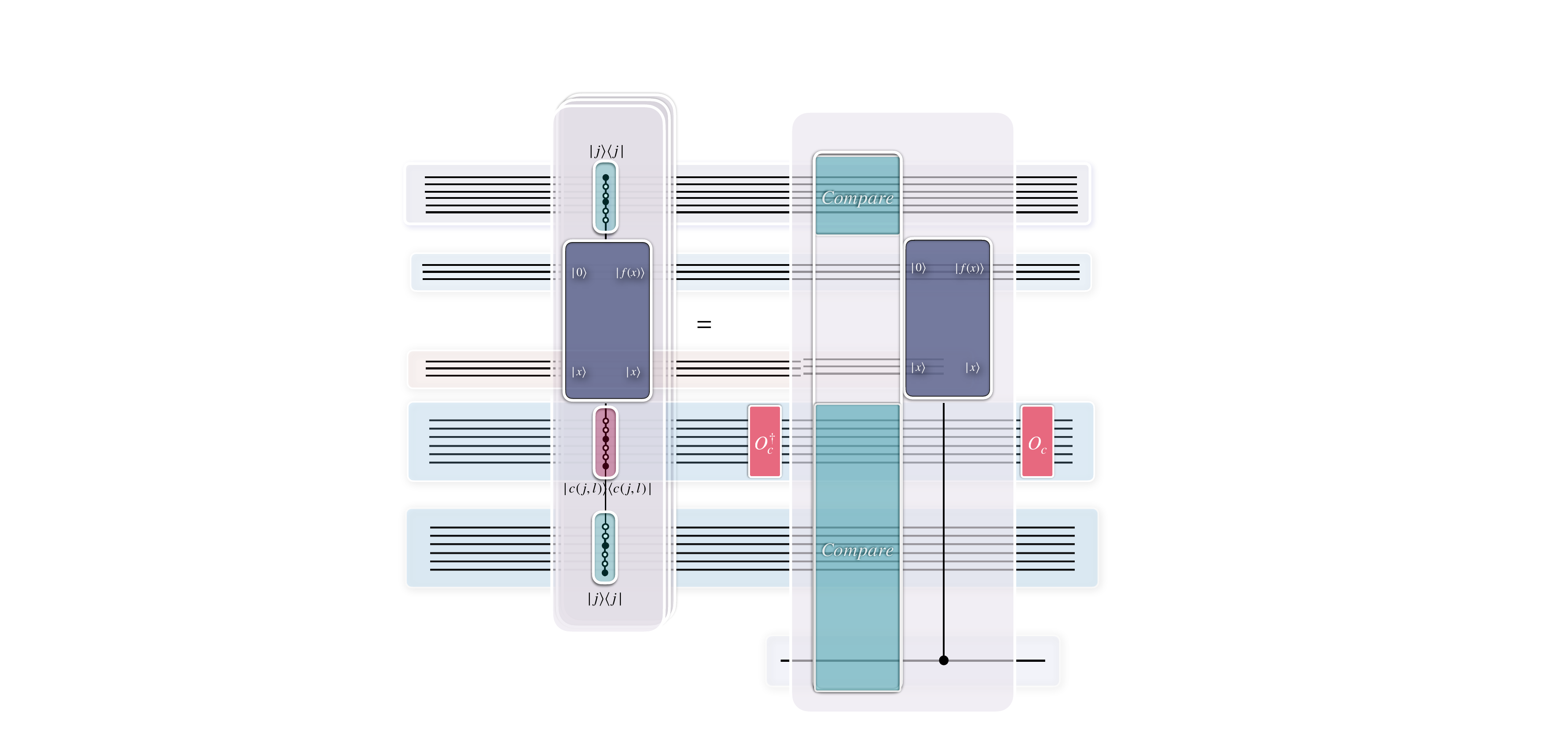}
   \caption{The multiple multi-controlled unitaries for the selective copying can be implemented by a circuit with constant depth.}
    \label{a1}
\end{figure}

In this section, we show that the selective copying operation can be implemented by a circuit with constant depth, as depicted in Fig.\ref{a1}.

\begin{figure}[h!]
    \centering
    \includegraphics[width=0.9\linewidth]{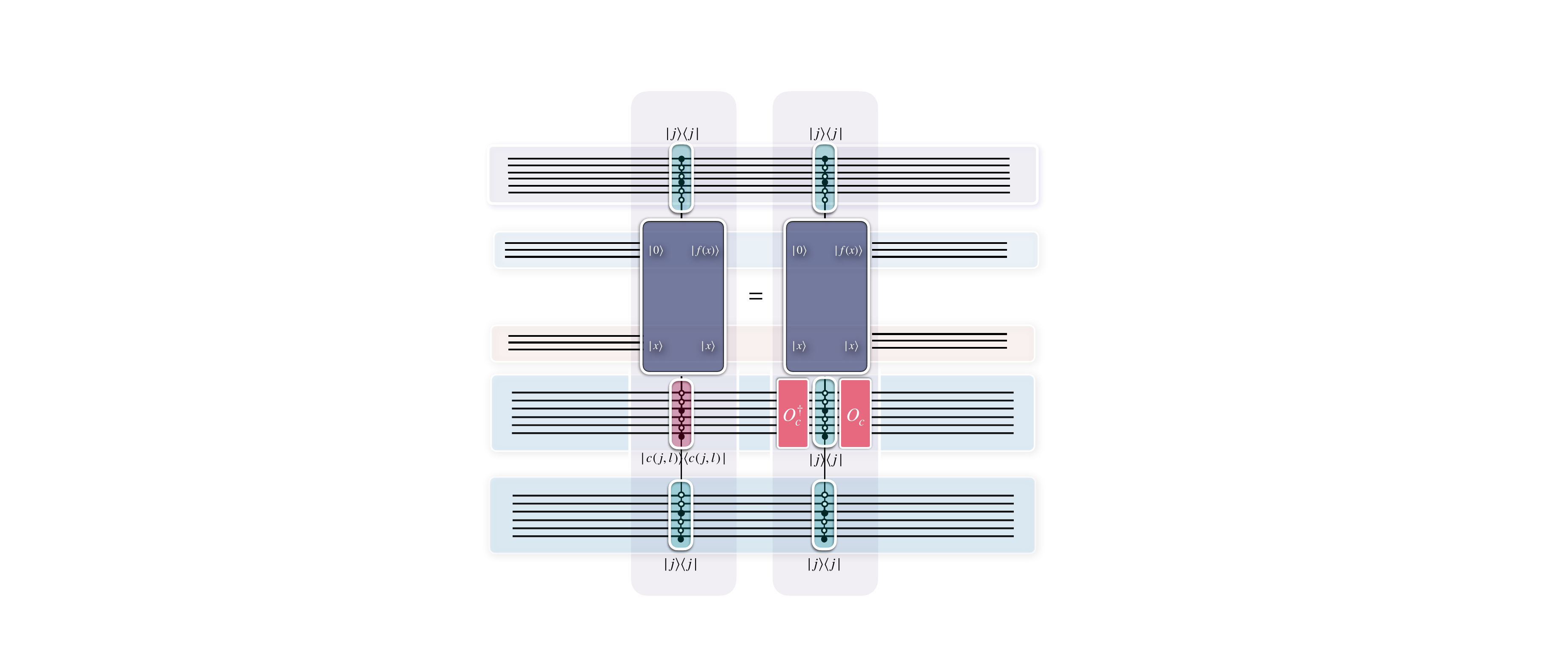}
   \caption{For each $j$, the multi-controlled unitaries can be rewritten as in this figure.}
    \label{a2}
\end{figure}

First, for each $j$, the multi-controlled unitaries can be rewritten as in Fig.\ref{a2}.

\begin{figure}[h!]
    \centering
    \includegraphics[width=0.9\linewidth]{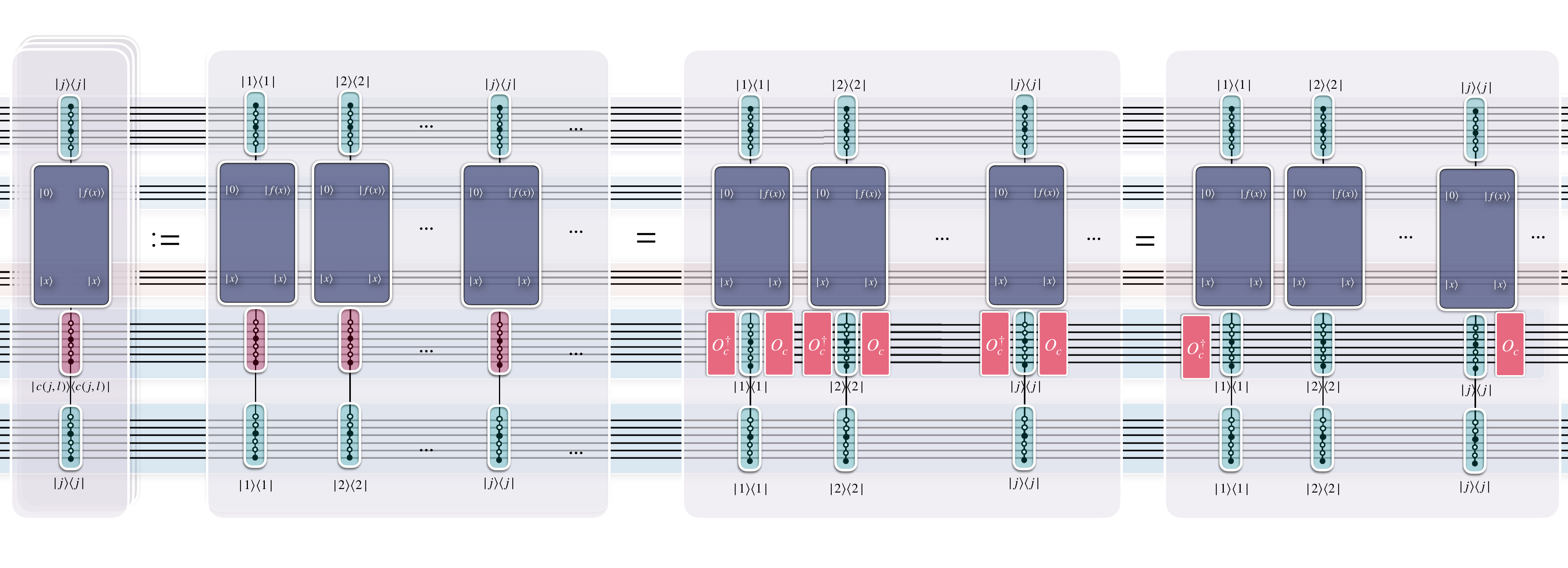}
   \caption{By piling up all the multi-controlled unitaries we see that cancellation happens in the middle.}
    \label{a3}
\end{figure}

\begin{figure}[h!]
    \centering
    \includegraphics[width=0.9\linewidth]{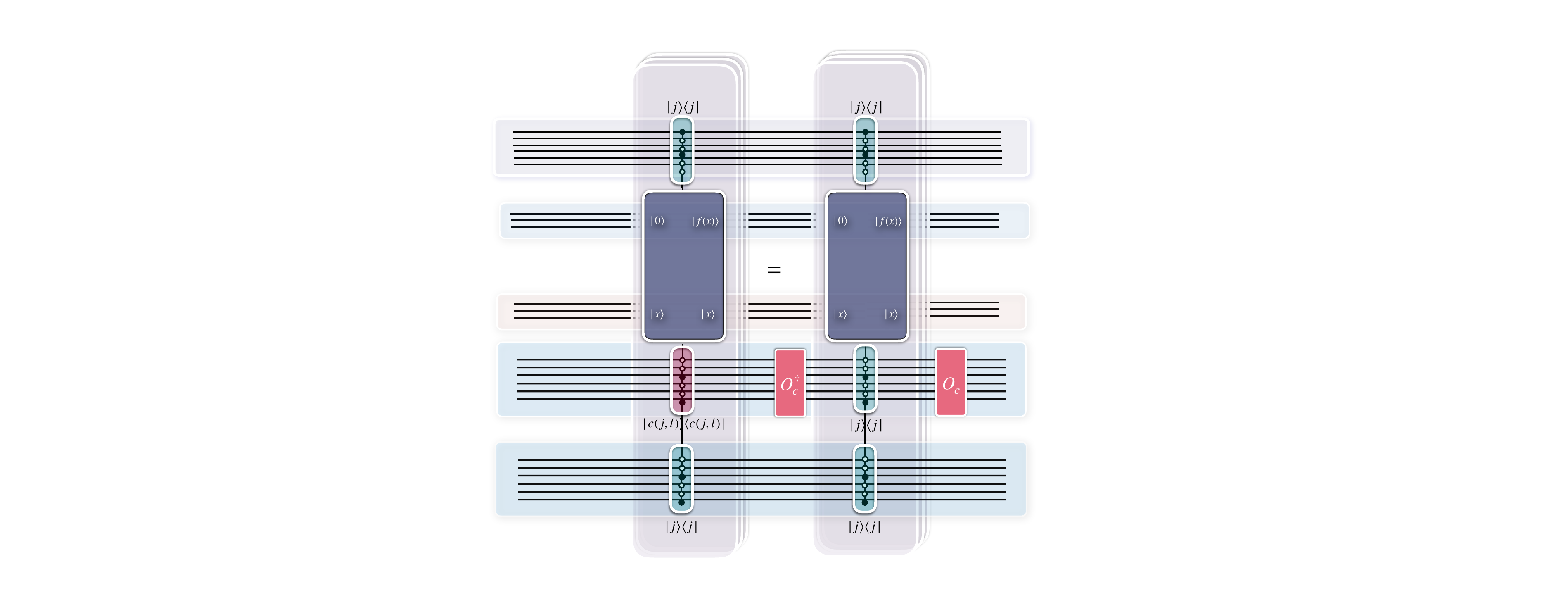}
    \caption{Result of piling up all the multi-controlled unitaries, cancellation happens as in Fig.\ref{a3}}
    \label{a4}
\end{figure}

Then by piling up all the multi-controlled unitaries, we see that cancellation happens in the middle as in Fig.\ref{a3} and we have the result depicted in Fig.\ref{a4}

\begin{figure}[h!]
    \centering
    \includegraphics[width=0.9\linewidth]{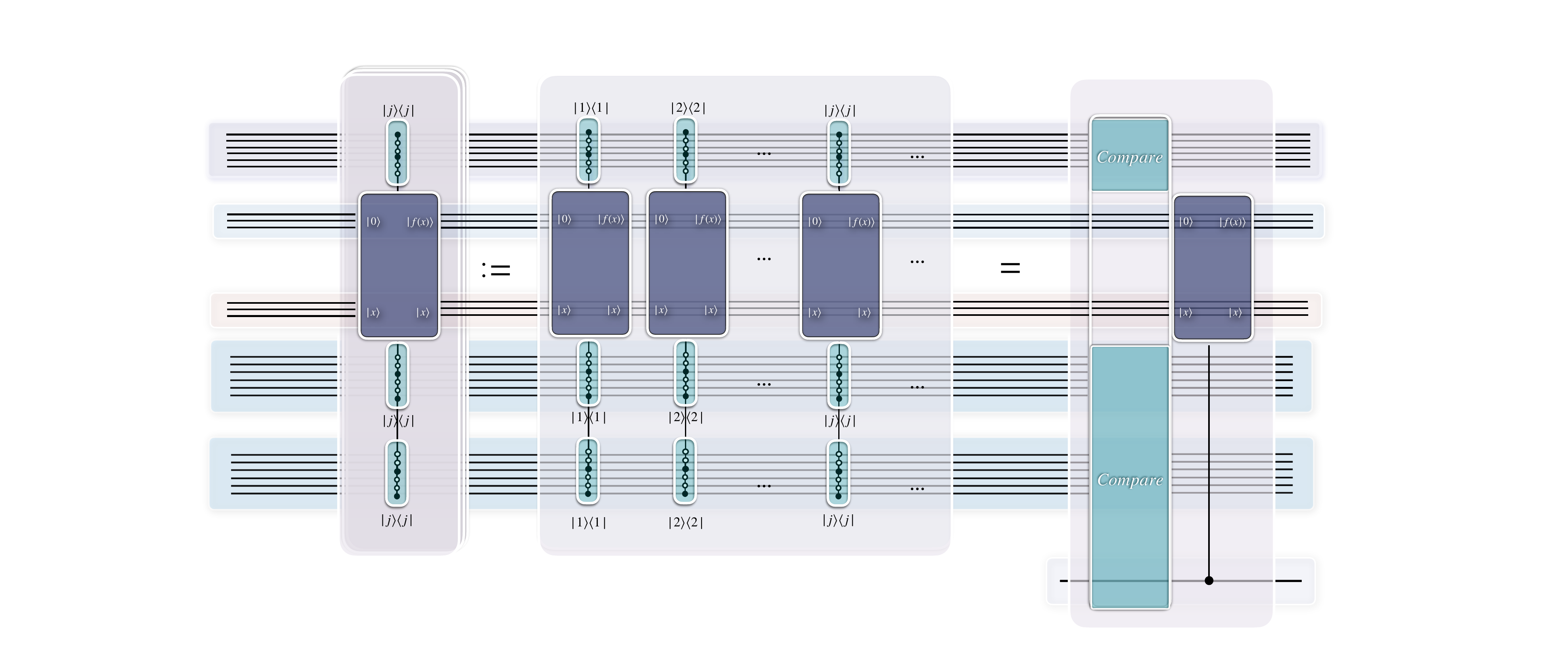}
    \caption{The stack on the right side in Fig.\ref{a4} can be implemented by a comparing unitary followed by a controlled copy}
    \label{a5}
\end{figure}

The stack on the right side can be implemented by a comparing unitary followed by a controlled copy, as depicted in Fig.\ref{a5}.

\section{Quantum Attention Mechanism}\label{attentionsec}

The quantum attention mechanism \cite{liao2024gpt} aims to coherently evaluate and store attention score $a(\bold{x}_i,\bold{x}_j)$ for each pair of the nodes, which can be defined as a quantum oracle $O_{\textsf{attention}}$ such that:
\begin{equation}
  O_{\textsf{attention}}  \ket{i}\ket{j}\ket{0} \to \ket{i}\ket{j}\ket{a(\bold{x}_i,\bold{x}_j)}
  \label{attentionoracle}
\end{equation}

In this section, we present the construction of the quantum attention oracle consisting of the following two steps:

\subsection{Evaluating Attention score in superposition}

The Attention score $a(\bold{x}_i,\bold{x}_j)$ in our Quantum Attention Mechanism can take one of the standard forms in classical literature \cite{ghojogh2020attention} --- the inner product of the linearly transformed feature vectors of each pair of nodes

\begin{equation}
	a(\bold{x}_i,\bold{x}_j)=\bold{x}_i^T W_K^T W_Q \bold{x}_j
	\label{attention-score }
	\end{equation}
, in which $W_K, W_Q$ are trainable linear transformations. \\

In terms of Dirac notation, this can be written as:

\begin{equation}
	a(\bold{x}_i,\bold{x}_j)=\bra{\bold{x}_i}U_K^{\dagger} U_Q \ket{\bold{x}_j}\end{equation}

in which $U_K, U_Q$ are trainable unitaries.\\

\begin{figure}[h!]
    \centering
    \includegraphics[width=\linewidth]{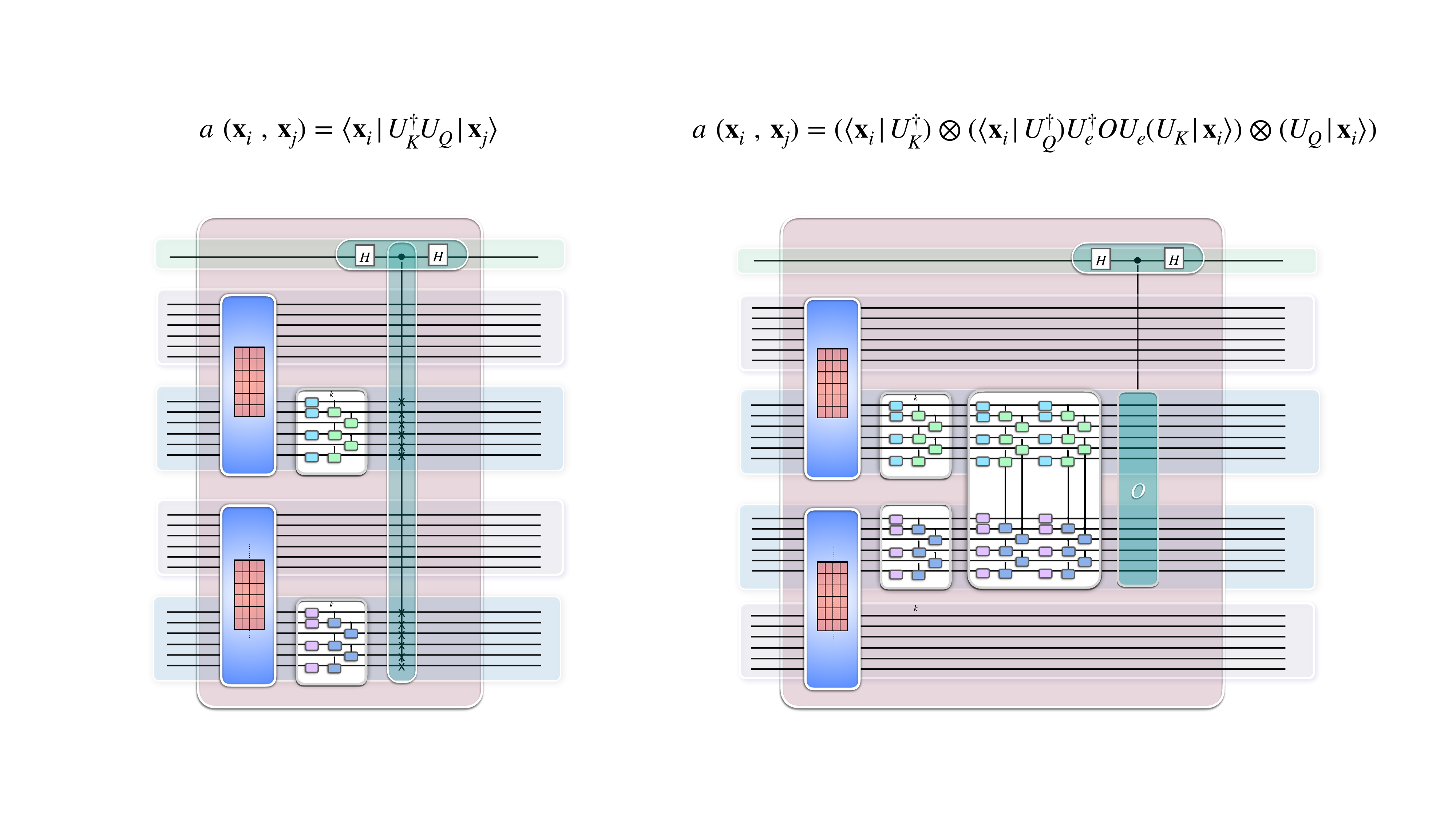}
   \caption{\textit{Quantum Attention Mechanism} The Attention score $a(\bold{x}_i,\bold{x}_j)$ in our Quantum Attention Mechanism can take the form of the inner product of the linearly transformed feature vectors of each pair of nodes $a(\bold{x}_i,\bold{x}_j)=\bold{x}_i^T W_K^T W_Q \bold{x}_j$, in which $W_K, W_Q$ are trainable linear transformations. In terms of Dirac notation, this can be written as: $a(\bold{x}_i,\bold{x}_j)=\bra{\bold{x}_i}U_K^{\dagger} U_Q \ket{\bold{x}_j}$, in which $U_K, U_Q$ are trainable unitaries. In our Quantum Attention Mechanism, this attention score can be evaluated in superposition on quantum circuit by parallel Swap Test, depicted as the left side of this figure. On the left side of this figure, we illustrate an alternative form of the Attention score, which can be evaluated by parallel Hadamard Test. }
    \label{att}
\end{figure}

In our Quantum Attention Mechanism, this attention score can be evaluated on quantum circuit by parallel Swap Test as depicted in Fig.~\ref{att} which we will discuss in detail below.\\

We denote the unitary for the parallel swap test circuit, as circled by the pink box on the left side of Fig.~\ref{att}, as $U$.
The input to $U$, $\ket{\Psi_{0}}$, can be written as (note here and throughout the paper, we omit the normalization factor):

\begin{equation}
\left|\Psi_{0}\right\rangle=\ket{0}\otimes(\sum_{i}\left|i\right\rangle)\otimes \left|0\right\rangle^{n}_\mathtt{K} \otimes(\sum_{j}\left|j\right\rangle)\otimes\left|0\right\rangle^{n}_\mathtt{Q} 
\end{equation}

where $\left|0\right\rangle^{n}_\mathtt{K}, \left|0\right\rangle^{n}_\mathtt{Q}$ are the initial states of two copies of data registers on which the node features  $a(\bold{x}_i,\bold{x}_j)$ will be loaded. The data encoding via Controlled Quantum State Preparation\cite{yuan2023optimal}, depicted as the blue boxes in Fig.\ref{att}, can be written as $\sum_{i}\left|i\right\rangle \bra{i}\otimes U_{\bold{x}_i}$ where $U_{\bold{x}_i}\ket{0}=\left|\bold{x}_i\right\rangle$.

Applying this data encoding on the two copies of data registers yields the overall state:

\begin{equation}
\left|\Psi_{1}\right\rangle=\ket{0}\otimes(\sum_{i}\left|i\right\rangle\otimes \left|\bold{x}_i\right\rangle^{n} )\otimes(\sum_{j}\left|j\right\rangle\otimes\left|\bold{x}_j\right\rangle^{n} )
\end{equation}

Node-wise linear transformation $U_K, U_Q$(trainable unitaries) implemented by PQC are then applied to the node feature registers, yielding the following state:

\begin{equation}
\left|\Psi_{2}\right\rangle=\ket{0}\otimes(\sum_{i}\left|i\right\rangle\otimes U_K \left|\mathbf{x}_i\right\rangle^{n} )\otimes(\sum_{j}\left|j\right\rangle\otimes U_Q\left|\bold{x}_j\right\rangle^{n} )
\end{equation}

We further define $\mathcal{K}_i,\mathcal{Q}_j$ and corresponding state $\ket{k_i}, \ket{q_j} $ as
$\mathcal{K}_i\ket{0}_\mathtt{K}^{n}=U_K \left|\mathbf{x}_i\right\rangle=\ket{k_i},\mathcal{Q}_j\ket{0}_\mathtt{Q}^{n}=U_Q\left|\mathbf{x}_j\right\rangle=\ket{q_j}$. Then $U$ can be written explicitly as

\begin{multline}
U\coloneqq[H\otimes I\otimes I\otimes I\otimes I]\cdot\\
[\ket{0}\bra{0}\otimes (\sum_{i}\sum_{j}\left|i \left>\right<i\right|\otimes \mathcal{K}_{i}\otimes\left|j \left>\right<j\right|\otimes \mathcal{Q}_{j} )+\ket{1}\bra{1}\otimes (\sum_{i}\sum_{j}\left|i \left>\right<i\right|\otimes \mathcal{Q}_{j}\otimes\left|j \left>\right<j\right|\otimes  \mathcal{K}_{i} )]\\\cdot[H\otimes I\otimes I\otimes I\otimes I],
\end{multline}

which can be rewritten as

\begin{align}
 U=\sum_{i}\sum_j\left|i \left>\right<i\right|\otimes \left|j \left>\right<j\right|\otimes U_{ij},
\end{align}

where

\begin{equation}
U_{ij}\coloneqq[H\otimes I\otimes I]\cdot
[\ket{0}\bra{0}\otimes  \mathcal{K}_{i}\otimes \mathcal{Q}_{j} +\ket{1}\bra{1}\otimes  \mathcal{Q}_{j}\otimes  \mathcal{K}_{i}\ ]\cdot[H\otimes  I\otimes I],
\end{equation}
\\

Define $\ket{\phi_{ij}} \coloneqq U_{ij}\ket{0}\left|0\right\rangle^{n}_{\mathtt{K}} \left|0\right\rangle^{n}_{\mathtt{Q}}$ and we have:
\begin{equation}
  |\phi_{ij}\rangle=\frac{1}{\sqrt{2}}(\ket{+}\ket{k_i}\ket{q_j}+\ket{-}\ket{q_j}\ket{k_i}).
  \label{phiij}
\end{equation}

Expanding and rearranging the terms in 
Eq.~\ref{phiij} we have
\begin{equation}
  |\phi_{ij}\rangle=\frac{1}{2}\left(|0\rangle\otimes(|k_i\rangle\ket{q_j}+|q_j\rangle\ket{k_i})+|1\rangle\otimes(|k_i\rangle\ket{q_j}-|q_j\rangle\ket{k_i}\right).
\label{sw}
\end{equation}

Denote $|u_{ij}\rangle$ and $|v_{ij}\rangle$ as the normalized states of $|k_i\rangle\ket{q_j}+|q_j\rangle\ket{k_i}$ and $|k_i\rangle\ket{q_j}-|q_j\rangle\ket{k_i}$ respectively.
Then there is a real number $\theta_{ij}\in[{\pi}/{4},{\pi}/{2}]$ such that
\begin{align}\label{amplitudeencoding}
|\phi_{ij}\rangle=\sin\theta_{ij}|0\rangle|u_{ij}\rangle+\cos\theta_{ij}|1\rangle|v_{ij}\rangle.
\end{align}
 $\theta_{ij}$ satisfies $\cos\theta_{ij}=\sqrt{1- |\langle k_i|q_j\rangle|^2}/\sqrt{2}$, $\sin\theta_{ij}=\sqrt{1+ |\langle k_i|q_j\rangle|^2}/\sqrt{2}$.\\

The final output state from $U$,  $\left|\Psi_{3}\right\rangle=U\left|\Psi_{0}\right\rangle $, can then be written as

\begin{equation}
\left|\Psi_{3}\right\rangle=\sum_{i}\sum_{j}|i\rangle|j\rangle (\underbrace{\left.\left.\sin \theta_{ij}\left|u_{ij}\right\rangle\ket{0}+\cos \theta_{ij}\left|v_{ij}\right\rangle|1\right\rangle\right)}_{\mid \phi_{ij}\rangle} = \sum_{i}\sum_{j}|i\rangle|j\rangle  \left|\phi_{ij}\right\rangle
\label{amplitudeencoding1}
\end{equation}

Note that $\langle k_i|q_j\rangle=\bra{\bold{x}_i}U_K^{\dagger} U_Q \ket{\bold{x}_j}=a(\bold{x}_i,\bold{x}_j)$ being the attention scores are encoded in the amplitudes of the output state $\left|\Psi_{3}\right\rangle$ of swap test as $|\langle k_i|q_j\rangle|^2=-\cos{2\theta_{ij}}$.

\subsection{Storing Attention score}\label{step2}

The second step is to use amplitude estimation \cite{2000quant.ph..5055B} to extract and store the attention scores into an additional register which we call the ``amplitude register''. \newline

After step 1, we introduce an extra register $\ket{0}^{t}_{\textsf{\textit{amplitude}}}$ and
the output state $\ket{\Psi_{3}}$ (using the same notation) becomes
\begin{align}
  \left|\Psi_{3}\right\rangle=\sum_{i}\sum_{j}|i\rangle|j\rangle  \left|\phi_{ij}\right\rangle\ket{0}^{t}_{\textsf{\textit{amplitude}}},
\end{align}
where $\ket{\phi_{ij}}$ can be decomposed as
\begin{equation}
\ket{\phi_{ij}}=\frac{-i}{\sqrt{2}}\left(e^{i \theta_{ij}}\ket{\omega_{+}}_{ij}-e^{i(-\theta_{ij})}\ket{\omega_{-}}_{ij}\right).
\end{equation}
Hence, we have
\begin{equation}
\left|\Psi_{3}\right\rangle=\sum_{i}\sum_{j}\frac{-i}{\sqrt{2}}\left( e^{i \theta_{ij}}|i\rangle\left|j\right\rangle\ket{\omega_{+}}_{ij}-e^{i(-\theta_{ij})}|i\rangle\left|j\right\rangle\ket{\omega_{-}}_{ij}\right) \ket{0}^{t}_{\textsf{\textit{amplitude}}}.
\label{phi1}
\end{equation}

\begin{figure}[h!]
    \centering
    \includegraphics[width=0.86\linewidth]{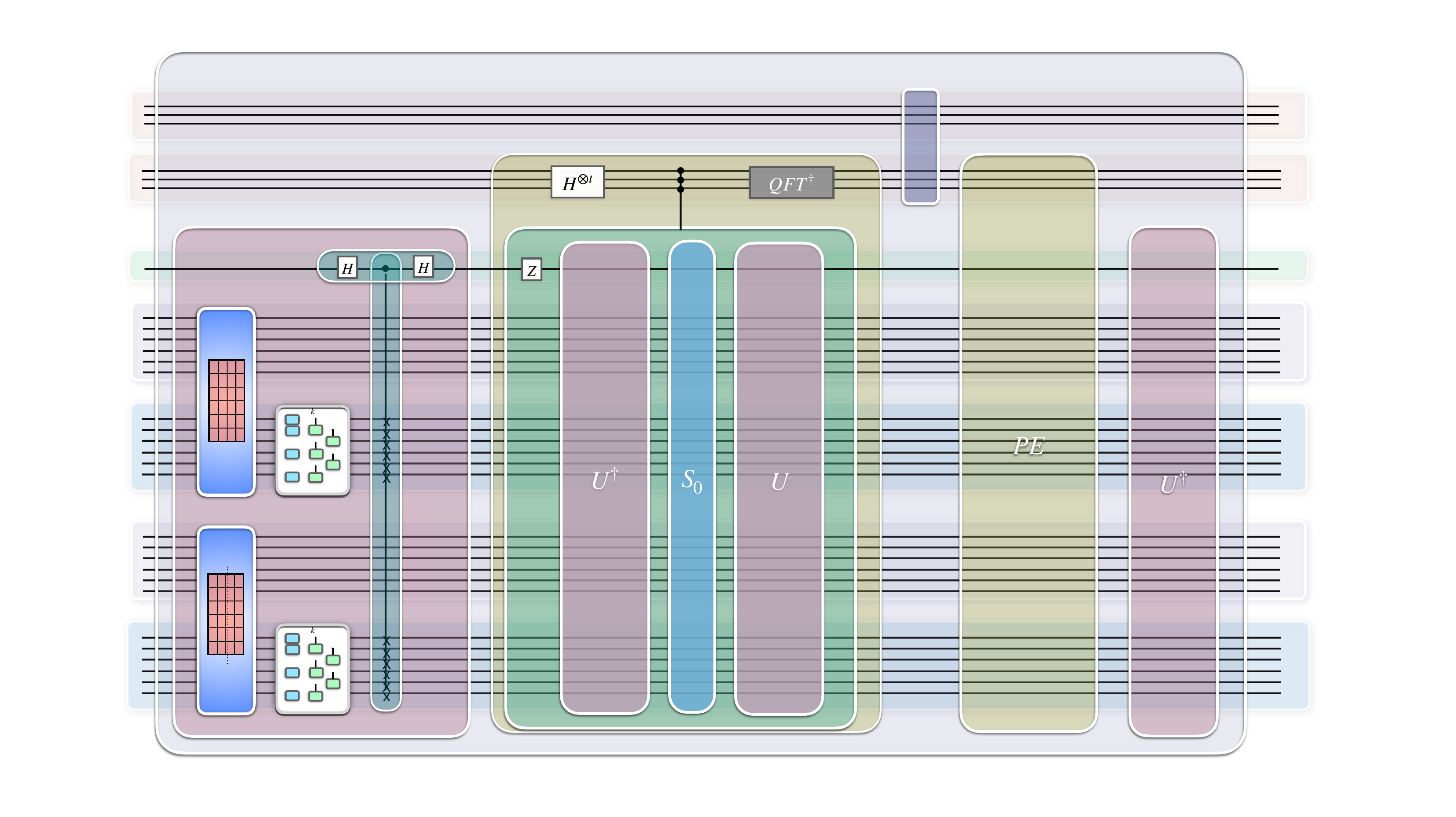}
   \caption{\textit{Quantum  attention oracle $O_{\textsf{attention}}$} The quantum attention mechanism aims to coherently evaluate and store attention score $a(\bold{x}_i,\bold{x}_j)$ for each pair of the nodes, which can be defined as a quantum oracle $O_{\textsf{attention}}$ such that $
  O_{\textsf{attention}}  \ket{i}\ket{j}\ket{0} \to \ket{i}\ket{j}\ket{a(\bold{x}_i,\bold{x}_j)}
$. The construction of the quantum attention oracle, depicted in this figure, is detailed in Appendix \ref{attentionsec}.}
    \label{att23}
\end{figure}

The overall Grover operator $G$ is defined as
\begin{equation}
G \coloneqq UC_{2}U^{\dagger}C_{1},
\label{grovero}
\end{equation}
where $C_{1}$ is the $Z$ gate on the swap ancilla qubit, and $C_{2}=I-2|0\rangle\langle0|$ is the ``flip zero state'' on registers other than the two registers hosting indices $i,j$ (represented as $S_0$ in Fig.\ref{att23}).
It can be shown that $G$ can be expressed as
\begin{align}
    G=\sum_{i}\sum_{j}\ket{i}\left|j \left>\right<j\right|\bra{i}\otimes G_{ij},
\end{align}
where $G_{ij}$ is defined as

\begin{equation}
    G_{ij}= (I-2|\phi_{ij}\rangle\langle\phi_{ij}|))C_{1}\end{equation}

It is easy to check that $|w_{\pm}\rangle_{ij}$ are the eigenstates of $G_{ij}$, that is,
\begin{align}\label{eq_G2}
G_{ij}|w_{\pm}\rangle_{ij}=e^{\pm\bm i  2\theta_{ij}}|w_{\pm}\rangle_{ij}.
\end{align}
The overall Grover operator $G$ possess the following eigen-relation:
\begin{align}
 G\ket{i}\left|j\right\rangle\ket{\omega_{\pm}}_{ij}=   e^{i (\pm 2\theta_{ij})}\ket{i}\left|j\right\rangle\ket{\omega_{\pm}}_{ij}.
\end{align}

Next, we apply phase estimation of the overall Grover operator $G$ on the input state
$\left|\Psi_{3}\right\rangle$. The resulting state $\left|\Psi_{4}\right\rangle$ can be written as
\begin{equation}\label{statephase2}
\left|\Psi_{4}\right\rangle=\sum_{i}\sum_{j}\frac{-i}{\sqrt{2}}\left( e^{i \theta_{ij}}\ket{i}\left|j\right\rangle\ket{\omega_{+}}_{ij}\ket{2\theta_{ij}}-e^{i(-\theta_{ij})}\ket{i}\left|j\right\rangle\ket{\omega_{-}}_{ij}\ket{-2\theta_{ij}}\right).
\end{equation}

Note here in Eq.~\ref{statephase2}, $\ket{\pm 2\theta_{ij}}$ denotes the eigenvalues $\pm 2\theta_{ij}$ being stored in the amplitude register with some finite precision.\newline

Next, we apply an oracle $U_{O}$ on the amplitude register and an extra ancilla register, which acts as
\begin{equation}
 U_{O}\ket{0}\ket{\pm 2\theta_{ij}}=\ket{a(\bold{x}_i,\bold{x}_j)}\ket{\pm2\theta_{ij}},
 \label{threshold}
\end{equation}

The state after the oracle can be written as

\begin{equation}
\left|\Psi_{5}\right\rangle=\sum_{i}\sum_{j}\frac{-i}{\sqrt{2}}\ket{a(\bold{x}_i,\bold{x}_j)}\left( e^{i \theta_{ij}}\ket{i}\left|j\right\rangle\ket{\omega_{+}}_{ij}\ket{2\theta_{ij}}-e^{i(-\theta_{ij})}\ket{i}\left|j\right\rangle\ket{\omega_{-}}_{ij}\ket{-2\theta_{ij}}\right).
\end{equation}

Then we perform the uncomputation of Phase estimation, the resulting state is
\begin{align}
\left|\Psi_{6}\right\rangle=\sum_{i}\sum_{j}\frac{-i}{\sqrt{2}}\ket{a(\bold{x}_i,\bold{x}_j)}\left( e^{i \theta_{ij}}\ket{i}\left|j\right\rangle\ket{\omega_{+}}_{ij}\ket{0}^{t}_{\textsf{\textit{amplitude}}}-e^{i(-\theta_{ij})}\ket{i}\left|j\right\rangle\ket{\omega_{-}}_{ij}\ket{0}^{t}_{\textsf{\textit{amplitude}}}\right)\\ =\sum_{i}\sum_{j}\ket{a(\bold{x}_i,\bold{x}_j)}|i\rangle|j\rangle  \left|\phi_{ij}\right\rangle\ket{0}^{t}_{\textsf{\textit{amplitude}}}
\end{align}

Finally, we perform the uncomputation of the swap test and the resulting state is
\begin{equation}
\left|\Psi_{7}\right\rangle=\sum_{i}\sum_{j}\ket{a(\bold{x}_i,\bold{x}_j)}|i\rangle|j\rangle  \left|0\right\rangle\ket{0}^{t}_{\textsf{\textit{amplitude}}}.
\label{dec}
\end{equation}

The above steps, as illustrated in Fig.~\ref{att23}, implemented the quantum attention oracle $O_{\textsf{attention}}$ such that:
\begin{equation}
  O_{\textsf{attention}}  \ket{i}\ket{j}\ket{0} \to \ket{i}\ket{j}\ket{a(\bold{x}_i,\bold{x}_j)}
\end{equation}

\section{Proof of the Layer-wise linear transformation for multi-channel GCN}\label{proofappendix}

From $H'^{(l)}=\hat{A} H^{(l)} W^{(l)}$ we have

\begin{equation}
   {H'^{(l)}}^T= {W^{(l)}}^T {H^{(l)}}^T {\hat{A}}^T 
   \label{proofe}
\end{equation}

Using $\textit{vec}(ABC)=(C^T\otimes A)\textit{vec}(B)$ (A, B,C are matrices), we have

\begin{equation}
     \textit{vec}({H'^{(l)}}^T)= \textit{vec}({W^{(l)}}^T {H^{(l)}}^T {\hat{A}}^T )
     \\=({\hat{A}}\otimes {W^{(l)}}^T)\textit{vec}({H^{(l)}}^T)
     \label{vec11}
\end{equation}

For an arbitrary matrix $M$, define vectors $\vec{\psi_{M}}=\textit{vec}(M)$, and Eqn.\ref{vec11} becomes

\begin{equation}
    \vec{\psi_{{H'^{(l)}}^T}}=({\hat{A}}\otimes {W^{(l)}}^T) \vec{\psi_{{H^{(l)}}^T}}
  \label{vec22}  
\end{equation}

Similar to Eqn.\ref{multistate}, we can define the quantum state on the two quantum registers $Reg(i)$ and $Reg(k)$ for $H^{(l)} $ as

\begin{equation}
\left|\psi_{H^{(l)}}\right\rangle=\sum_{i=1}^{N} |i\rangle\otimes\ket{\bold{x}^{(l)}_{i}}
\label{multistate2}
\end{equation}

and for $H'^{(l)}$:

\begin{equation}
\left|\psi_{H'^{(l)}}\right\rangle=\sum_{i=1}^{N} |i\rangle\otimes\ket{\bold{x'}^{(l)}_{i}}
\label{multistate3}
\end{equation}

Writing the quantum states in Eqn.\ref{multistate2} and \ref{multistate3} as vectors we note that

\begin{equation}
\vec{\psi_{{H^{(l)}}^T}}=\left|\psi_{H^{(l)}}\right\rangle
\label{tensor11}
\end{equation}

\begin{equation}
\vec{\psi_{{H'^{(l)}}^T}}=\left|\psi_{H'^{(l)}}\right\rangle
\label{tensor112}
\end{equation}

\begin{figure}[h!]
    \centering
    \includegraphics[width=\linewidth]{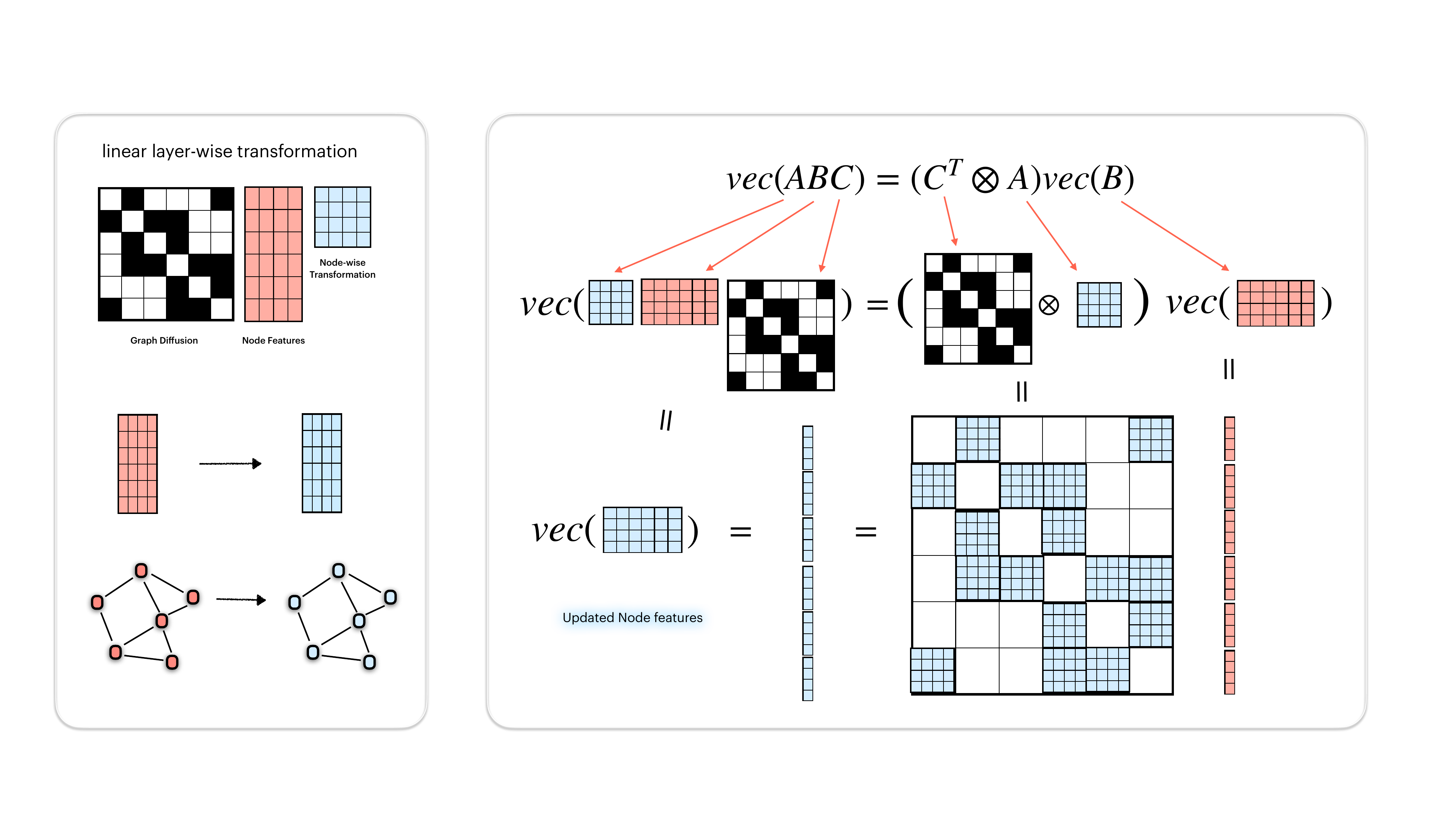}
   \caption{\textit{Proof of our Quantum implementation of linear layer-wise transformation for multi-channel GCN} The linear layer-wise transformation for multi-channel GCN (i.e. the layer-specific trainable weight matrix and adjacency matrix multiplied on the node feature matrix), can be implemented by applying the block-encoding of the normalized adjacency matrix and a parametrized quantum circuit on the two quantum registers $Reg(i)$ and $Reg(k)$ respectively. The figure summarises the proof from Eqn.\ref{proofe} to \ref{v11}. Note that the schematics in this figure are for illustration purposes only, e.g. the normalized adjacency matrix depicted here does not include the added self-connections.}
    \label{proof}
\end{figure}

From Eqn.\ref{multistate2}, \ref{multistate3}, \ref{tensor11}, \ref{tensor112} and \ref{vec22} we have

\begin{equation}
  \left|\psi_{H'^{(l)}}\right\rangle=({\hat{A}}\otimes {W^{(l)}}^T) \left|\psi_{H^{(l)}}\right\rangle
  \label{v11}  
\end{equation}

in which $({\hat{A}}\otimes {W^{(l)}}^T)$
corresponds to applying the block-encoding of ${\hat{A}}$ and a parameterized quantum circuit implementing ${W^{(l)}}^T$ on the two quantum registers $Reg(i)$ and $Reg(k)$ respectively. That is, $H'^{(l)}=\hat{A} H^{(l)} W^{(l)}$ --- the layer-specific trainable weight matrix and normalized adjacency matrix multiplied on the node feature matrix can be implemented by applying the block-encoding of the normalized adjacency matrix and a parameterized quantum circuit on the two quantum registers $Reg(i)$ and $Reg(k)$ respectively. The proof can be summarised in Fig.~\ref{proof}.

\begin{figure}[h!]
    \centering
    \includegraphics[width=\linewidth]{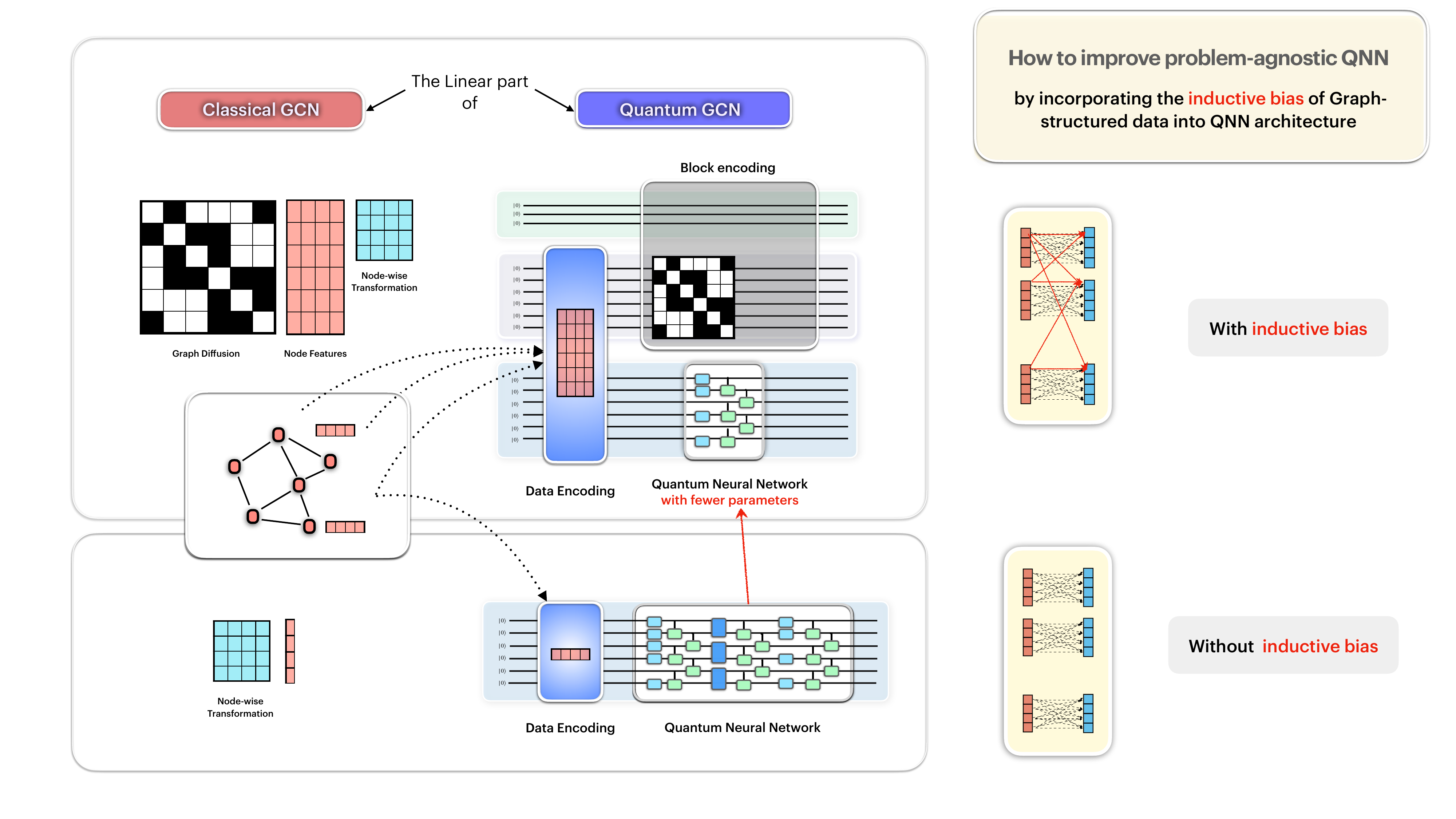}
    \caption{Our work also falls within the emerging field of Geometric Quantum Machine Learning (GQML) \cite{ragone2022representation,larocca2022group, meyer2022exploiting, zheng2021speeding, sauvage2022building}, which aims to create quantum machine learning models that respect the underlying structure and symmetries of the data they process. To illustrate how our frameworks align with the principles of GQML, we present an overview of our approach for Quantum Graph Convolutional Networks, summarized in this figure. This example demonstrates how our Quantum GNNs incorporate inductive biases to process graph-structured data, potentially leading to improvements compared to problem-agnostic quantum machine learning models. We start with the scenario where neural networks (classical and quantum) process data without inductive bias, depicted as the lower part of this figure. In this scenario, the classical and quantum neural networks process each  data point individually without acknowledging the connections between them. Here for a classical neural network, we depicted a linear layer represented as a matrix acting on a single data point as a vector. For the quantum neural network, we depicted a parametrized quantum circuit for implementing the linear layer. In the upper part of this figure, we illustrate the scenario where classical and quantum GNNs process data with inductive bias of graph-structured data. In this scenario, the classical and quantum GNN process all the data points for every node on a graph, with cross-node connections between them. Here for classical GNN, we depicted the layer-wise linear transformation for multi-channel Graph Convolutional Networks: the trainable weight matrix(for node-wise transformation) and the normalized adjacency matrix(for Graph diffusion) multiplied on the node feature matrix. In our Quantum GNN Architecture, this layer-wise linear transformation is implemented by applying the block-encoding of the normalized adjacency matrix and a parameterized quantum circuit following a data encoding procedure. By incorporating the inductive bias into the architecture, our Quantum GNN can potentially operate with fewer parameters than its problem-agnostic counterpart. This can potentially lead to more efficient training and less overfitting, improving the problem-agnostic QNNs. Note that the schematics in this figure are for illustration purposes only, e.g. 1) the normalized adjacency matrix depicted here does not include the added self-connections; 2) the ancillary qubits used in the quantum state preparation for the data encoding is not depicted in this figure.}
    \label{frgnn}
\end{figure}

\section{Brief Introduction of Quantum Neural Networks and Block-encoding}\label{Preliminaries}

\tocless\subsection{Quantum Neural Networks}\label{para}

Classical neural networks are fundamentally built on the structure of multi-layer perceptrons which involve layers of trainable linear transformations and element-wise non-linear transformations (activation functions such as ReLU, sigmoid, or tanh).\footnote{we assume the readers of this paper are familiar with classical neural networks. For reference of classical neural networks, see e.g. \cite{goodfellow2016deep}}  On the other hand, Quantum Neural Networks (QNNs), which are often defined as parametrized quantum circuits with a predefined circuit ansatz, do not naturally exhibit this kind of structure. In QML literature, a QNN, denoted as $U(\thv)$, often have a has an  $L$-layered structure of the form \cite{larocca2021theory}
\begin{equation}\label{eq:PSA_ansatz}
    U(\thv)=\prod_{l=1}^LU_l(\thv_l)\,, \quad U_l(\thv_l)=\prod_{k=1}^K e^{-i \theta_{lk}H_k}\,,    
\end{equation}
where the index $l$ represents the layer, and the index $k$ covers the Hermitian operators $H_k$ that generates the unitaries in the circuit ansatz, $\thv_{l}=(\theta_{l1},\ldots\theta_{lK})$ represents the parameters in a single layer, and $\thv=\{\thv_{1},\ldots,\thv_L\}$ represents the collection of adjustable parameters in the QNN. Examples of circuit ansatz represented by Eq.~\ref{eq:PSA_ansatz} include: the hardware-efficient ansatz~\cite{kandala2017hardware}, quantum alternating operator ansatz~\cite{hadfield2019quantum}, and quantum optimal control Ansatz~\cite{choquette2020quantum}, among others. \newline

The emulation of classical perceptrons with non-linearities in quantum circuits is an area of active research. There are several proposals for how this might be achieved \cite{Cao2017QuantumNeuron, Torrontegui2019UnitaryQuantumPerceptron, Schuld2015QuantumPerceptron}, they often involve intricate methods of encoding information into quantum states and performing measurements. The difficulty arises from the need to replicate the non-linear characteristics of classical neural networks within the linear framework of quantum mechanics.\newline

In this paper, we utilize the conventional QNN as in Eq.~\ref{eq:PSA_ansatz} for implementing some trainable linear transformations in classical neural networks, and we
extend the conventional notion of a QNN to a broadly structured quantum circuit containing parameterized quantum circuits (represented by Eq.~\ref{eq:PSA_ansatz})  as its components, that is, parameterized quantum circuits (which contain only parameterized gates) sit within a broader non-parameterized quantum circuit. \newline
\tocless\subsection{Block-encoding}\label{blocke}

\begin{figure}[h!]
    \centering
    \includegraphics[width=\linewidth]{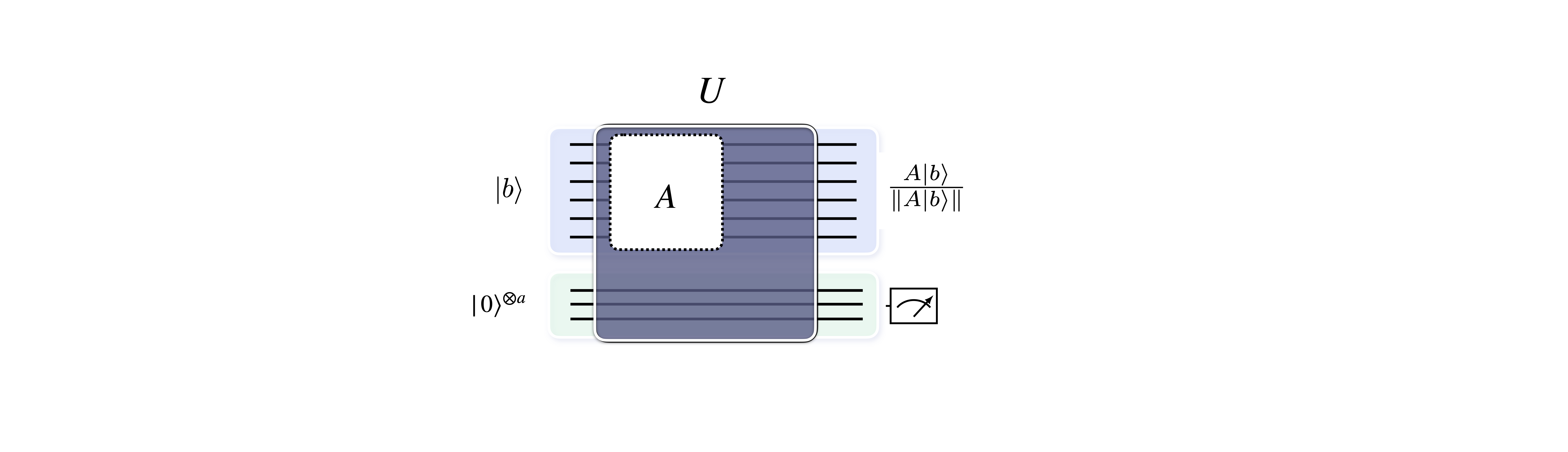}
   \caption{\textit{Block-encoding } \(U\), the Block-encoding of a matrix A, can be considered as a probabilistic implementation of \(A\): applying the unitary \(U\) to a given input state \(|0\rangle^{\otimes a}|b\rangle\), measuring the first \(a\)-qubit register and post-selecting on the outcome \(|0\rangle^{\otimes a}\), we get state proportional to \( A|b\rangle\) in the second register.}
    \label{Block}
\end{figure}

Block encoding is a powerful modern quantum algorithmic technique that is employed in a variety of quantum algorithms for solving linear algebra problems on a quantum computer \cite{sunderhauf2024block}.  A unitary $U$ is a block encoding of a not-necessarily-unitary square matrix $A$ ($A$ is scaled to satisfy $\|A\|_2\leq 1$) if $A$ is encoded in the top-left block of the unitary $U$ as:
$$
U=\left[\begin{array}{cc}
A & . \\
\cdot & \cdot
\end{array}\right],
$$

where the $\cdot$ symbol stands for a matrix block. Equivalently, we can write 

\begin{equation}
    A=\left(\left\langle 0\right|^{\otimes a} \otimes I\right) U\left(|0\rangle^{\otimes a} \otimes I\right) 
\end{equation}

where $a$ is the number of ancilla qubits used for the block encoding of $A$. $U$ can be considered as a probabilistic implementation of $A$: by applying the unitary $U$ to an input state $ |0\rangle^{\otimes a} |b\rangle$, measuring the first $a$-qubit register and post-selecting on the outcome $ |0\rangle^{\otimes a}$, we obtain a state that is proportional to $A |b\rangle$ in the second register. This can be illustrated in Fig.~\ref{Block}. \newline

\section{Comparisons with some related works}\label{relatedgcn}

In the niche of quantum graph \textit{convolutional} neural networks, we can compare our work (specifically, quantum GCN/SGC/LGC) with three other related works:\newline

Hu et al.~\cite{hu2022novel} designed a quantum graph convolutional neural network for semi-supervised node classification. While both works design quantum circuits to implement the graph convolutional neural network, there are significant differences in the approaches. For data encoding, Hu et al. use $N$ separate quantum circuits ($N$ is the number of nodes), with each circuit encoding the features of a single node. In contrast, our work coherently encodes all $N$ node features into a single quantum state on two entangled registers. For node-wise transformations, Hu et al. apply $N$ subsequent parameterized quantum circuits (PQCs) acting on each of the separate circuits to implement the trainable weight matrix. For aggregation over neighborhood nodes, they first perform measurements on all $N$ separate circuits to obtain the transformed node features, then regroup the features into different channels. For each channel, an $N$-dimensional vector is encoded into the amplitudes of a quantum state, resulting in $C$ separate quantum circuits ($C$ is the number of features/channels per node). They then utilize Givens rotations to perform aggregation over neighborhood nodes. In contrast, thanks to our data encoding scheme, we are able to simultaneously apply the block encoding of the normalized adjacency matrix (and further, QSVT for spectral convolution) and a single PQC for node-wise transformation on the two entangled registers, achieving both node-wise transformation and aggregation over neighborhood nodes simultaneously. Furthermore, their complexity analysis focuses on time complexity, while our analysis reveals a trade-off between time and space complexity.\newline

Zheng et al. \cite{zheng2021quantum} proposed a quantum graph convolutional neural network model to accomplish graph-level classification tasks. While both works aim to develop quantum versions of GCNs, there are several key differences. For data encoding, Zheng et al. use separate quantum circuits for each node, whereas our work coherently encodes all node features into a single quantum state on two entangled registers. Moreover, Zheng et al. focus on graph classification tasks, while our work focuses on node classification tasks (although our architectures are also well suited for graph classification tasks). Furthermore, our work provides complexity analysis demonstrating potential quantum advantages, while Zheng et al. focus on numerical simulations without theoretical analysis.\newline

Chen et al. \cite{chen2022novel} proposed a parameterized quantum circuit architecture for quantum graph convolutional networks. Although both works design quantum circuits for implementing the adjacency matrix and the learnable weight matrix, the approaches differ. For aggregation over neighborhood nodes, Chen et al. use LCU to implement the adjacency matrix. In contrast, we utilize block encoding for the normalized adjacency matrix which enables the usage of QSVT for spectral graph convolutions and the corresponding higher order neighborhood propagation. (Although LCU effectively implements certain block-encoding, but a general block-encoding can accommodate more matrix--without the limitations of the approach used by Chen et al.) Notably, for cost-function evaluation, we only perform measurements on a single ancillary qubit, whereas they perform measurements on all the qubits. Another differentiation is that for implementation of the nonlinear activation function, we utilize NTCA for a two-layer GCN.  While Chen et al. evaluate their QGCN on certain benchmarking dataset, our work focuses on the theoretical aspects and performed rigorous complexity analysis, revealing potential quantum advantages in terms of time and space complexity.\newline

\bibliography{Ref.bib}

\end{document}